\documentclass[longauth]{aa} 

\usepackage{lscape}
\usepackage[varg]{txfonts}
\usepackage[normalem]{ulem}

\usepackage{graphicx}
\usepackage{longtable,lscape}
\usepackage{natbib}
\usepackage{amsmath}
\newcommand*\tess{\textit{TESS }}

\newcommand{\logrhk}{$\rm log\,R^{\prime}_\mathrm{HK}$}
\newcommand{\kms}{\,km\,s$^{-1}$} 
\newcommand{\ms}{\,m\,s$^{-1}$} 

\newcommand{\mstar}{M$_{\star}$}
\newcommand{\rstar}{R$_{\star}$}
\newcommand{\lstar}{L$_{\star}$}

\newcommand{\lsun}{$L_{\odot}$}

\newcommand{\vsini}{$v$\,sin\,$i_\star$}

\newcommand{\teff}{$T_{\rm eff}$}
\newcommand{\logg}{log\,{\it g$_\star$}}
\newcommand{\feh}{[Fe/H]}
\newcommand{\tih}{[Ti/H]}

\newcommand{\prot}{$P_{\rm rot}$}

\begin{document} 

\title{The GAPS Programme with HARPS-N at TNG.}
   \subtitle{XXXVII. A precise density measurement of the  young ultra-short period planet TOI-1807~b\thanks{Based on observations made with the Italian {\it Telescopio Nazionale Galileo} (TNG) operated by the {\it Fundaci\'on Galileo Galilei} (FGG) of the {\it Istituto Nazionale di Astrofisica} (INAF) at the {\it  Observatorio del Roque de los Muchachos} (La Palma, Canary Islands, Spain).}}


\author{D.~Nardiello\inst{1,2}
\and
L.~Malavolta\inst{3,1}            
\and
S.~Desidera\inst{1}
\and
M.~Baratella\inst{4}
\and
V.~D'Orazi\inst{1}
\and
S.~Messina\inst{5}
\and
K.~Biazzo\inst{6}
\and
S.~Benatti\inst{7}
\and
M.~Damasso\inst{8}
\and
V.~M.~Rajpaul\inst{9} 
\and
A.~S.~Bonomo\inst{8}
\and
R.~Capuzzo~Dolcetta\inst{10}
\and
M.~Mallonn\inst{4}
\and
B.~Cale\inst{11}
\and
P.~Plavchan\inst{11}
\and
M.~El Mufti\inst{11}
\and
A.~Bignamini\inst{12}
\and
F.~Borsa\inst{13}
\and
I.~Carleo\inst{14}
\and
R.~Claudi\inst{1}
\and
E.~Covino\inst{15}
\and
A.~F.~Lanza\inst{5}
\and
J.~Maldonado\inst{7}
\and
L.~Mancini\inst{16,17,8}
\and
G.~Micela\inst{7}
\and
E.~Molinari\inst{18}
\and
M.~Pinamonti\inst{8}
\and
G.~Piotto\inst{3,1}
\and
E.~Poretti\inst{19,13}
\and
G.~Scandariato\inst{5}
\and
A.~Sozzetti\inst{8}
\and
G.~Andreuzzi\inst{19,6}
\and
W.~Boschin\inst{19,20,21}
\and
R.~Cosentino\inst{19}
\and
A.~F.~M.~Fiorenzano\inst{19}
\and
A.~Harutyunyan\inst{19}
\and
C. Knapic\inst{12}
\and
M.~Pedani\inst{19}
\and
L.~Affer\inst{7}
\and
A.~Maggio\inst{7}
\and
M.~Rainer\inst{13}
}

   \institute{
INAF -- Osservatorio Astronomico di Padova, Vicolo dell'Osservatorio 5, 35122 -- Padova, Italy\\
\email{domenico.nardiello@inaf.it}
\and
Aix Marseille Univ, CNRS, CNES, LAM, Marseille, France 
\and
Dipartimento di Fisica e Astronomia ``Galileo Galilei'' – Universit\`{a} di Padova, Vicolo dell’Osservatorio 3, 35122 -- Padova, Italy 
\and
Leibniz-Institut f\"ur Astrophysik Potsdam (AIP), An der Sternwarte 16, 14482 Potsdam, Germany
\and
INAF -- Osservatorio Astrofisico di Catania, via S. Sofia 78, 95123 -- Catania, Italy
\and
INAF -- Osservatorio Astronomico di Roma, Via Frascati 33, 00078 -- Monte Porzio Catone (Roma), Italy
\and
INAF -- Osservatorio Astronomico di Palermo, Piazza del Parlamento, 1, 90134 -- Palermo, Italy 
\and
INAF -- Osservatorio Astrofisico di Torino, via Osservatorio 20, 10025 -- Pino Torinese, Italy
\and
Astrophysics Group, Cavendish Laboratory, JJ Thomson Avenue, Cambridge CB3 0HE, UK
\and
Dipartimento di Fisica - Universit\`{a} di Roma La Sapienza, P.le A.~Moro 5, 00185 -- Roma, Italy
\and
Department of Physics and Astronomy, George Mason University, 4400 University Drive, Fairfax, VA 22030, USA
\and
INAF -- Osservatorio Astronomico di Trieste, via Tiepolo 11, 34143 -- Trieste, Italy
\and
INAF -- Osservatorio Astronomico di Brera, Via E. Bianchi 46, 23807 -- Merate (LC), Italy
\and
Astronomy Department, Indiana University, Bloomington, IN 47405-7105, USA
\and
INAF -- Osservatorio Astronomico di Capodimonte, Salita Moiariello 16, 80131 -- Naples, Italy
\and
Department of Physics, University of Rome ``Tor Vergata'', Via della Ricerca Scientifica 1, 00133 -- Rome, Italy
\and
Max Planck Institute for Astronomy, K\"{o}nigstuhl 17, 69117 -- Heidelberg, Germany
\and
INAF -- Osservatorio Astronomico di Cagliari, via della Scienza 5, 09047 -- Selargius (CA), Italy
\and
Fundaci\'{o}n Galileo Galilei - INAF, Rambla Jos\'{e} Ana Fernandez P\'{e}rez 7, 38712 Bre\~{n}a Baja, TF, Spain
\and
Instituto de Astrofísica de Canarias, C/V\'ia Lactea s/n, E-38205 La Laguna (Tenerife), Spain
\and
Departamento de Astrof\'isica, Univ. de La Laguna, Av. del Astrof\'isico F. S\'anchez, s/n, E-38205 La Laguna (Tenerife), Spain
}
\date{Received April 08, 2022; accepted June 03, 2022}

 
\abstract
{Great strides have been made in recent years in the
    understanding of the mechanisms involved in the formation and
    evolution of planetary systems; despite this, many observational
    facts still do not have an explanation. A great contribution to
    the study of planetary formation processes comes from the study of
    young, low-mass planets, with short orbital periods ($\lesssim$100
    days).   In the last three
  years, the NASA/\tess satellite has identified many planets of this
  kind, and their characterization is clearly mandatory to understand
  how they formed and evolved.}
   {Within the framework of the Global Architecture of Planetary
     System (GAPS) project, we performed the validation and
     characterization (radius and mass) of the ultra-short period
     planet TOI-1807~b, orbiting its young host star BD+39~2643 ($\sim
     300$~Myr) in only 13~hours. This is the youngest ultra-short
     period planet discovered so far.}
   {Thanks to a joint  modeling of the stellar activity and
     planetary signals in the \tess light curve and in new HARPS-N
     radial-velocity measurements, which were combined with accurate
     estimation of stellar parameters, we validated the planetary
     nature of TOI-1807~b and measured its orbital and physical
     parameters.}
   {By using astrometric, photometric, and spectroscopic observations
     we found that BD+39~2643 is a young, active K dwarf star, member
     of a $300\pm 80$~Myr old moving group and that it
     rotates in $P_{\rm rot}=8.8 \pm 0.1$~days. This star hosts an
     ultra-short period planet showing an orbital period of only
     $P_{\rm b}=0.54937 \pm 0.00001$~days. Thanks to the exquisite
     photometric and spectroscopic series, and the accurate
     information on the stellar activity, we measured both the radius
     and the mass of TOI-1807~b with high precision, obtaining $R_{\rm
       P,b}=1.37\pm0.09~R_{\oplus}$ and   $M_{\rm
       P,b}=2.57\pm0.50~M_{\oplus}$. These planet parameters
     correspond to a rocky planet with an Earth-like density
      ($\rho_{\rm b}=1.0\pm0.3~\rho_{\oplus}$) and no extended H/He
     envelope. From the analysis of the age-$R_{\rm P}$ distribution
     for planets with well measured ages, we inferred
      that TOI-1807~b may have already lost a large
     part of its atmosphere during its 300~Myr life.  }
   {}

   \keywords{ planets and satellites: fundamental parameters -- stars: fundamental parameters -- stars: individual: BD+39~2643 -- technique: photometric -- technique: spectroscopic -- technique: radial velocities      }

   \maketitle
%

\section{Introduction}

In the last decade, the space missions {\it Kepler}
(\citealt{2010Sci...327..977B}), {\it K2}
(\citealt{2014PASP..126..398H}) and \tess (Transiting Exoplanet Survey
Satellite, \citealt{2015JATIS...1a4003R}) have allowed us to identify
a large number of transiting planets and planetary systems with ages
younger than 1 Gyr. The study of these young planets is important to
better understand the mechanisms that come into play in the early
stages of planetary formation and evolution, such as migration,
planetary impacts, atmospheric evaporation, etc. (see, e.g.,
\citealt{2007ApJ...654.1110T, 2010ApJ...719..810I,
  2012ApJ...751..158H, 2013ApJ...776....2L, 2013ApJ...775..105O,
  2015Icar..247...81S, 2018MNRAS.479.5012O,
  2018haex.bookE.141S,2019NatAs...3..416B}). Identifying and
characterizing planets around young stars is not always an easy task,
because of the magnetic activity of the host star. Indeed, the
  strong starspots' activity on the stellar surface (typical of young
  stars with ages $\lesssim 1$~Gyr) generates (periodical) important
  variations of the star's flux both in the photometric and
  spectroscopic time series, that can mask the planets' signals if not
  appropriately modeled. Despite this, in recent years, the
population of young planets has been growing significantly; planets
have been found around young stellar cluster members like M44 and the
Hyades
(\citealt{2012ApJ...756L..33Q,2014ApJ...787...27Q,2016A&A...588A.118M,
  2016ApJ...818...46M,2018AJ....156..195R, 2018AJ....156...46V}),
young stellar associations and moving groups
(\citealt{2019AA...630A..81B, 2020AJ....160..179M,
  2020AJ....160...33R, 2021AJ....161..171T, 2021AJ....161...65N,
  2021arXiv211009531M}), and single young stars
(\citealt{2019ApJ...885L..12D, 2020Natur.582..497P,
  2021A&A...645A..71C, 2021arXiv211214776B}).  However, only a
  few sample of young exoplanets with ages $<0.5-1$~Gyr have
  well-constrained ages, (upper-limit) masses, and radii, showing a
  large variety of planets with densities included between
  $<$0.5-1~$\rho_{\oplus}$ (see, e.g., \citealt{2021A&A...650A..66B})
  and 2-3~$\rho_{\oplus}$ (see, e.g,
  \citealt{2019MNRAS.490..698B,2022MNRAS.tmp..699B}).

Ultra-short period (USP) planets are planetary objects with orbital
periods $P\lesssim 1$~day (\citealt{2006Natur.443..534S}). These 
  ``exotic'' objects can provide many information on the processes of
formation and evolution of planetary systems, especially if observed
orbiting young stars. Even if this kind of planets is easy to detect
because of the short orbital period, only about a hundred of them have
been discovered so far, with an occurrence rate of 0.4-0.5 \%
(\citealt{2014ApJ...787...47S, 2018NewAR..83...37W,
  2021ARA&A..59..291Z}). The large part of USP planets have radii
$R_{\rm P} < 2 R_{\oplus}$, and they could be the remnant rocky cores
of gaseous giant planets after processes of atmospheric evaporation or
could be rocky planets migrated inward the planetary systems
(\citealt{2013MNRAS.431.3444C, 2014ApJ...793L...3V,
  2017ApJ...842...40L, 2019AJ....157..180P}).

To date, we have not yet identified USP planets around very young
stars (<100 Myr), and therefore we still lack information on the very
early stages of formation and evolution of this kind of planets. In
the context of the Global Architecture of Planetary Systems (GAPS) -
Young Objects project (see \citealt{carleo2020}), we are monitoring
with HARPS-N (\citealt{2012SPIE.8446E..1VC}) at Telescopio Nazionale
Galileo (TNG) young stars (2-600 Myr) with the aim of identifying and
characterizing young planets around them, including USP planets. In
this paper we present the results obtained for the youngest USP planet
discovered so far: with an age of $\sim 300$ Myr, TOI-1807~b orbits
the K dwarf BD+39~2643 (TOI-1807) in about 0.55~day. The manuscript we
present is structured as follows: Section~\ref{sec:obs} describes the
photometric and spectroscopic data used in this work to characterize
the host star and the planet; Section~\ref{sec:stellar} reports the
stellar properties. The procedures adopted to identify and confirm
TOI-1807~b in the \tess light curve are explained in
Section~\ref{sec:planetdet}. The modeling of photometric and RV
series and the planet's properties are reported in
Section~\ref{sec:toi1807prop}. In Section~\ref{sec:sum} we discuss and
summarize the results obtained in this work.

\begin{figure}
  \centering
  \frame{\includegraphics[width=0.485\textwidth]{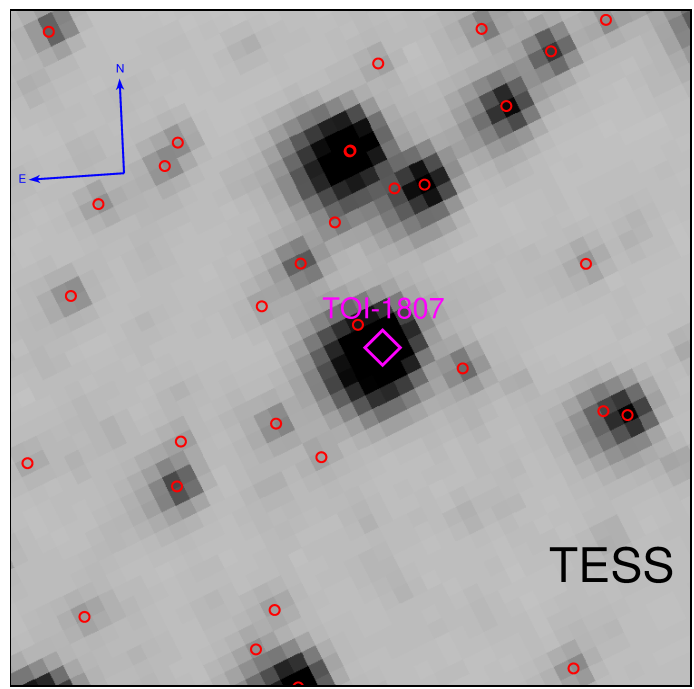}} \\ 
  \caption{Finding chart of TOI-1807 (magenta diamond) from the
      TESS FFI \texttt{tess2020090105920-s0023-2-2-0177-s\_ffic}:
      red circles are the neighbor stars in the Gaia eDR3 catalog with $G
      \leq 17$. The field of view is
      15$\times$15~arcmin$^2$, and it is oriented with north up and east left. . \label{fig:1r}}
\end{figure}

\begin{figure*}
  \centering
  \includegraphics[width=0.93\textwidth]{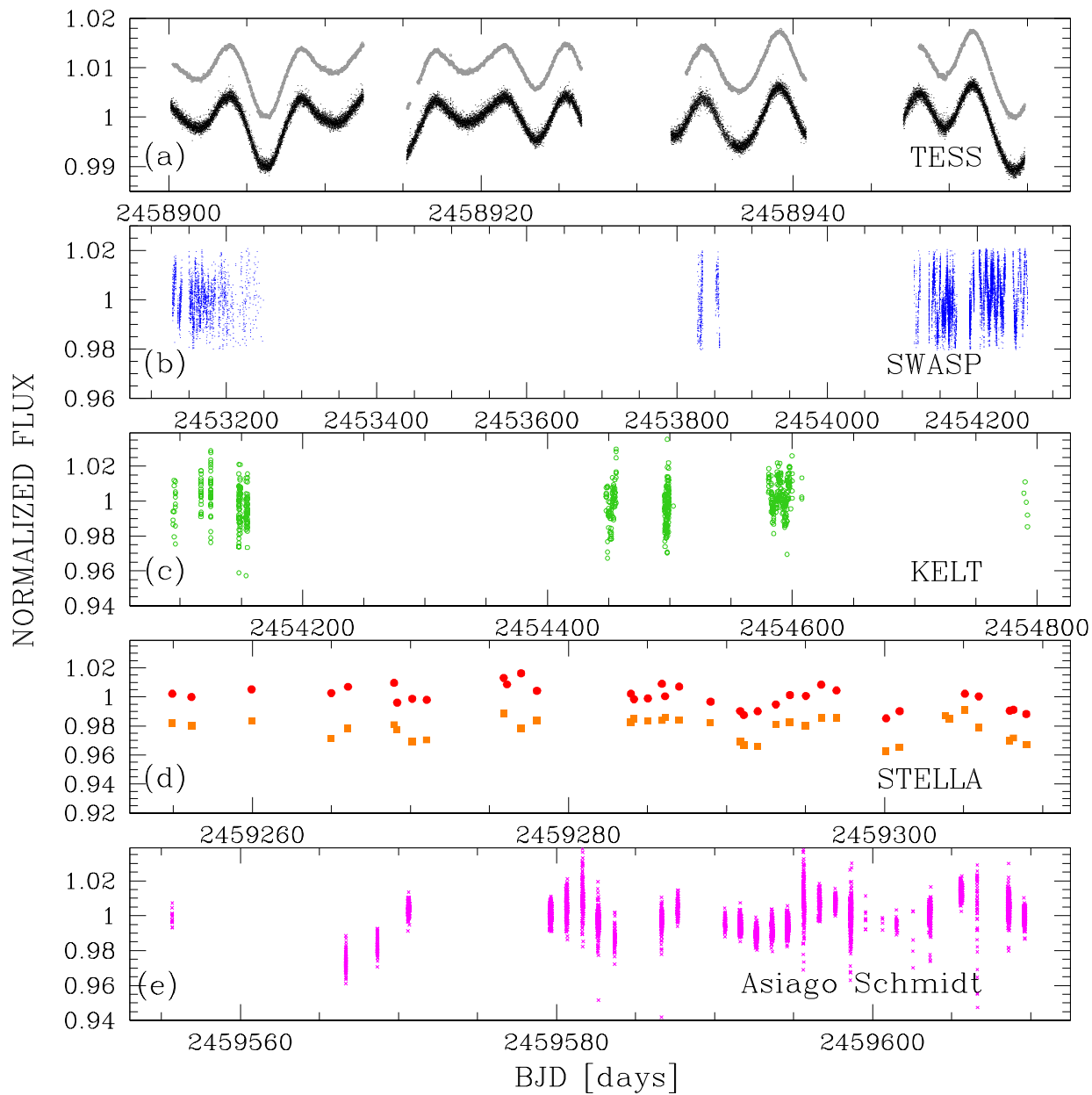} \\ 
  \caption{Light curves of TOI-1807 used in this work. Panel
      (a) shows the long- (gray points) and the short-cadence (black
    points) \tess light curve; in panel (b) and (c) are illustrated
    the SuperWASP and KELT light curves, respectively. Panel (d) shows
    the STELLA light curves obtained in $V$ (orange points) and $I$
    (red points) bands. Panel (e) is the light curve obtained with the
    Asiago Schmidt 67/92~cm telescope. \label{fig:2r}}
\end{figure*}

\begin{figure*}
  \centering
  \includegraphics[width=0.93\textwidth, bb=23 311 579 717]{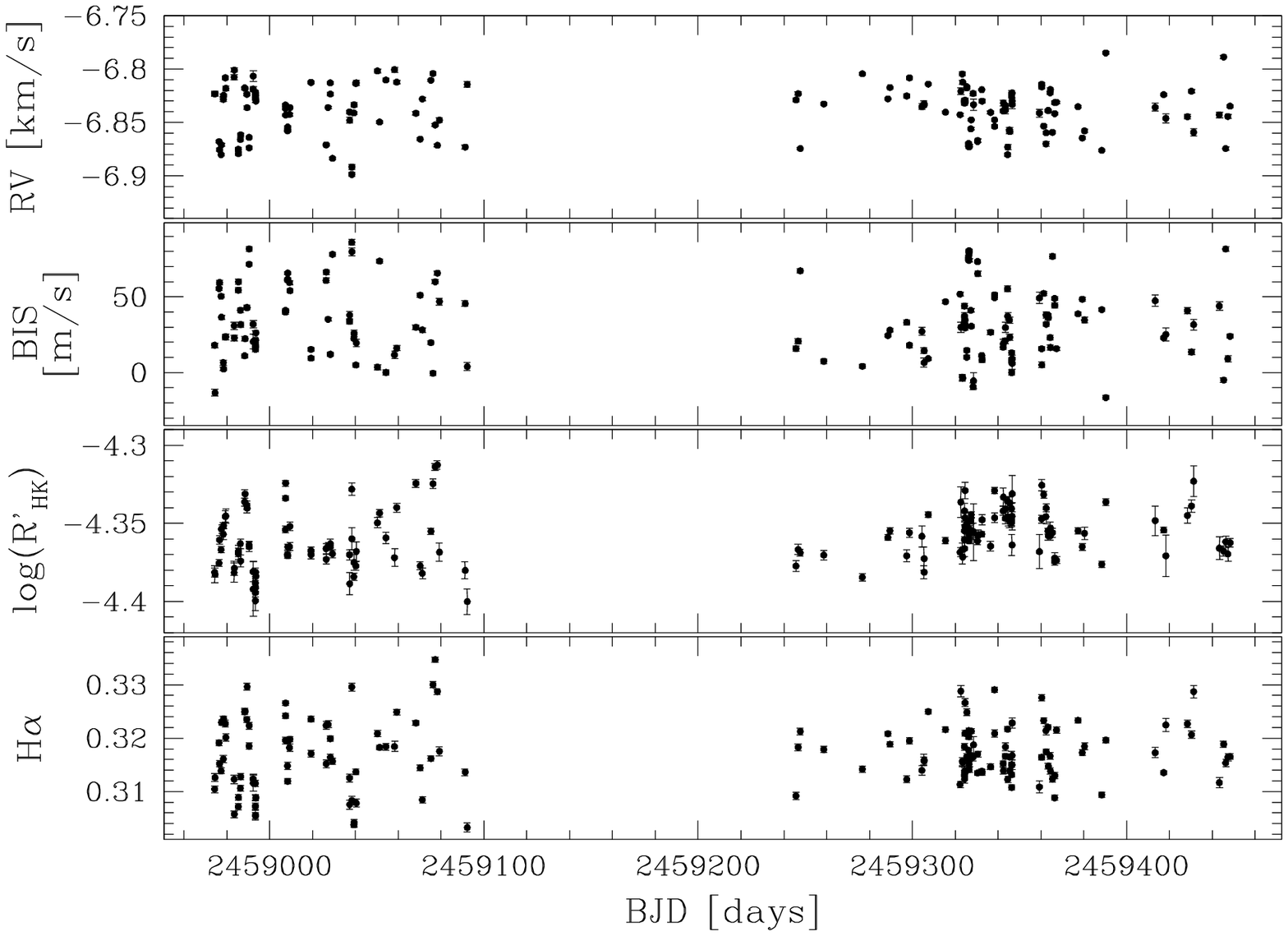} \\ 
  \caption{Spectroscopic time series obtained with HARPS-N and
      used in this work. From the top to the bottom panels the RV,
      BIS, \logrhk, and H$\alpha$ time series are reported. See text
      for details. \label{fig:3r}}
\end{figure*}

\section{Observations and data reduction}
\label{sec:obs}

\subsection{\tess Photometry}
\label{sec:tessphot}
The star TOI-1807 was observed by \tess both in short- (2-minutes) and
long-cadence (30-minutes) mode in Sectors 22 and 23
(Table~\ref{tab:1}); in this work we used the light curves extracted
from both the data-sets.

For the long-cadence data, we extracted the light curves from \tess
Full Frame Images (FFIs, see Fig.~\ref{fig:1r}) adopting the software \texttt{IMG2LC}
developed by \citet{2015MNRAS.447.3536N} and used for the extraction
of light curves of stars in stellar clusters from images obtained with
ground-based instruments (\citealt{2016MNRAS.455.2337N}), {\it Kepler}
(\citealt{2016MNRAS.456.1137L,2016MNRAS.463.1780L,2016MNRAS.463.1831N}),
and \tess
(\citealt{2019MNRAS.490.3806N,2020MNRAS.498.5972N,2021MNRAS.505.3767N}). The
routine allows to minimize the neighbor contamination and extract
high-precision photometry even for very faint stars (see, e.g.,
\citealt{2021ApJ...906...64A}). We corrected the light curves for
systematic effects by fitting to them the Cotrending Basis Vectors
(CBVs) extracted by \citet{2020MNRAS.495.4924N}.

For the short-cadence data, we corrected the Simple Aperture
Photometry (SAP) light curves by using CBVs obtained with the routines
developed by \citet{2020MNRAS.495.4924N} and all the SAP light curves
in the same Camera and CCD in which TOI-1807 was located. This
correction was mandatory because the Pre-search Data Conditioning
Simple Aperture Photometry (PDCSAP) light curves
(\citealt{2012PASP..124.1000S,2012PASP..124..985S,2014PASP..126..100S})
of the target suffered of many systematic effects due to
over-corrections and/or injection of spurious signals. In
Appendix~\ref{sec:light_curve_corr} we report the differences between
PDCSAP light curves and those corrected by us.  For the analysis
carried out in this paper, for both long and short cadence light
curves, we used the points flagged with \texttt{DQUALITY}=0.

We also excluded from the long cadence light curve the points
  associated to values of local background $>4 \sigma_{\rm sky}$ above
  the mean value of the local background. After the selections, the
  short and long cadence light curves contain 27866 and 1784 points,
  respectively, that span 55.5~days. We calculated the simple RMS and
  the P2P RMS as done in \citet{2020MNRAS.498.5972N}; the first
  parameter is sensitive to the stellar variability, while the P2P RMS
  is not sensitive to the variations of the light curve. We obtained
  an RMS of 4 parts-per-thousand (ppt) both for the short- and the
  long-cadence light curve; we measured a P2P RMS of 0.9~ppt for the
  short cadence light curve and 0.2~ppt for the long-cadence light
  curve. The short- and long-cadence light curves are shown in panel (a) of Fig.~\ref{fig:2r}.

\subsection{SuperWASP}
We downloaded the publicly available light curves obtained with
SuperWASP observations
(\citealt{2006PASP..118.1407P,2010A&A...520L..10B}), and we removed
from them all the points with the quality flag \texttt{FLAG=0}
(not-corrected photometric points) and all the outliers. The light
curve of TOI-1807 we used contains 13038 points collected
between May 2004 and June 2007 (1138.2~days, see panel (b) of Fig.~\ref{fig:2r}). 
The simple and P2P RMS are 9~ppt and 6~ppt, respectively.

\subsection{KELT}
We used the light curves obtained from data collected during the
Kilodegree Extremely Little Telescope (KELT) survey
(\citealt{2007PASP..119..923P}), and available
online\footnote{\url{https://exoplanetarchive.ipac.caltech.edu/docs/KELT.html}}. We
did not do any selection on these light curves (556
points, panel (c) of Fig.~\ref{fig:2r}). Observations of TOI-1807 
were carried out between December 2006 and November 2008 (698~days).
The KELT light curve has an RMS of 10~ppt and a P2P RMS of 7~ppt.

\subsection{STELLA}
We collected data of TOI-1807 with the WiFSIP imager mounted at the
robotic STELLA telescope (\citealt{2004AN....325..527S}) between
February and April 2021 (52~days). Observations were carried
out in $V$-Johnson ($t_{\rm exp}=8$~s) and $I$-Cousin bands ($t_{\rm
  exp}=5$~s). Differential light curves were extracted as described by
\citet{2015A&A...580A..60M,2018A&A...614A..35M}. The $V$ and $I$ light
curves contain 35 and 36 points, respectively, and are shown
  in panel (d) of Fig~\ref{fig:2r}; each point is the average of five
individual exposures. The RMS of the light curves is about 10
  ppt, while the P2P RMS is 8~ppt.

\subsection{Asiago Schmidt 67/92 cm}
We collected images of TOI-1807 with the Asiago Schmidt 67/92 cm
telescope. Observations were carried out between December 2021 and
January 2022 (54~days) in $i$-sloan filter with an exposure time of 7s. We
extracted the light curve of TOI-1807 by adopting the pipeline
developed by \citet{2015MNRAS.447.3536N}. The light curve contains
4771 points (panel (e) of Fig.~\ref{fig:2r}), and is characterized 
by an RMS of 9~ppt and a P2P RMS of 6~ppt.

\subsection{HARPS-N data} \label{sec:harpsn}
We monitored TOI-1807 with HARPS-N at TNG within the GAPS - Young
Objects program (\citealt{carleo2020}).  We obtained 161 spectra
between May 2020 and August 2021, with exposure times of 1200~s.  We
followed an observational strategy specifically suited for ultra-short
period planets which consists of gathering at least two observations
every night (\citealt{2021MNRAS.501.4148L}). We collected 161
  spectra in 474~nights. The spectra have, on average, a
  signal-to-noise ratio (S/N) $\sim 68$. Details about the number of
spectra collected, the total time-span of the observations and the
typical  S/N  are also
reported in Table \ref{tab:1}. In this work, we excluded measurements
collected during the night between May 5th and 6th, 2020, because
obtained during bad weather conditions, and resulting as outliers in
the time series.

The spectra were reduced by using the HARPS-N Data Reduction Software
(DRS), providing the RV extraction through the Cross-Correlation
Function (CCF) method (see \citealt{2002A&A...388..632P} and
references therein). With this technique, the scientific spectra are
cross-correlated with a binary mask depicting the typical features of
a star with a selected spectral type. We used a K5 mask for
TOI-1807. The resulting CCFs provide a proxy for the mean line profile
changes of each spectrum. To improve the fitting of the continuum of
the CCF, we reprocessed our data by enlarging the width of the CCF
evaluation window by using the offline version of the HARPS-N DRS
which is implemented at the INAF Trieste
Observatory\footnote{\url{https://www.ia2.inaf.it/}} through the YABI
workflow interface \citep{YABI}.  The resulting RVs have a typical
dispersion of a few tens of \ms ($\sim 23$~m\,s$^{-1}$). RV
internal errors are of the order of a few \ms ($\sim
  1.7$~m\,s$^{-1}$). A summary of these values is also reported in
Table \ref{tab:1}. We compared the RVs obtained by the HARPS-N
  DRS with those provided by the TERRA pipeline
  \citep{2012ApJS..200...15A}, for an independent extraction with a
  template matching method, generally employed in case of active
  stars. Since we found no differences between the two data sets (same
  RV dispersion and RV error bars), we decided to adopt the DRS RVs
  data sets for our analysis.

We also extracted the time series of a set of activity indices, which
are useful to evaluate the jitter in the RV series due to stellar
activity.  The HARPS-N DRS provides the value of the CCF bisector span
(BIS), while the \logrhk\,index from the Ca II H\&K lines was obtained
by using a procedure available on YABI (described in
\citealt{2011arXiv1107.5325L} and references therein). Finally, we
used the ACTIN
Code\footnote{\url{https://github.com/gomesdasilva/ACTIN}}
\citep{2018JOSS....3..667G} to extract the H$\alpha$
index. The spectroscopic time series are illustrated in
  Fig.~\ref{fig:3r}.

\begin{table} 
  \centering
  \caption{Summary of the observations.}
  \begin{tabular}{lc}
    \hline \hline
     TOI            &  1807 \\
    \hline 
    \multicolumn{2}{l}{\textit{TESS observations}} \\
    Sectors         & 22,23 \\ 
    \hline
    \noalign{\smallskip}
    \multicolumn{2}{l}{\textit{HARPS-N RV monitoring}} \\
       Nr. spectra & 161   \\
       Time-span [d]     & 474 \\
       <RV$_{\rm err}$> [m s$^{-1}$]   & 1.74  \\
       $\sigma_{\rm RV}$ [m s$^{-1}$]  & 23.0  \\
       <S/N>                           & 67.9 \\
       \hline \hline
\end{tabular}

  \label{tab:1}
\end{table}

\begin{table}
  \caption{Stellar parameters}
  \resizebox{0.49\textwidth}{!}{
    \begin{tabular}{l c c}
\hline
\hline
Parameter     &  TOI-1807 & Reference \\
\hline
\multicolumn{3}{c}{\textit{Other Target Identifiers}} \\
TIC           &     180695581       & (1)\\
2MASS         &   J13250800+3855210 & (2)\\
Gaia~DR2      & 1476485996883837184 & (3)\\
\hline
\multicolumn{3}{c}{\textit{Astrometric information}} \\
$\alpha$(J2016.0)~[deg.]  & 201.28260370719   &   (4)    \\
$\delta$(J2016.0)~[deg.]  & $+$38.92236336733 &   (4)     \\ 
$\mu_{\alpha^{\star}}$~[mas\,yr$^{-1}$] & $-124.608 \pm 0.008$   &     (4)         \\
$\mu_{\delta}$~[mas\,yr$^{-1}$] &  $-27.300\pm 0.009$     &  (4)    \\
Parallax~[mas]  &  $23.4804\pm 0.0142$ &    (4) \\
Distance~[pc] &   $42.58\pm 0.06$ & (5) \\
\hline
\multicolumn{3}{c}{\textit{Photometric information}} \\
$T$~[mag]    & $9.036  \pm	0.006$  & (1)            \\
$G$~[mag]    & $9.6752 \pm	0.0003$ & (3)        \\
$B$~[mag]    & $11.082 \pm	0.057$  & (1)            \\
$V$~[mag]    & $10.000\pm	0.030$  & (1)   \\
$J$~[mag]    & $7.646\pm 0.037$     & (2)         \\
$H$~[mag]    & $7.605 \pm	0.018 $ & (2)	        \\
$K$~[mag]    & $<7.568$             & (2)	        \\
$W_1$~[mag]  & $7.395 \pm 	0.032$  & (6)    \\
$W_2$~[mag]  & $7.508 \pm	0.020$  & (6)              \\
$W_3$~[mag]  & $7.445 \pm	0.017 $ & (6)      \\
$W_4$~[mag]  & $7.368 \pm	0.115$  & (6)        \\
\hline
\multicolumn{3}{c}{\it Fundamental parameters}   \\
RV [\kms]      & $ -7.33\pm 0.59 $    & (3) \\
RV [\kms]      & $ -6.8380\pm0.0031 $ & (7)\\
U\ [\kms]      & $-16.40\pm0.02$      & (7)\\
V\ [\kms]      & $-20.90\pm0.05$      & (7) \\
W\ [\kms]      & $-2.07\pm0.19$       & (7)\\
\teff\ (spectroscopic) [K]     &   $4730\pm75  $  & (7)\\
\teff\ (photometric) [K]       &   $4830\pm100 $  & (7) \\  
\logg\ [cgs]                   &   $ 4.55\pm0.05$ & (7) \\
\feh\ [dex]                    &  $ -0.04\pm0.02$ & (7) \\
\tih\ [dex]                    &  $0.01\pm0.08$   & (7) \\
\lstar\ [\lsun]                &  $0.215\pm0.016$ & (7) \\   
\mstar\ [$\mathrm{M_\odot}$]   &  $0.76\pm0.03  $ &     (7) \\ 
\rstar\ [$\mathrm{R_\odot}$]   &  $0.690\pm0.036$ & (7) \\ 
Age [Myr]                      &  $300\pm80$  & (7) \\ 
$E(B-V)$ [mag]                 &  $0.002^{+0.014}_{-0.002}$  & (7) \\
\vsini\ [\kms]                 & $4.2\pm0.5$ & (7) \\
\prot\ [d]                     & $8.8\pm 0.1 $ & (7) \\
S-index (MW)                    & $ 0.918\pm0.005 $  & (7) \\
\logrhk\                       & $-4.363\pm0.002 $  & (7) \\
$\log L_{X}$ [erg s$^{-1}$]    & $ 28.36^{+0.16}_{-0.26}$   & (7) \\   
$\log L_{X}/L_{bol}$           & $ -4.55\pm0.25$   & (7) \\
$EW_{\rm Li}$ [\AA]            & $ 100.0\pm2.5  $ & (7) \\
A(Li)$_{\rm NLTE}$             & $1.67 \pm0.09  $ & (7) \\

\hline
\multicolumn{3}{l}{{\it Notes:} } \\
\multicolumn{3}{l}{$^{(1)}$~\tess Input Catalogue v8 (\citealt{2018AJ....156..102S})} \\
\multicolumn{3}{l}{$^{(2)}$~Two Micron All Sky Survey (2MASS,\citealt{2006AJ....131.1163S})} \\
\multicolumn{3}{l}{$^{(3)}$~Gaia~DR2 (\citealt{2018A&A...616A...1G})} \\
\multicolumn{3}{l}{$^{(4)}$~Gaia~eDR3 (\citealt{2020arXiv201201533G})} \\
\multicolumn{3}{l}{$^{(5}$~\citet{2018AJ....156...58B}} \\
\multicolumn{3}{l}{$^{(6)}$~Wide-field Infrared Survey Explorer (WISE,\citealt{2010AJ....140.1868W})} \\
\multicolumn{3}{l}{$^{(7)}$~This work} \\
\end{tabular}

  }
  \label{tab:2}
\end{table}

\begin{figure*}
  \centering
  \includegraphics[width=0.495\textwidth,bb= 64 594 600 1119]{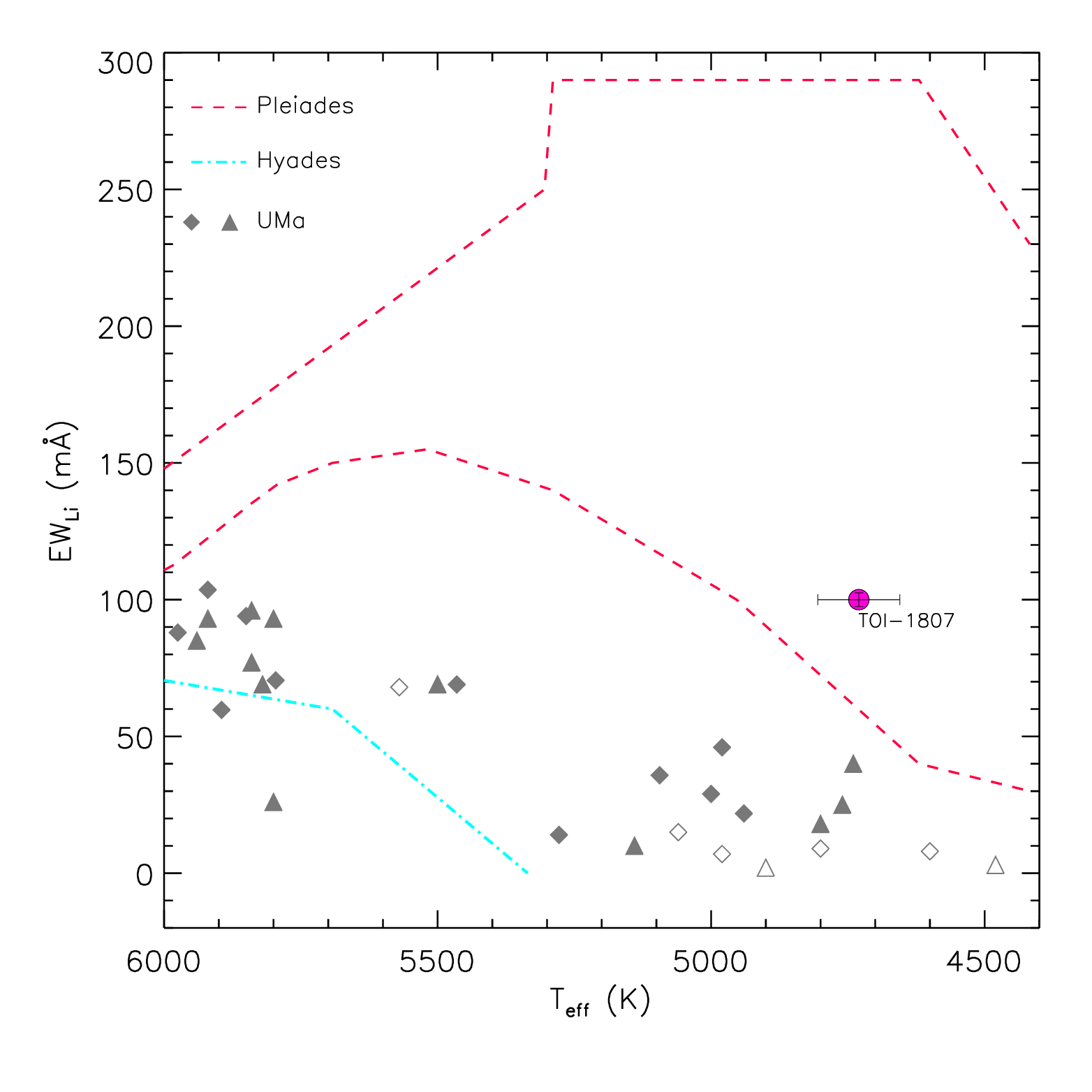}
  \includegraphics[width=0.495\textwidth,bb= 64 594 600 1119]{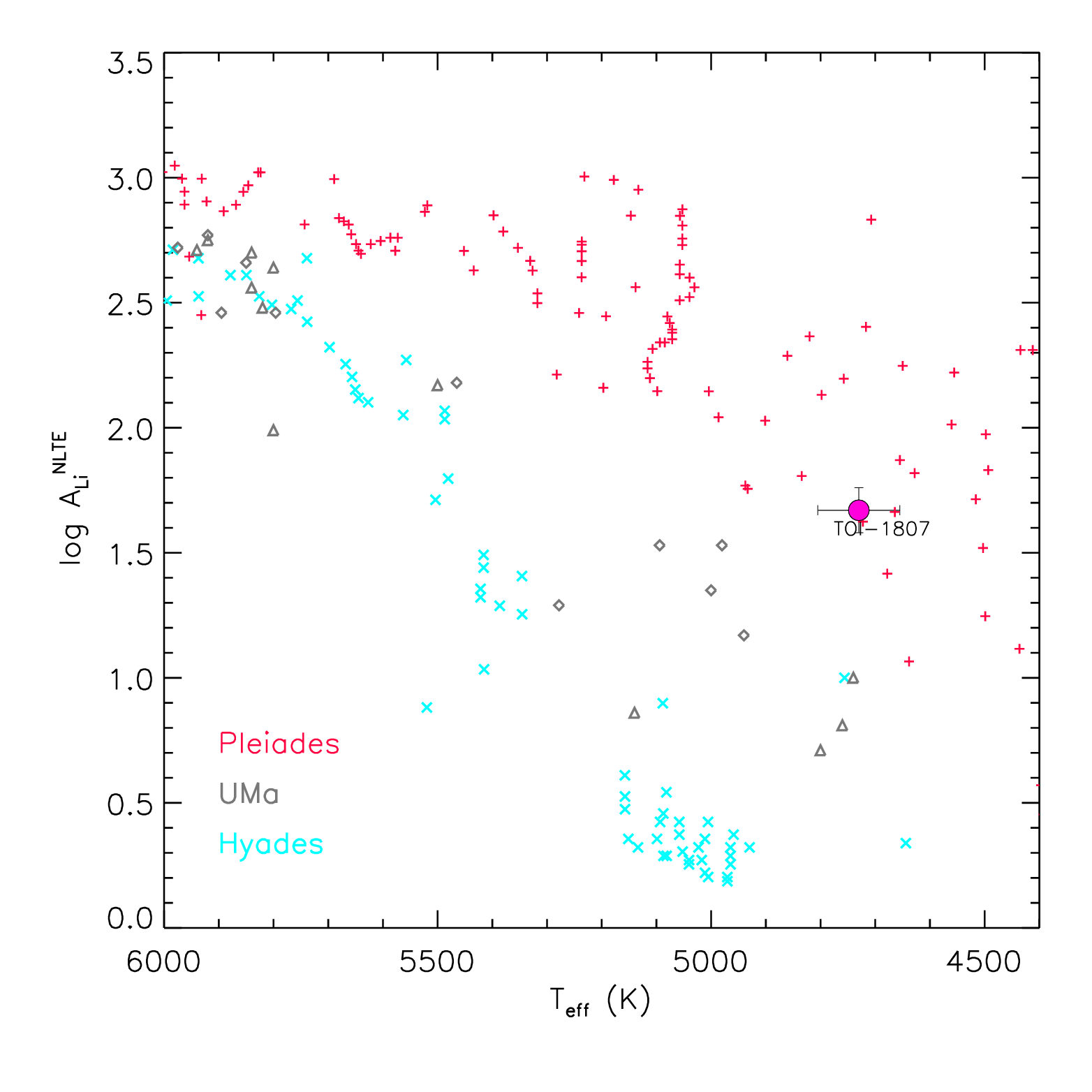}
  \caption{{\it Left panel:} Equivalent width of the \ion{Li}{i}
    $\lambda$6707.8\,\AA\,line plotted as a function of the
    spectroscopic effective temperature. The lines show the upper boundary for
    Hyades (dash-dotted; \citealt{sestitorandich2005}) and the
    lower/upper envelopes of the Pleiades cluster (dashed;
    \citealt{soderblometal1993a}). The position of targets in the UMa
    group is represented by diamonds and triangles (open symbols are
    upper limits), as found by \cite{ammlerguenther2009} and
    \cite{soderblometal93b}, respectively. {\it Right panel:} Lithium
    abundance as a function of $T_{\rm eff}$. Plus, diamond/triangle,
    and cross symbols represent the position of targets in Pleiades,
    UMa, and Hyades clusters, respectively, by the same
    authors.\label{fig:3}}
\end{figure*}

\begin{figure*}
  \centering
  \includegraphics[width=0.95\textwidth, bb=20 144 554 710]{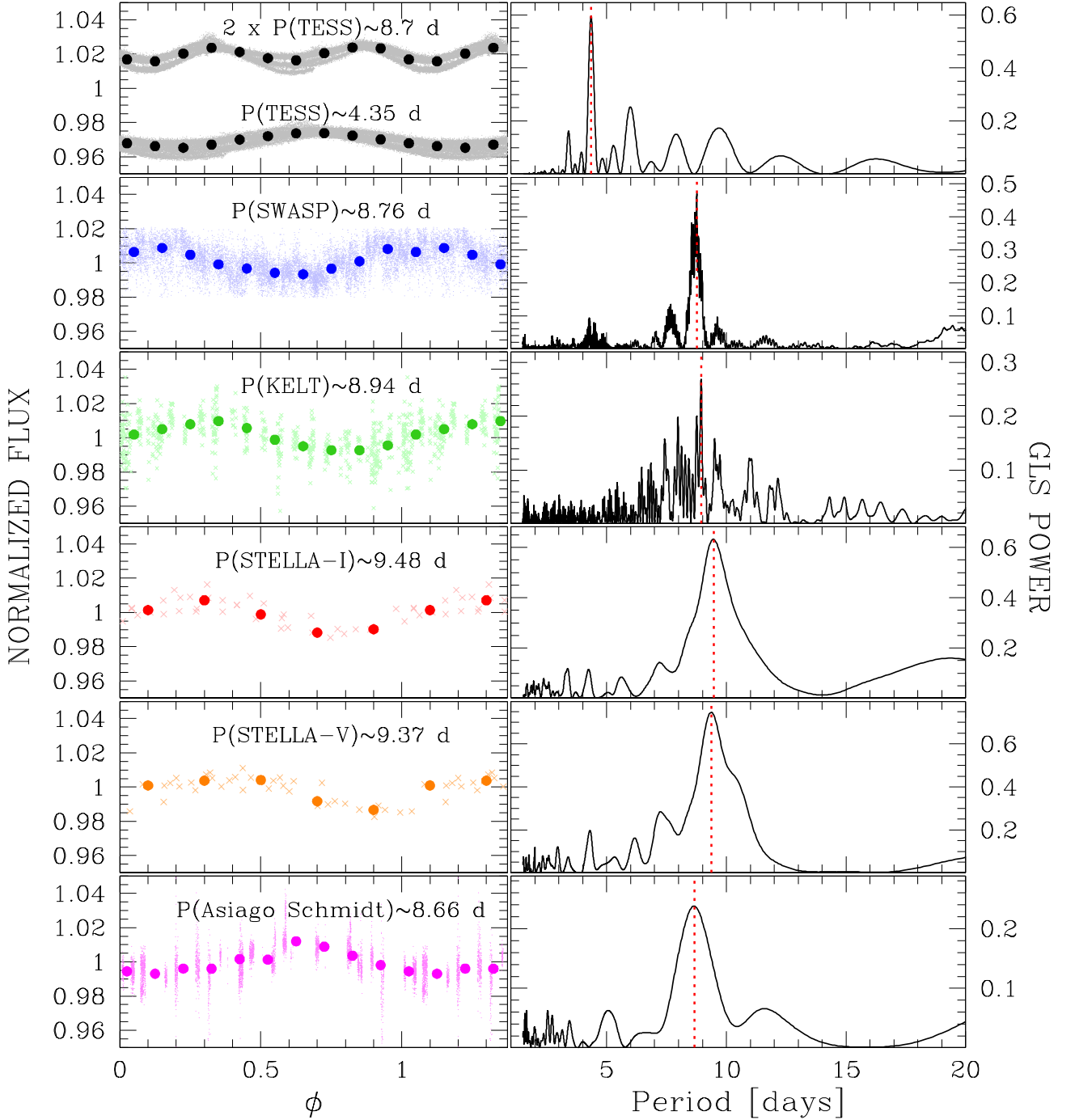} \\ 
  \caption{Left panels: Phased light curves obtained with
    SuperWASP (blue), KELT (green), \tess (gray), Asiago Schmidt 67/92
    cm Telescope (magenta), and STELLA (red and orange for I- and
    V-band, respectively) by using the periods found by
    GLS. Right panels: GLS periodograms obtained from the
      corresponding light curves in the left panels; the red dashed
      line indicates the period of the peak.\label{fig:1}}
\end{figure*}
\begin{figure}
  \centering
  \includegraphics[width=0.5\textwidth, bb=34 70 550 693 ]{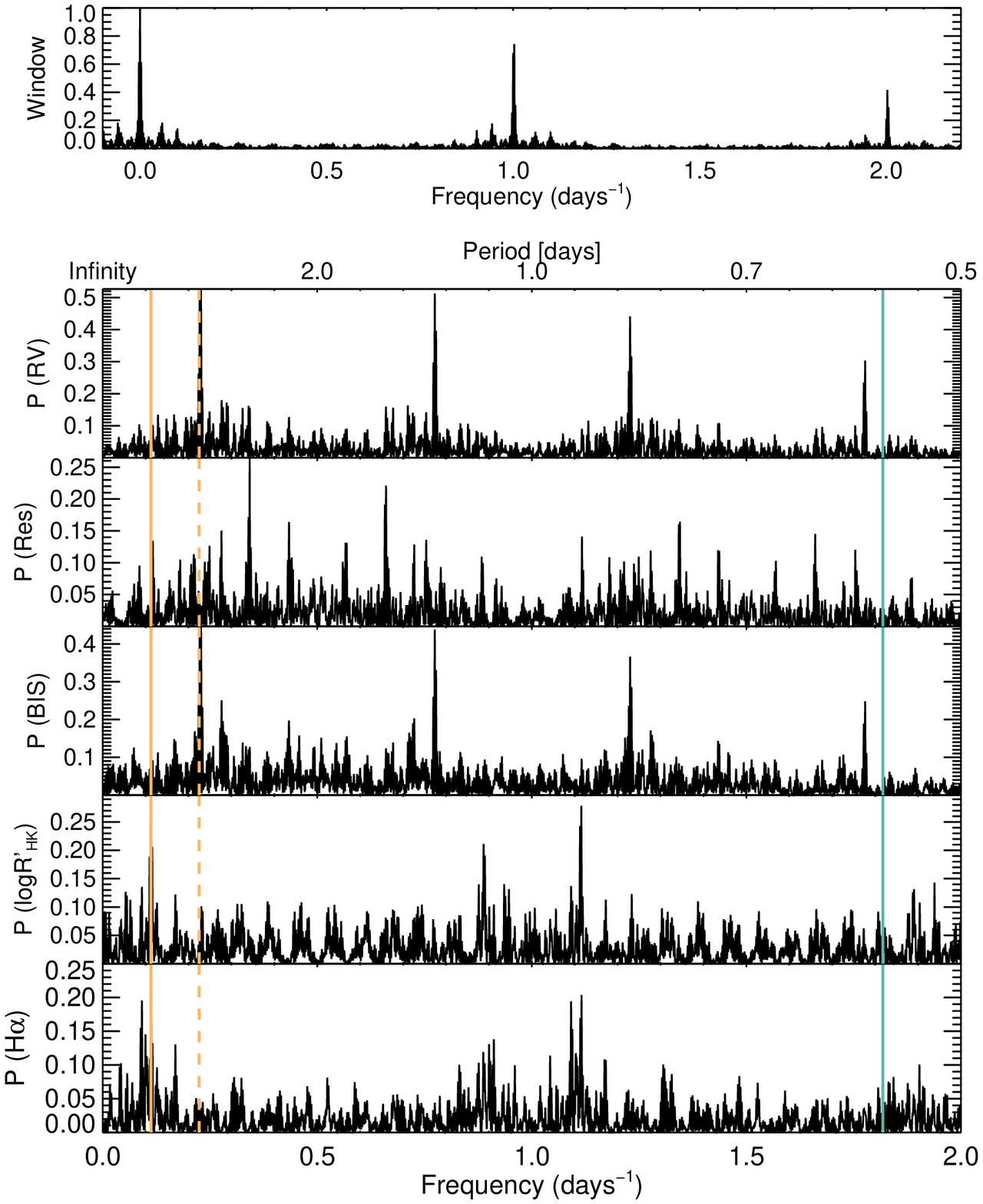} 
  \caption{GLS periodograms of the spectroscopic time series from
    HARPS-N data. The window function is depicted on the top panel
    while the periodograms of the RVs, the corresponding residuals,
    the bisector, the \logrhk and the H$\alpha$ time series are
    reported in the following panels. The orange vertical lines
    represent the location of the rotation period (solid) and its
    first harmonic (dashed), while the cyan solid line indicates the
    position of the planet orbital period.  \label{fig:2}}
\end{figure}

\section{Stellar parameters}
\label{sec:stellar}


We present in this section the methods we used for the determination
of stellar parameters, combining photometric, spectroscopic,
astrometric data and additional information, with special
focus on stellar age.

\subsection{Kinematics and membership to moving groups}
\label{sec:kinematics}
U, V, and W space velocities were derived for our target using the
kinematic data from Table~\ref{tab:2} and the formalism by
\citet{johnson1987}. We also searched for membership to known groups
exploiting the BANYAN $\Sigma$ on-line tool
\citep{gagne2018} \footnote{
  \url{http://www.exoplanetes.umontreal.ca/banyan/banyansigma.php} }
and the literature regarding TOI-1807. Finally, we performed our own
search for comoving objects, as described in Appendix
\ref{sec:comoving}.

TOI-1807 was not previously identified as a member of known groups.
As previously done by \citet{2021AJ....162...54H}, we noticed that the young
planet host TOI-2076 has very similar kinematic parameters (their
space velocities differ by just 0.5 km/s and their separation on the
sky is of 12.46 deg, 9.2 pc at their common distance). We then
searched for additional comoving objects, as detailed in Appendix
\ref{sec:comoving}; 76 comoving objects were identified.  Among them,
we found 26 objects which result to have similar age. All these
objects but three are also rather close on the sky and at a similar
distance (median distance 42 pc, rms 7 pc), making the case for a
common origin stronger.  These objects were considered to refine the
age of TOI-1807, assuming that they are coeval and formed together.
The case of TOI-2076 system will be discussed in more detail in a
forthcoming paper.

\subsection{Photometric \teff~ and reddening}
\label{sec:photteff}
Reddening toward our target was estimated from interpolation of the 3D
reddening maps of \citet{lallement2018}. Reddening is negligible for
TOI-1807 as expected from its close distance from the Sun ($E(B-V)
\approx 0.002$~mag, $d\sim 42.6$~pc).

Photometric \teff~ using the tables by
\citet{pecaut2013} 
results in 4830$\pm$100 K, corresponding to K3V spectral type.

\subsection{Atmospheric parameters}
\label{sec:atmo_param}

The star TOI-1807 has a relatively young age ($\tau \sim$ 300 Myr), as
further discussed below. We derived atmospheric parameters applying
the standard spectroscopic method (i.e., using only Fe lines) and the
new spectroscopic approach described in
\cite{2020baratellaA,2020baratellaB}. The spectroscopic analysis of
intermediate-age and young dwarf stars may be hampered by the presence
of intense magnetic fields that alter the structure of the upper
layers of the photosphere. These affect the formation of strong
spectral lines, for which the abundances show a trend with optical
depths \citep{2020baratellaB}. As a consequence, when deriving the
micro-turbulence velocity ($\xi$) by imposing that weak and strong Fe
lines have the same abundance, one has lead to increase its value
until it reaches $\sim$ 2-2.5 \kms. An over-estimation of $\xi$
results in an under-estimation of the iron abundance ([Fe/H]) and all
the abundance ratios that re-scale accordingly. Following the same
strategy as in \cite{2020baratellaA}, we applied a new method that
consists of using a combination of Ti and Fe lines to derive
\teff\,(by imposing the excitation equilibrium), and using only Ti
lines to derive \logg\,and $\xi$\,(by imposing the ionization
equilibrium and by zeroing the trend between individual abundances and
strength, or equivalent width EW, of the lines, respectively). We use
the driver abfind of the code MOOG \citep{1973sneden,2011sobeck}. The
EWs have been measured with the software ARESv2.0 \citep{2015sousa}:
we discard those lines with errors larger than 10$\%$ and with
EW$>120$\,m\AA.\, We use 1D Local Thermodynamic Equilibrium (LTE)
model atmospheres linearly interpolated from the ATLAS9 grid of
\cite{2003castelli}, with new opacities (ODFNEW).

As input parameters for the analysis, we used the photometric
\teff\,derived in Sect.~\ref{sec:photteff}, and we estimated the
gravity from the classical equation using the Gaia parallaxes ($\log
g_{\rm{trig}}$). Initial values of $\xi$ are derived following
\cite{2015ferreira}.

The final spectroscopic values of \teff, \logg, [Fe/H] and [Ti/H]
derived with the new approach are reported in Table \ref{tab:2}. We
obtained a \teff=4730$\pm$75 K, which is 100 K cooler than the
photometric temperature. Nonetheless, our spectroscopic \teff\,has
been confirmed also using the standard analysis. Regarding the
\logg\,values, the spectroscopic estimates (obtained with both
methods) are slightly smaller than the input estimates obtained from
the Gaia parallaxes. The difference between the spectroscopic
\logg\,and the input $\log g_{\rm{trig}}$ is $\Delta (\log g_{\star} -
\log g_{\rm{trig}})=-0.07$\,dex. Such discrepancies are expected for
such cool stars \citep[see e.g.][]{2015maldonado}. The final $\xi$
values obtained with the new spectroscopic approach confirm the input
estimates ($\xi=0.67$\,\kms). Overall, the star has solar [Fe/H] and
[Ti/H].

\subsection{Lithium content}
Lithium equivalent width ($EW_{\rm Li}$) at $\sim \lambda=6707.8$
\AA\,of TOI-1807 was estimated using three measurements performed with
the IRAF\footnote{https://iraf-community.github.io/} task {\tt splot},
and its error was computed from the standard deviation of the
resulting values. The lithium abundance ($\log A{\rm (Li)}^{\rm
  NLTE}$) was derived from the measured $EW_{\rm Li}$, the
spectroscopic parameters ($T_{\rm eff}$, $\log g$, [Fe/H]) and using
the non-LTE (NLTE) prescriptions given by \citet{lindetal2009}. Errors in $\log
n{\rm (Li)}$ were estimated considering the errors in the $EW_{\rm
  Li}$ measurement and adding quadratically the uncertainties on
spectroscopic parameters. The values of $EW_{\rm Li}$ and $\log A{\rm
  (Li)}^{\rm NLTE}$ are reported in Table\,\ref{tab:2} and plotted in
Fig.\,\ref{fig:3} as a function of effective temperature:
TOI-1807 seems to be close to the lower envelope of the Pleiades
cluster.

\subsection{Coronal and chromospheric activity}
\label{sec:corchr}
The chromospheric activity was measured on the HARPS-N spectra
following the dedicated tool implemented in the YABI environment. The
procedure mirrors that developed for HARPS spectra
(\citealt{lovis2011}).  The S-index tabulated in Table~\ref{tab:2} is
then calibrated in the M. Wilson scale \citep{baliunas1995}.  The
value \logrhk\ = -4.36 is intermediate between Hyades and Pleiades.

We also searched for X-ray emission on ROSAT and other public X-ray
catalogs.  TOI-1807 has X-ray detection (source 1RXS J132508.6+385517)
within $\sim 8.1$ arcsec from the optical position on ROSAT Faint
Source Catalog \citep{rosatfaint}.  The corresponding RX = $\log
L_{X}/ L_{bol}=-4.55$ is in agreement with that of the Hyades members
of similar color \citep{desidera2015}.

\subsection{Projected rotational velocity}
\label{sec:vsini}
We derived the projected rotational velocity \vsini\,through spectral
synthesis of two wavelength regions around 6200 and 6700\,\AA\,and
fixing the macro-turbulence velocity ($v_{\rm macro}$), as previously
done by our team (see, e.g., \citealt{barbatoetal2020}, and references
therein). Using the same grid of model atmospheres and codes as those
considered to derive the stellar atmospheric parameters in
Sect.\,\ref{sec:atmo_param} and fixing $v_{\rm macro}$ at the value of
1.6\,km/s from the relations by \cite{Breweretal2016}, we found a
rotational velocity of \vsini$=4.2 \pm 0.5$\,km/s (see
Table\,\ref{tab:2}).

\subsection{Rotation and activity}
\label{sec:rotation}

\subsubsection{Rotation from light curves}
\label{sec:rotlc}
We used ground-based photometric series and \tess light curves to
obtain a first guess estimate of the rotation period of TOI-1807. In
order to identify the best rotation period, we extracted the
Generalized Lomb-Scargle (GLS) periodograms
(\citealt{2009A&A...496..577Z}) of the light curves and we identified
the period associated to the  most powerful peak in the
periodogram and the associated power ($Pw$).

We used the light curves of TOI-1807 obtained with SuperWASP, KELT,
STELLA, Asiago Schmidt 67/92 cm and \tess surveys. In the case of
SuperWASP we measured a single strong peak in the periodogram ($Pw
\sim 0.47$) at $P=8.7633 \pm 0.0004$~d. From the KELT light curve we
obtained a period $P = 8.945 \pm 0.006$~d ($Pw\sim 0.27$). We measured
the rotation period of TOI-1807 from the $V$- and $I$-band
observations obtained with STELLA, finding a single strong peak ($Pw
\sim 0.75$) at $P = 9.37\pm 0.11$~d and $P=9.48 \pm 0.16$; however, we
highlight that the small number of points (35-36) spread out on 54
days can bias the detection of the real rotation period. By using the
light curve obtained with the Asiago Schmidt 67/92 cm telescope, we
found a peak in the periodogram at $P= 8.66 \pm 0.04$~d ( $Pw \sim
0.24$). Finally, we measured the rotation period from the \tess light
curve, obtaining a period $P=4.350 \pm 0.001$~d ($Pw\sim 0.60$);
anyway, looking the light curve and to its features, it appears more
likely that the period is twice the detected one, i.e. $P\sim 8.7$~d,
in correspondence with the first harmonic in the periodogram ($Pw \sim
0.35$) and in agreement with the other measurements. Indeed the light
curve is double dip, i.e. the variability is due to the presence of
spots that rotate on opposite sides of the star. Figure~\ref{fig:1}
shows the phased light curves of TOI-1807 obtained with the different
instruments and the periodograms associated with them.

\subsubsection{Frequency analysis of the HARPS-N data and stellar activity}\label{sec:rvfreq}
We evaluated the GLS periodogram for the HARPS-N time series described
in Sect. \ref{sec:harpsn}, aiming to obtain the activity-related
periodicities. Figure~\ref{fig:2} shows the window function of our
sampling and the periodograms of the RVs, the corresponding residuals
after removing the most prominent signal, and the spectroscopic
activity indices.  The GLS of both the RVs and BIS shows a very well
defined and isolated peak at the first harmonic of the rotation period
(4.36 days, dashed orange line) while the GLS of both the \logrhk and
H$\alpha$ shows a periodicity of 8.8 days, in agreement with the
photometric time series and the adopted rotation period (solid orange
line). The periodicities at the first and the second harmonic are
expected when the RV variations are dominated by the flux
perturbations due to dark spots
(\citealt{2018NewAR..83...37W,2011A&A...528A...4B}).  We removed a
sinusoidal model from the RVs and obtained the GLS of the RV residuals
(second panel of the second block of periodograms). Besides a marginal
signal at 8.8 days, we found a large periodicity at 2.9 days,
corresponding to the second harmonic of the rotation period.  The high
level of the stellar activity of our target is then well represented
in the GLS periodograms. The rotational frequency $f_{\rm rot}$=1/8.7
d = 0.12 d$^{-1}$ generates an alias at $2-f_{\rm rot}$ = 1.88
d$^{-1}$, close to the expected orbital frequency $f_{\rm orb}$=1/0.55
d = 1.82 d$^{-1}$. The two peaks should be clearly separated thanks to
the very long time baseline of our monitoring, but the stellar
activity strongly increases the noise level (up to 4 \ms) around
$f_{\rm orb}$. Therefore, no peak can be seen at the orbital frequency
of the planet (cyan solid line).  \\

\begin{figure*}
  \centering
  \includegraphics[width=12cm, angle=90]{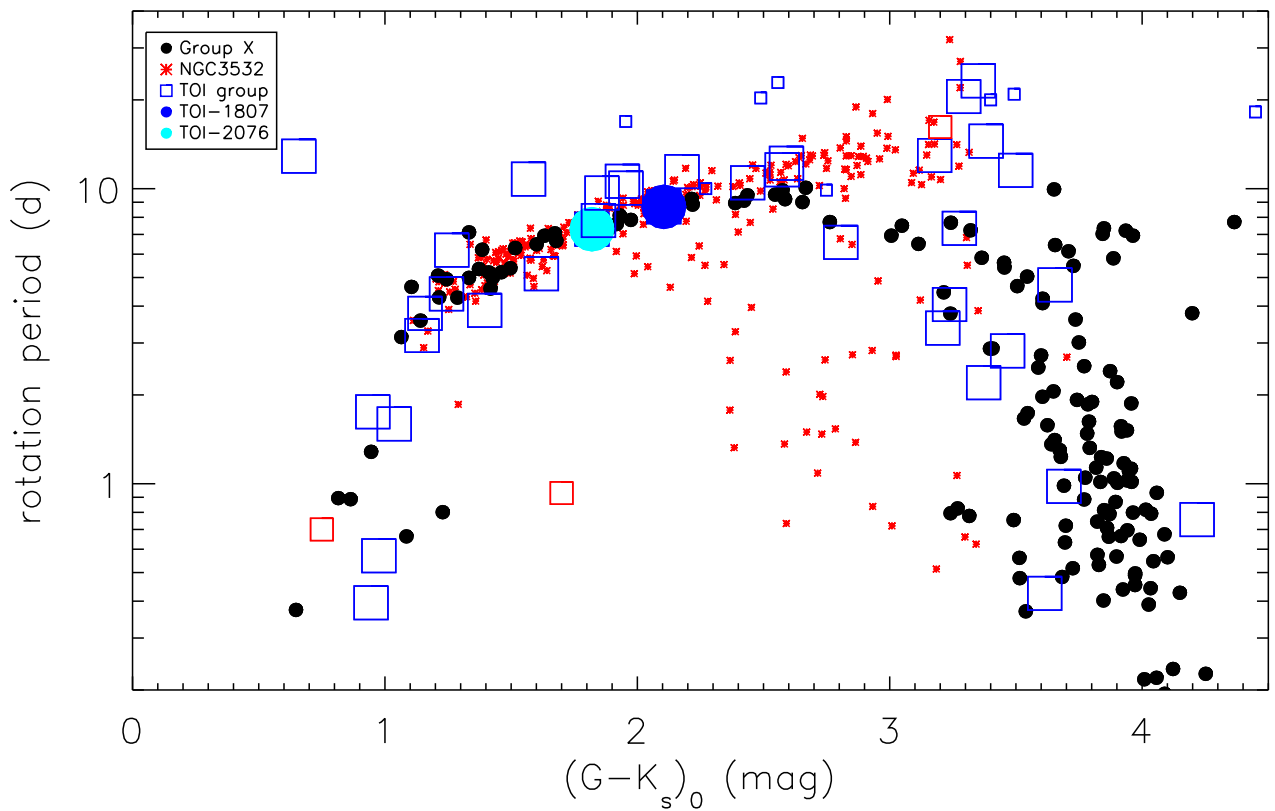}
  \caption{Rotation period distribution of members of Group X (black
    points) and NGC\,3532 (red starred points) with a quoted age of
    300\,Myr and of the wide companions and comoving stars identified
    in this study (blue squares). Big bullets indicate
    TOI-1807 and TOI-2076. Small squares are used to indicate grade B
    rotation periods, whereas the three red squares are periods from
    the literature (see Appendix \ref{sec:comoving}).\label{fig:4}}
\end{figure*}

\subsubsection{Rotation period from \logrhk}
In Sect.~\ref{sec:rotlc} and \ref{sec:rvfreq} we obtained,
  in the periodograms of the large part of the analyzed time series, a
  peak corresponding to a rotation period between 8.7 and 8.9
  days. However, the analysis of the GLS periodograms of the \tess
  light curve, the RVs and BIS shows a strong peak at $\sim
  4.35$~d. In order to confirm that the real rotation period of TOI-1807 is
  $P_{\rm rot}=8.8 \pm 0.1$~days, we calculated the expected rotation
  period by using the equation and the coefficients for K stars
  reported by \citet{2016A&A...595A..12S} and the value of \logrhk
  calculated in Sect.~\ref{sec:corchr}. We obtained a $P_{\rm rot}=8.7
  \pm 1.2$~days, in agreement with the period of $\sim 8.8$~days found
  in this work for TOI-1807.

\subsection{Stellar age, radius, and mass}

\subsubsection{Stellar age}
\label{sec:age}

In order to determine the stellar age of TOI-1807 we followed the
approach by \citet{desidera2015}, combining the results of several
indirect methods (rotation, chromospheric emission, coronal emission,
lithium content) and identification of comoving and coeval members
(Appendix \ref{sec:comoving}).  Isochrone fitting does not provide
significant age constraints in this case.

The indicators agree quite consistently on an age intermediate between
Pleiades and Hyades (125~Myr and 625~Myr, respectively), although the Li
EW of TOI-1807 is close to the lower envelope of Pleiades members.
The rotation period vs color for TOI-1807, TOI-2076 and the other
comoving objects (see Fig.\,\ref{fig:4}) matches well the
sequence of Group X recently obtained by \citet{Messina2022}.  Group X
is a sparse association overlapping on the sky with the Coma Ber open
cluster, but with distinct kinematics and age.  \citet{Messina2022}
derived an age of 300$\pm$60 Myr for this group, considering both the
rotation sequence and isochrone fitting of turn-off members, making it
an ideal reference for objects with age intermediate between Hyades
and Pleiades.  From the overall similitude to Group X age, we adopt
300$\pm$80 Myr for TOI-1807.

\subsubsection{Stellar mass, radius, and luminosity}
\label{sec:massradius}

Stellar luminosity and radius were derived as in \citet{carleo2020},
exploiting Stefan-Boltzmann law and \citet{pecaut2013} tables (updated
version 2021.03.02), and adopting the spectroscopic \teff.  The
stellar radius results of 0.690$\pm$0.036~$R_{\odot}$ and the
luminosity $0.215\pm 0.016$ $L_{\odot}$.  For the stellar mass we
exploited the PARAM \citep{dasilva2006} bayesian interface \footnote{
  \url{http://stev.oapd.inaf.it/cgi-bin/param_1.3} }, selecting the
range of ages allowed by indirect methods. A stellar mass of
0.76$\pm$0.03~$M_{\odot}$ is derived.  These values are used as priors
in the transit fitting, where stellar density is derived again.

\subsection{System inclination}

Coupling the rotation period from Sect. \ref{sec:rotation}, and the
stellar radius from Sect.~\ref{sec:massradius}, an equatorial velocity
of 4.0$\pm$0.2 km\,s$^{-1}$ is derived.  This is fully compatible with
the observed \vsini from Sect. \ref{sec:vsini} (4.2$\pm$0.5
km\,s$^{-1}$), and suggesting that the star is seen very close to
equator-on and that likely the stellar rotation and planetary orbit
are aligned.


\begin{figure*}
  \centering
  \includegraphics[width=0.8\textwidth, bb=24 484 589 717]{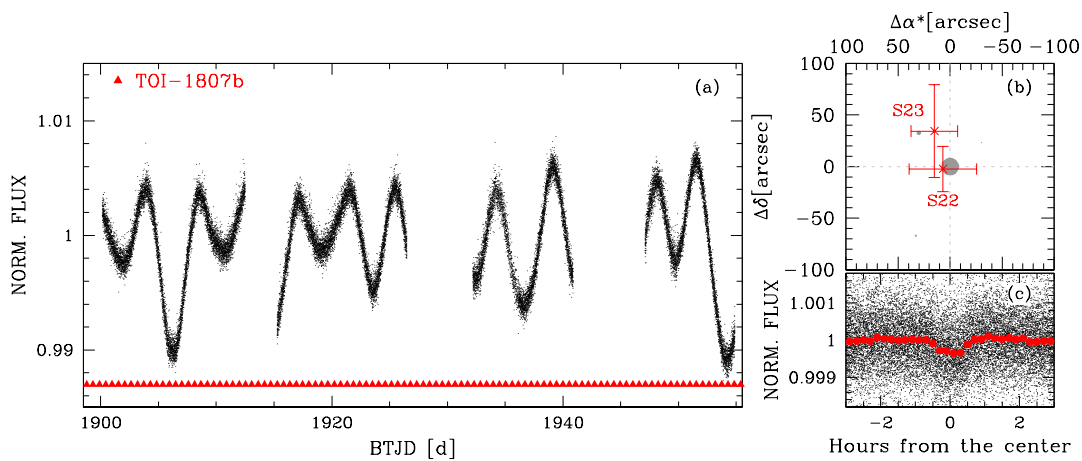} \\
  \caption{Overview of  some validation tests of
    TOI-1807~b. Panel (a) shows the flux normalized short-cadence
    light curve of TOI-1807; red triangles indicate the central time
    of the TOI-1807~b's transits. Panel (b) illustrates the
    in-/out-of-transit centroid test, for Sector 22 and 23; within the
    errors, the transits are associated to the central star. Panel (c)
    is the flattened light curve, folded with the period of TOI-1807~b
    ($\sim 0.549$~d).  \label{fig:5}}
\end{figure*}

\section{Planet detection and vetting}
\label{sec:planetdet}
The candidate exoplanet orbiting the star TOI-1807 has been identified
by the \tess official pipelines.  In particular, the Science
Processing Operations Center (SPOC) pipeline
(\citealt{2016SPIE.9913E..3EJ}) detected one candidate ultra-short
period exoplanet around TOI-1807 ($P \sim 0.55$~d). The planetary
nature of the object has also been confirmed by
\citet{2021AJ....162...54H}.

We used the short-cadence light curve corrected by us to further
confirm the candidate exoplanet. In order to detect the transit
signals, we followed the procedure adopted by
\citet{2020MNRAS.495.4924N}. Briefly, we modeled and removed the
variability of the star interpolating to the light curve a 5th-order
spline defined over a grid of $N_{\rm knots}$ knots at intervals of
13-hours. We removed bad quality measurements clipping away all the
points of the light curve above $4 \sigma$ and below $20 \sigma$ the
mean value of the flattened light curve. We extracted the Transit
Least Squares (TLS) periodograms (\citealt{2019A&A...623A..39H}) of
the flattened light curve, searching for transit signals with periods
between 0.5~d and $T_{\rm LC}/2$, where $T_{\rm LC}$ is the maximum
temporal interval covered by the light curve. We confirmed the
presence of one periodic signal of period $P_{\rm b} \sim 0.549$~d
(with a signal detection efficiency, SDE $\sim 30$). The position of
the transits and the folded light curve of this candidate are shown in
panels (a) and (c) of Fig.~\ref{fig:5}, respectively. We also looked
for other transiting exoplanets in the \tess light curve as follows:
we removed the transits of TOI-1807~b from the light curve and
extracted again the TLS periodogram of the light curve searching for
transit signals with periods between 1.0~d and $T_{\rm LC}/2$. We
obtained a (weak) peak in the periodogram at $\sim 24.986$~d,
corresponding to a very low SDE$\sim 5$. From a visual inspection of
the phased light curve we did not see any transit feature. Also after
inspecting the entire light curve looking for single transits, we did
not detect any presence of a second transiting exoplanet around
TOI-1807 in the \tess light curve of Sectors 22 and 23.

\begin{figure}
  \centering
  \includegraphics[width=0.4\textwidth, bb=196 408 435 706]{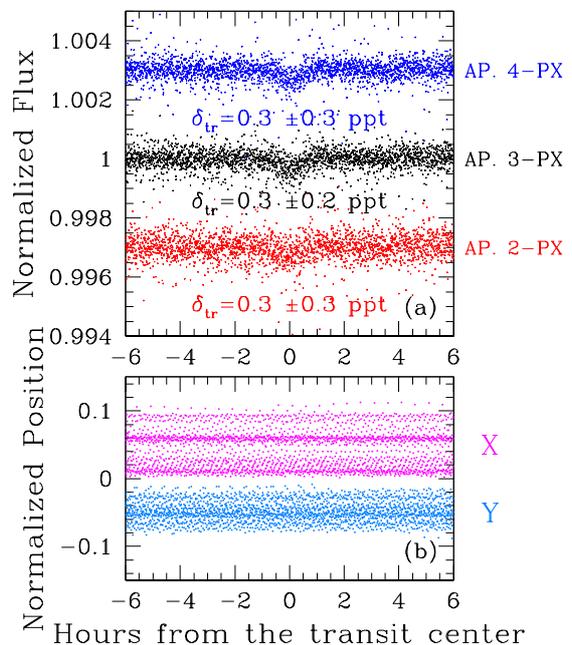} \\
  \caption{Two validation tests for confirming the planetary nature of
    the transits. Panel (a) shows the phased long-cadence, flattened
    light curves of TOI-1807 obtained with different photometric
    apertures (2-pixel in red, 3-pixel in black, and 4-pixel in
    blue). The mean transit depth ($\delta_{\rm tr}$, in
      parts-per-thousand, ppt) is the same in all the cases within the
    errors. Panel (b) shows the normalized position X/Y folded with
    the period found for TOI-1807~b. No particular features correspond
    to the transit events.\label{fig:6}}
\end{figure}

TOI-1807 is an isolated star with few very faint neighbor
stars. However, we performed a series of vetting tests to check if the
transit signals belong to the star subject of our study or are due to
contamination and/or systematic effects. By using the long-cadence
light curves, for which we have photometries extracted with 5
different photometric methods, we checked if the transit depths vary
changing the aperture. As demonstrated in panel (a) of
Fig.~\ref{fig:6}, there is no dependence of the transit depth on the
photometric aperture.\footnote{For this test we excluded PSF-fitting
  and 1-pixel aperture photometries (that work better for stars with
  $T>12$) because the light curves extracted with these techniques
  show scatters larger than the transit depths.}. We also confirm that
there is no correlation between the (X,Y)-positions of the star
(measured with PSF-fitting on FFIs) during the photometric series and
the transit signals (see panel (b) of Fig.~\ref{fig:6}). We checked
the mean depths of odd and even transits, finding that, within the
errors, all the transits have the same depth. Finally, we performed an
analysis of the in-/out-of-transit centroid to be sure that the
transits are not due to contaminant neighbors. For a detailed
description of this analysis we refer the reader to
\citet{2020MNRAS.495.4924N} and \citet{2020MNRAS.498.5972N}. The
results are reported in panel (b) of Fig.~\ref{fig:5}: within the
errors, the in-/out-of-transit mean centroids, calculated for each
sector, are in agreement with the position of our target star.

\section{Properties of the planetary system TOI-1807}
\label{sec:toi1807prop}

We used
\texttt{PyORBIT}\footnote{\url{https://github.com/LucaMalavolta/PyORBIT}}
(\citealt{2016ascl.soft12008M,2016A&A...588A.118M,2018AJ....155..107M})
to model both the stellar activity and the planetary signal in both
the \tess light curve and the RV series. \texttt{PyORBIT} is a package
for modeling planetary transits and radial velocities, taking also
into account the effects of stellar activity. It is mainly based on
the combined use of the optimization algorithm
\texttt{PyDE}\footnote{\url{https://github.com/hpparvi/PyDE}}
(\citealt{1997..............S}), and the affine invariant Markov Chain
Monte Carlo (MCMC) sampler \texttt{emcee}
(\citealt{2013PASP..125..306F}). For this analysis, we first attempted
to retrieve the semi-amplitude of the planet by applying the floating
chunk offset method (\citealt{2014A&A...568A..84H}); however the
prominent stellar activity prevented a precise detection of the
planetary signal, despite our dedicated observational strategy. An
attempt to substitute the nightly offset with a nightly linear trend
(considering only the nights with at least three observations) did not
success either. We therefore tested several approaches to
  model stellar activity though the use of Gaussian Processes (GP,
  \citealt{RasmussenW06, 2014MNRAS.443.2517H}), characterized by
  different forms for the model covariance function (or kernel) and
  different combination of data sets to constraint the GP
  hyper-parameters, and considering different planetary system
architectures (one or more planets).  In this manuscript, we report
the three most significant cases:
\begin{description}
\item{\bf Case 1:} Single planet system and activity modeling trained on photometry 
\item{\bf Case 2:} Single planet system and activity modeling trained on spectroscopy 
\item{\bf Case 3:} Two-planet system  
\end{description}
These three cases are described in the following sections.

\subsection{Case 1: Single planet system and activity modeling trained on photometry}
\label{sec:case1}
First, we tested the model of a planetary system formed by only the
transiting USP TOI-1807~b. For this case, we considered both the
photometric and the spectroscopic observations to constrain all the
planet parameters.

Simultaneously to the activity, \texttt{PyORBIT} modeled the signals
of TOI-1807~b both in the light curve and in the RV series. Parameters
of the host star, like stellar radius ($R_{\star}$) and mass
($M_{\star}$), effective temperature ($T_{\rm eff}$), and gravity
($\log g$), come from the analysis performed in
Sect.~\ref{sec:stellar}. For the transit modeling, we included the
central time of the first transit ($T_{\rm 0,b}$), the orbital period
($P_{\rm b}$), the impact parameter ($b_{\rm b}$), the
planetary-to-stellar-radius ratio ($R_{\rm P,b}/R_{\star}$), and the
stellar density ($\rho_{\star}$). On the basis of the $\log g$ and
$T_{\rm eff}$, we calculated the limb darkening (LD) coefficients by
using the grid of values reported by \citet{2018A&A...618A..20C},
adopting the LD parametrization by \citet{2013MNRAS.435.2152K}, and we
used them as priors of the fitting\footnote{We also tested the case
  with no priors on LD parameters, finding no differences on the final
  planet parameters obtained.}. Transit modeling was carried out by
using the package \texttt{batman} (\citealt{2015PASP..127.1161K}) and
taking into account the 2-minute cadence of \tess data
(\citealt{2010MNRAS.408.1758K}). For the RV modeling, we included the
RV semi-amplitude $K_{\rm \star,b}$ and we considered a circular orbit
($e=0$, a legitimate choice for an USP planet). Uniform priors on
$P_{\rm b}$ and $T_{\rm 0,b}$ and a Gaussian prior on $\rho_{\star}$
were used in the model procedure.

Since the periodogram of the RV variations analyzed in
Sect.~\ref{sec:rvfreq} suggests that the activity-induced variations
are dominated by the flux effect (\citealt{2010A&A...520A..53L}), we
begin our analysis by considering the photometric variations as a
proxy for those RV intrinsic variations.  We modeled the stellar
activity in the light curve, in the RV, and in the \logrhk series
simultaneously, through Gaussian Process (GP) regression. We employed
two different kernels having in common the value of the stellar
rotation period $P_{\rm rot}$. For the RV and \logrhk series we used a
quasi-periodic kernel as defined by \citet{2015ApJ...808..127G},
having as hyper-parameters the characteristics decay time scale
$P_{\rm dec}$, the coherence scale $w$, and the amplitudes of the
covariance matrix $h_{\rm RV}$ and $h_{\rm act}$. GP regression
  was performed with the package \texttt{george}
  (\citealt{2015ITPAM..38..252A}) for the RV and \logrhk\, data-sets,
  and the package \texttt{celerite2} (\citealt{celerite1,celerite2})
  for the light curve. Activity in the light curve was treated with a
stochastically driven, damped harmonic oscillator term ({\tt SHOTerm})
to model the stellar granulation, together with a combination of two
{\tt SHOTerms} at the rotation period and its first harmonic ({\tt
  RotationTerm}) to model stellar rotation modulation, as proposed by
\cite{2020A&A...634A..75B}.  For the granulation {\tt SHOTerm} (grn)
we fixed the quality factor $Q=1/\sqrt{2}$, while leaving the undamped
period of the oscillator $P_{\rm grn}$ and the standard deviation of
the process $\sigma _{\rm grn}$ as free hyper-parameters
\citep{2017AJ....154..220F}; for the {\tt RotationTerm} we left as
free hyper-parameters the quality factor for the secondary oscillation
$Q_{\rm 0,rot}$, the difference between the quality factors of the
first and the second modes $\Delta Q_{\rm rot}$, the fractional
amplitude of the secondary mode compared to the primary $f^{\rm
  mix}_{\rm rot}$, and the standard deviation of the process $\sigma
_{\rm rot}$, while the rotation period $P_{\rm rot}$ is in common with
the quasi-periodic kernel
(\citealt{2019ApJ...885L..12D,2020MNRAS.492.1008G}).   A uniform prior
on the rotation period $P_{\rm rot}$ was assigned on the basis of the
analysis performed in Sect.~\ref{sec:rotation}\footnote{We also tested
  the modeling of the spectroscopic series only, while using larger
  uniform priors $\mathcal{U}( 1.0, 100.0)$ d, obtaining the same
  results described in the paper.}.  All the priors adopted for planet
and activity modeling are reported in Table~\ref{tab:3}.  We
  employed $8 n_{\rm dim}$ walkers for the chains, with $n_{\rm dim}$
  being the dimensionality of the model.  We ran the sampler
with the standard ensemble method from
  \citet{2010CAMCS...5...65G} for 200\,000 steps (excluding the first
20\,000 as burn-in), and using a thinning factor of 100 to reduce the
effect of the chain auto-correlation.  According to the
  Gelman-Rubin statistics \citep{1992StaSc...7..457G} and
  auto-correlation analysis of the chains (see
  \citealt{2021MNRAS.501.4148L} for more details), convergence was
  always reached well before the (conservative) number of steps chosen
  as burn-in. The same sampler configuration and methodology has been
  applied to all the following instances of \texttt{emcee}, unless
  explicitly stated otherwise.

\begin{figure*}
\centering
\includegraphics[width=0.9\textwidth, bb=53 147 547 709]{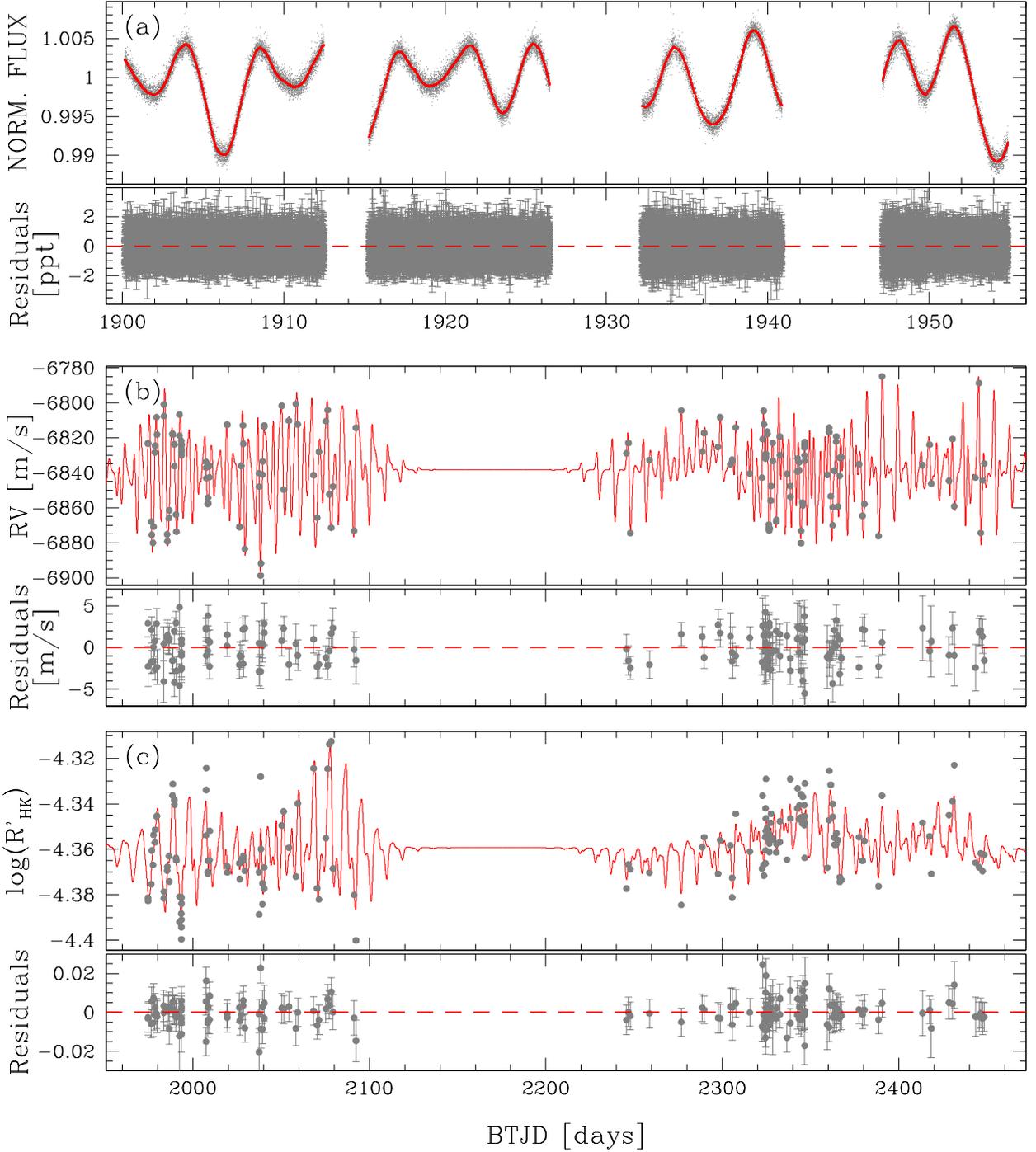} \\
\caption{Overview of the modeling of the stellar
  activity performed for the Case 1. Panel (a) shows the light curve
  (in gray) of the star obtained with \tess short cadence data; the
  model of the activity obtained with the GP fit is reported in
  red. The panel below is the plot of the residuals after the
  subtraction of the model to the light curve. Panels (b) and (c)
  represent the RV and \logrhk series of TOI-1807 (gray points): the
  stellar activity models are illustrated in red, while the residuals
  obtained after the subtraction of the model to the observed data are
  reported in the plots below each panel. \label{fig:7}}
\end{figure*}

\begin{figure*}
\centering
\includegraphics[width=0.49\textwidth, bb=29 483 461 710]{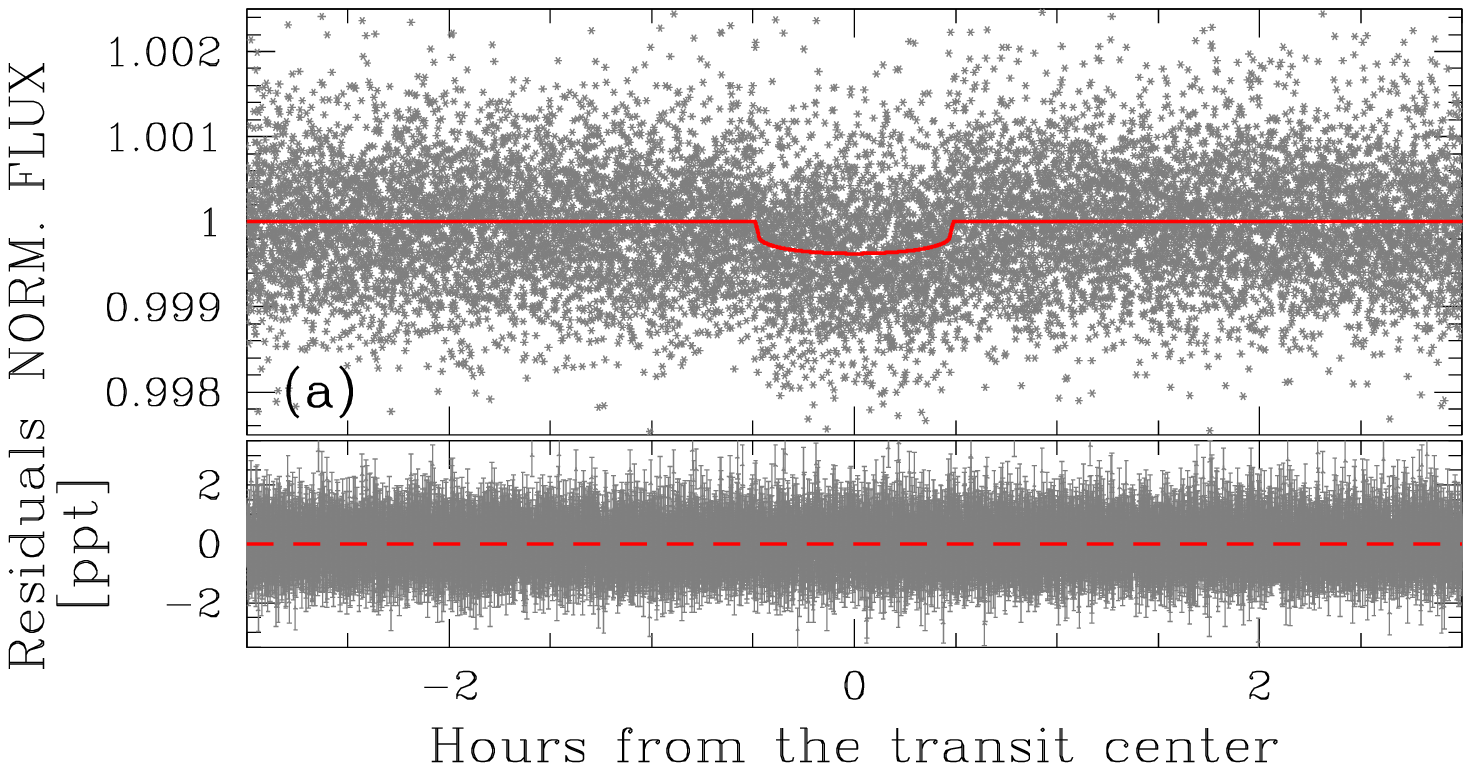} 
\includegraphics[width=0.49\textwidth, bb=29 483 461 710]{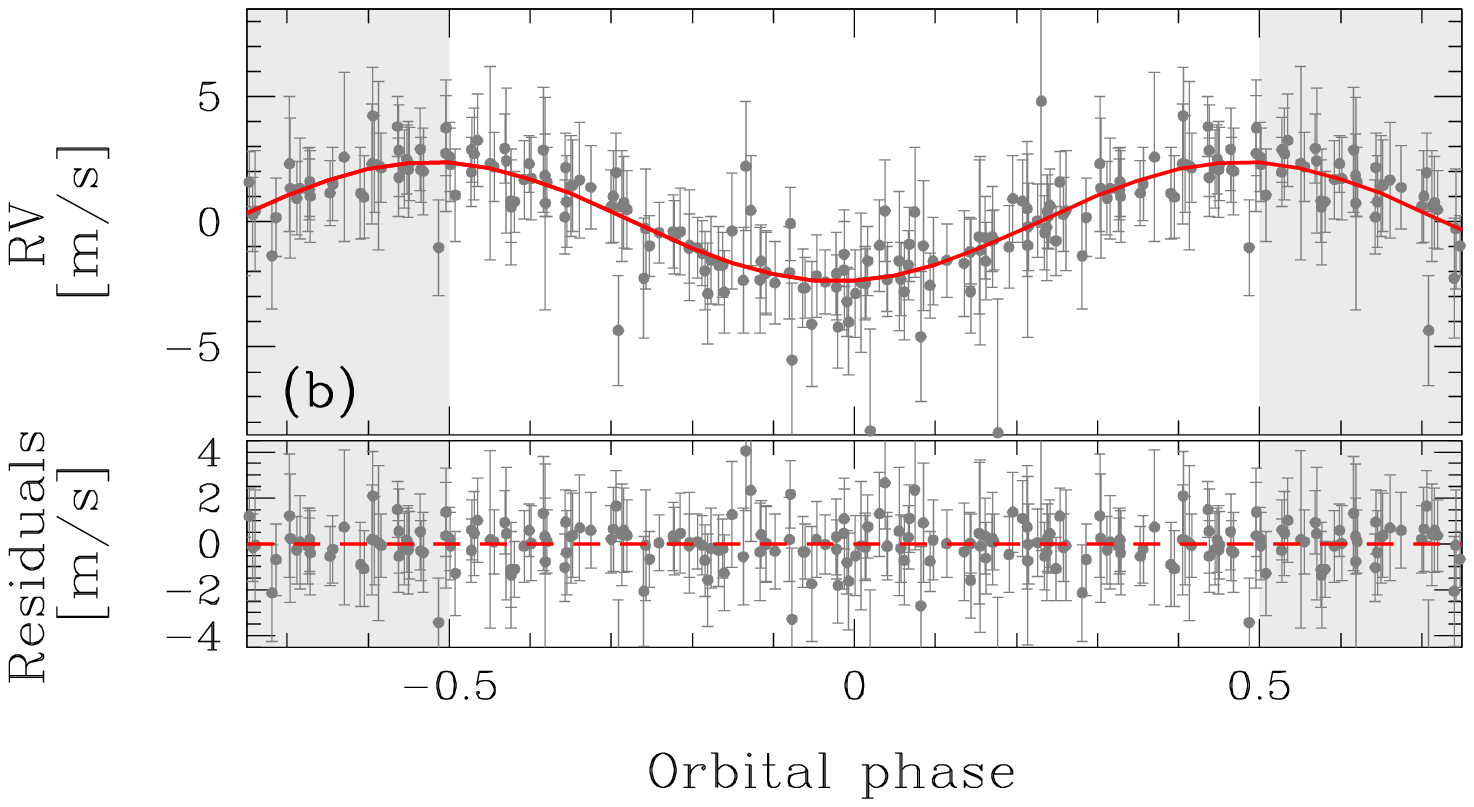} 
\caption{Photometric and RV modeling of TOI-1807~b planetary signal
  obtained in Case 1. Panel (a) shows the folded transits in the light
  curve of TOI-1807 after the subtraction of the stellar activity
  signal, and the model of the transits (red line). In the panel
  below, the residuals of the light curve after the subtraction of the
  planetary transit model. Panel (b) shows the RV curve of the star
  after removing the stellar activity contribution, phased with the
  period of TOI-1807~b; the red line represents the model used to
  measure the RV semi-amplitude of the curve.The region
    between the two gray shaded areas corresponds to one orbital
    phase. In the panel below the RV residuals after model
  subtraction are shown.  \label{fig:8}}
\end{figure*}

Figure~\ref{fig:7} shows the results obtained for the modeling of the
activity both in the photometric (panel (a)) and in spectroscopic
series (panels (b) and (c)). We obtained a rotation period of $P_{\rm
  rot} = 8.83 \pm 0.08$~d. Results of the modeling are reported in
Table~\ref{tab:3}, while Fig.~\ref{fig:8} shows an overview of the
planet modeling after the subtraction of the activity signals from the
light and RV curves. The resulting semi-amplitude $K_{\rm b}$
  is about 2.4 \ms\, only, and this explains why it cannot be seen in
  the noisy RV periodogram of the original data (Fig. \ref{fig:2}, see
  also discussion in Sect. \ref{sec:rvfreq}).

\begin{table*}
  \renewcommand{\arraystretch}{1.25}
  \caption{Priors and results of the model of planet b from the combined analysis of the light curve and RV series (Case 1)}
    \begin{tabular}{l l l l}
\hline
\hline
\multicolumn{4}{c}{{\bf Stellar activity}} \\
\hline
Parameter    &  Unit  &    Prior      &       Value  \\
\hline
\multicolumn{4}{l}{\it Common RV$+$\logrhk$+$light curve } \\
Rotational period ($P_{\rm rot}$)           & days       & $\mathcal{U}( 8.0,  10.0)$    &  $ 8.83 \pm 0.08$                    \\
\multicolumn{4}{l}{{\it  Quasi-periodic kernel}} \\
Decay Timescale of activity ($P_{\rm dec}$)                & days       & $\mathcal{U}(10.0, 100.0)$    &  $14.3^{+1.2}_{-1.1}$                    \\
Coherence scale ($w$)                 &            & ...                           &  $0.26^{+0.03}_{-0.02}$                 \\
Amplitude of the RV signal                  &  m/s       & $\mathcal{U}(0.01, 1500.00)$  &  $25.2^{+2.6}_{-2.2}$            \\
Uncorrelated RV jitter ($\sigma_{\rm jitter}^{\rm RV}$) &  m~s$^{-1}$ & ... & $0.69_{-0.45}^{+0.53}$ \\
RV offset ($\gamma^{\rm RV}$) &  m~s$^{-1}$ & ... & $-6838.3 \pm 4.2$ \\
\multicolumn{4}{l}{{\it  \texttt{SHOTterm}$+$\texttt{RotationTerm}~kernel}} \\
Standard deviation of the rotation       & ppt        & ...                           &  $5.2_{-1.0}^{+2.0}$     \\
Granulation period                          & days       & $\mathcal{U}(0.05, 5.00)$      &  $0.18^{+0.16}_{-0.06} $                    \\
Standard deviation of granulation           & ppt       & ...                            &    $0.067 \pm 0.010 $    \\
Photometric jitter ($\sigma_{\rm jitter}^{\rm LC}$) &  ppm & ... & $303 \pm 7$ \\
\hline
\hline
\multicolumn{4}{c}{{\bf Planet b}} \\
\hline
Parameter    &  Unit  &    Prior      &       Value  \\
\hline
Orbital Period ($P_{\rm b}$)                           & days   &  $\mathcal{U}(0.549, 0.550)$  & $0.549374_{-0.000013}^{+0.000010}$ \\
Central time of the first transit ($T_{\rm 0,b}$)      & BJD    &  $\mathcal{U}(2458899.3, 2458899.4)$  & $2458899.3449_{-0.0005}^{+0.0008}$ \\
Duration of the transit ($T_{\rm 14,b}$)  & hours & ... & $0.98_{-0.02}^{+0.03}$ \\
Limb darkening ($u_1$)         &   & $\mathcal{N}(0.47, 0.05)$ & $0.46 \pm 0.05$ \\
Limb darkening ($u_2$)         &   & $\mathcal{N}(0.18, 0.05)$ & $0.17 \pm 0.05$ \\
Impact factor ($b_{\rm b}$)          &        &  ...                              &  $0.53_{-0.11}^{0.09}$ \\
Orbital inclination ($i_{\rm b}$)  & deg & ... & $82.0 \pm 2.0$ \\
Orbital eccentricity ($e_{\rm b}$)  &   & fixed & 0 \\
Semi-major-axis-to-stellar-radius ratio ($(a_{\rm b}/R_{\star})$) &  & ... & $3.8\pm 0.2$ \\
Orbital Semi-major axis ($a_{\rm b}$) & au & ... & $0.0120 \pm 0.0003$ \\
Planetary-to-stellar-radius ratio ($(R_{\rm P,b}/R_{\star})$) &  & ... & $0.0182\pm 0.0006$ \\
Planetary radius ($R_{\rm P,b}$) & ${\rm R_{\oplus}}$ & ... & $1.37 \pm 0.09$  \\
RV semi-amplitude ($K_{\rm b}$)                              & m~s$^{-1}$ & $\mathcal{U}(0.01, 10)$ & $2.39_{-0.46}^{+0.45}$ \\
Planetary mass ($M_{\rm P,b}$) & ${\rm M_{\oplus}}$ & ... & $2.57 \pm 0.50$  \\
Planetary density ($\rho_{\rm b}$) & $\rho_{\oplus}$ & ... & $1.00 \pm 0.29$\\
Stellar density ($\rho_{\star}$) & $\rho_{\odot}$ & $\mathcal{N}(2.3, 0.4)$ & $2.4 \pm 0.4$\\
\hline
\end{tabular}

      \label{tab:3}
\end{table*}

\begin{figure*}
  \centering
  \includegraphics[width=0.85\textwidth, bb=24 219 587 694]{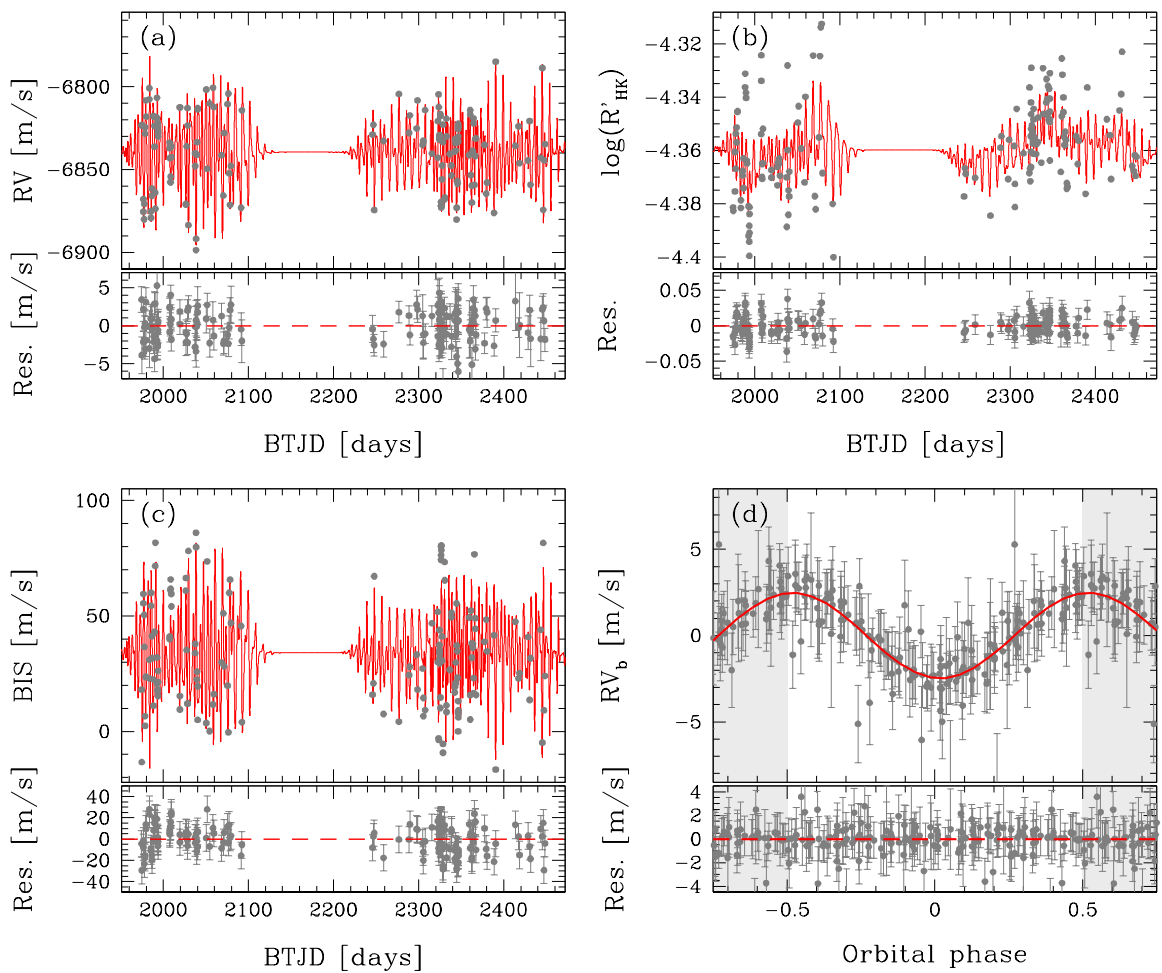} \\
  \caption{Overview of the activity+planet modeling in the
    spectroscopic series obtained in case 2. Panel (a), (b), and (c)
    show the RV, \logrhk, and BIS series, respectively (gray points);
    red lines represent the models of the stellar activity obtained in
    the GP framework we used. The panels below report the difference
    between the observed points and the stellar activity models. Panel
    (d) shows the RV curve of the star after the subtraction of the
    stellar activity and phased with the period of TOI-1807~b; model
    of the RV series is reported in red; the region
    between the two gray shaded areas corresponds to one orbital
    phase.. \label{fig:9}}
\end{figure*}

\begin{table*}
  \renewcommand{\arraystretch}{1.25}
  \caption{Priors and results of the model of planet b from the analysis of spectroscopic series with GP framework (Case 2)}
  \begin{tabular}{l l l l}
\hline
\hline
\multicolumn{4}{c}{{\bf GP framework parameters}} \\
\hline
Parameter    &  Unit  &    Prior      &       Value  \\
\hline
Uncorrelated RV jitter ($\sigma_{\rm jitter,0}^{\rm RV}$) &  m~s$^{-1}$ & ... & $0.79_{-0.49}^{+0.51}$ \\
RV offset ($\gamma_{\rm 0}^{\rm RV}$) &  m~s$^{-1}$ & ... & $-6839.6 \pm 1.1$ \\
Uncorrelated BIS jitter ($\sigma_{\rm jitter,0}^{\rm BIS}$) &  m~s$^{-1}$ & ... & $12.30_{-0.76}^{+0.83}$ \\
BIS offset ($\gamma_{\rm 0}^{\rm BIS}$) &  m~s$^{-1}$ & ... & $34.0 \pm 1.4$ \\
Uncorrelated \logrhk jitter ($\sigma_{\rm jitter,0}^{\rm \log{R'_{HK}}}$) &    & ... & $0.0132_{-0.0009}^{+0.0010}$ \\
\logrhk offset ($\gamma_{\rm 0}^{\rm \log{R'_{HK}}}$) &   & ... & $-4.360 \pm 0.003$ \\
$V_c$  & m~s$^{-1}$ & ... & $1.2_{-1.9}^{+1.9}$ \\
$V_r$  & m~s$^{-1}$ & ... & $27.8_{-3.4}^{+4.3}$ \\
$B_c$  & m~s$^{-1}$ & ... & $-1.9_{-1.8}^{+1.7}$ \\
$B_r$  & m~s$^{-1}$ & ... & $-23.8_{-3.9}^{+3.2}$ \\
$L_c$  &            & ... & $-0.0110_{-0.0020}^{+0.0019}$ \\
\hline
\hline
\multicolumn{4}{c}{{\bf Stellar activity}} \\
\hline
Parameter    &  Unit  &    Prior      &       Value  \\
\hline
Rotational period ($P_{\rm rot}$)           & days       & $\mathcal{U}( 8.0,  9.5)$    &  $ 8.84 \pm 0.08$                    \\
Decay Timescale of activity ($P_{\rm dec}$)                & days       & $\mathcal{U}(10.0, 1000.0)$    &  $12.85^{+1.15}_{-1.12}$                    \\
Coherence scale ($w$)             &            & $\mathcal{U}(0.01,0.60)$     &  $0.43_{-0.04}^{+0.04}$\\
\hline
\hline
\multicolumn{4}{c}{{\bf Planet b}} \\
\hline
Parameter    &  Unit  &    Prior      &       Value  \\
\hline
Orbital Period ($P_{\rm b}$)                           & days   &  $\mathcal{N}(0.549374, 0.00002)$  & $0.549380_{-0.000016}^{+0.000015}$ \\
Central time of the first transit ($T_{\rm 0,b}$)      & BJD    &  $\mathcal{N}(2458899.3449, 0.0008)$  & $2458899.3449_{-0.0008}^{+0.0008}$ \\
Orbital eccentricity ($e_{\rm b}$)  &   & fixed & 0 \\
Semi-major-axis-to-stellar-radius ratio ($(a_{\rm b}/R_{\star})$) &  & ... & $3.7\pm 0.2$ \\
Orbital Semi-major axis ($a_{\rm b}$) & au & ... & $0.0120 \pm 0.0002$ \\
RV semi-amplitude ($K_{\rm b}$)                              & m~s$^{-1}$ & $\mathcal{U}(0.01, 10)$ & $2.48_{-0.39}^{+0.38}$ \\
Planetary mass ($M_{\rm P,b}$, $i=82\pm 2$~deg) & ${\rm M_{\oplus}}$ & ... & $2.67 \pm 0.43$  \\
Planetary density ($\rho_{\rm b}$) & $\rho_{\oplus}$ & ... & $1.04 \pm 0.28$\\
\hline
\end{tabular}

  \label{tab:4}
\end{table*}

\subsection{Case 2: Single planet system and activity modeling trained on spectroscopy}
\label{sec:case2}

In the second case, we modeled stellar activity through the
multi-dimensional GP framework developed by
\citet{2015MNRAS.452.2269R}, re-implemented in \texttt{PyORBIT}
following the prescription in the paper (see also
\citealt{2022MNRAS.509..866B}). Given the high computational cost of
this approach, we modeled only the spectroscopic data-sets, i.e. RV,
\logrhk, and BIS series, following the same name conventions as in
\cite{2015MNRAS.452.2269R}.  Also in this case we assumed a circular
orbit for the USP planet. We assigned Gaussian priors on the orbital
period and central time of transit of TOI-1807~b based on the results
of Case 1\footnote{As these parameters are constrained by photometry,
  it is legitimate to use the results of the previous analysis as
  priors even if the RV and \logrhk data-sets are in common.}, and
uniform priors to all the other parameters, including the rotation
period $P_{\rm rot}$. We ran the sampler for 100\,000 steps,
considering the first 25\,000 as burn-in, and using a thinning factor
of 200. The results of the modeling are reported in Fig.~\ref{fig:9}
and in Table~\ref{tab:4}. The true mass of the planet has been derived
by taking into account the uncertainty on the stellar mass and the
inclination distribution obtained from Case 1. We found that stellar activity and planet
  parameters values are in agreement with those found in case 1;
  however the error on the RV semi-amplitude $K_{\rm b}$ is slightly
  smaller than the error found in case 1 ($\sim 38$~cm~s$^{-1}$
  versus $\sim 45$~cm~s$^{-1}$), proving again the advantages of this
approach on young and intermediate-age stars (e.g.,
\citealt{2019MNRAS.490..698B}). We noted that the
  residuals of the fit of the \logrhk series shown in panel (b) of
  Fig.~\ref{fig:9} present some significant variations and that the
  model does not fit perfectly the data. This problem has been already
  highlighted by \citet{2022MNRAS.tmp..699B}, and it could be ascribed
  to instrumental systematics or second-order astrophysical effects
  not included in the model.

Besides the multi-dimensional GP framework, we tested other three
approaches for the modeling of stellar activity, all involving the use
of Gaussian Processes trained on spectral indexes only: a
quasi-periodic kernel as the one of Scenario 1; a GP framework-like
approach where the quasi-periodic kernel and its first derivative are
employed for the RV and the BIS data-sets, but covariance matrices of
each data-set are computed independently, i.e., only the
hyper-parameters are shared rather than assuming a single underlying
GP; a quasi-periodic with cosine kernel as introduced by
\citet{2021A&A...645A..58P}. Priors on planetary parameters and MCMC
settings were the same as the GP framework analysis. The use of
different models for the stellar activity always resulted in a RV
semi-amplitude of TOI-1807~b and the rotational period of the star
matching well within 1$\sigma$ with respect to the previous results,
with error bars similar to the case of the photometrically trained
GP. The only parameter showing a significant change across different
GP kernels is the time scale of the decay of the active regions, which
has no impact on the modeling of the USP planet.

\subsection{Case 3: a search for the second planet}
Knowing that many USP planets belong to multi-planet systems, we
searched for the signature of a second planet in our RV data-set. We
employed the same GP activity models as illustrated in the previous
sections, and added a second planet with orbital period between 1 and
300 days in addition to the USP planet. We did not assume circularity
for the orbit of the second planet, nevertheless we imposed a Gaussian
prior on $e_{\rm b}$ centered around zero and with standard deviation
equal to 0.098, following \citet{2019AJ....157...61V}.  The outcome of
the MCMC was initially confusing, because in all cases we obtained a
significant ($> 5 \sigma$) detection of a planet, but with period and
semi-amplitude drastically changing depending on the activity model
employed in the analysis. Most notably, employing the GP framework
resulted in a period of the additional planet extremely close to the
stellar rotational period ($P_{c} \simeq 8.5$~d versus $P_{\rm rot}
\simeq 8.8$~d). We repeated the analysis using the dynamic nested
sampling algorithm \citep{2019S&C....29..891H, 2014arXiv1407.5459B}
implemented in the package \texttt{dynesty}
\citep{2020MNRAS.493.3132S}, with the double goal of checking the
posterior distributions for the parameters of the additional planet
and computing the Bayesian evidence of the two-planet models with
respect to the single-planet one. The analysis confirmed the presence
of a marked multi-modality in the posterior of the additional planet,
with contrasting results regarding the Bayesian evidence between models with the same activity treatment but different
number of planets (e.g.,  $ \Delta\ln\mathcal{Z} \approx 2$ when using a quasi-periodic with derivative kernel, $ \Delta\ln\mathcal{Z} \approx 6$ when using a quasi-periodic with cosine kernel).
We tentatively explain the initial detection as
the keplerian absorbing part of the strong activity signal, although a
detailed analysis of the behavior of the GP/Keplerian modeling is
beyond the scope of this paper. We note however how testing different
assumptions on the stellar activity and the use of different posterior
samplers averted the claim of a likely false positive detection.  More
importantly, in all cases the posterior distribution of the
semi-amplitude of the USP planet in the two-planet model closely
matched the one obtained with the corresponding single-planet model,
independently of the sampler used in the analysis, thus confirming the
robustness of our detection.

\begin{figure*}
\centering
\includegraphics[width=0.95\textwidth, trim=0 0 0 200,clip]{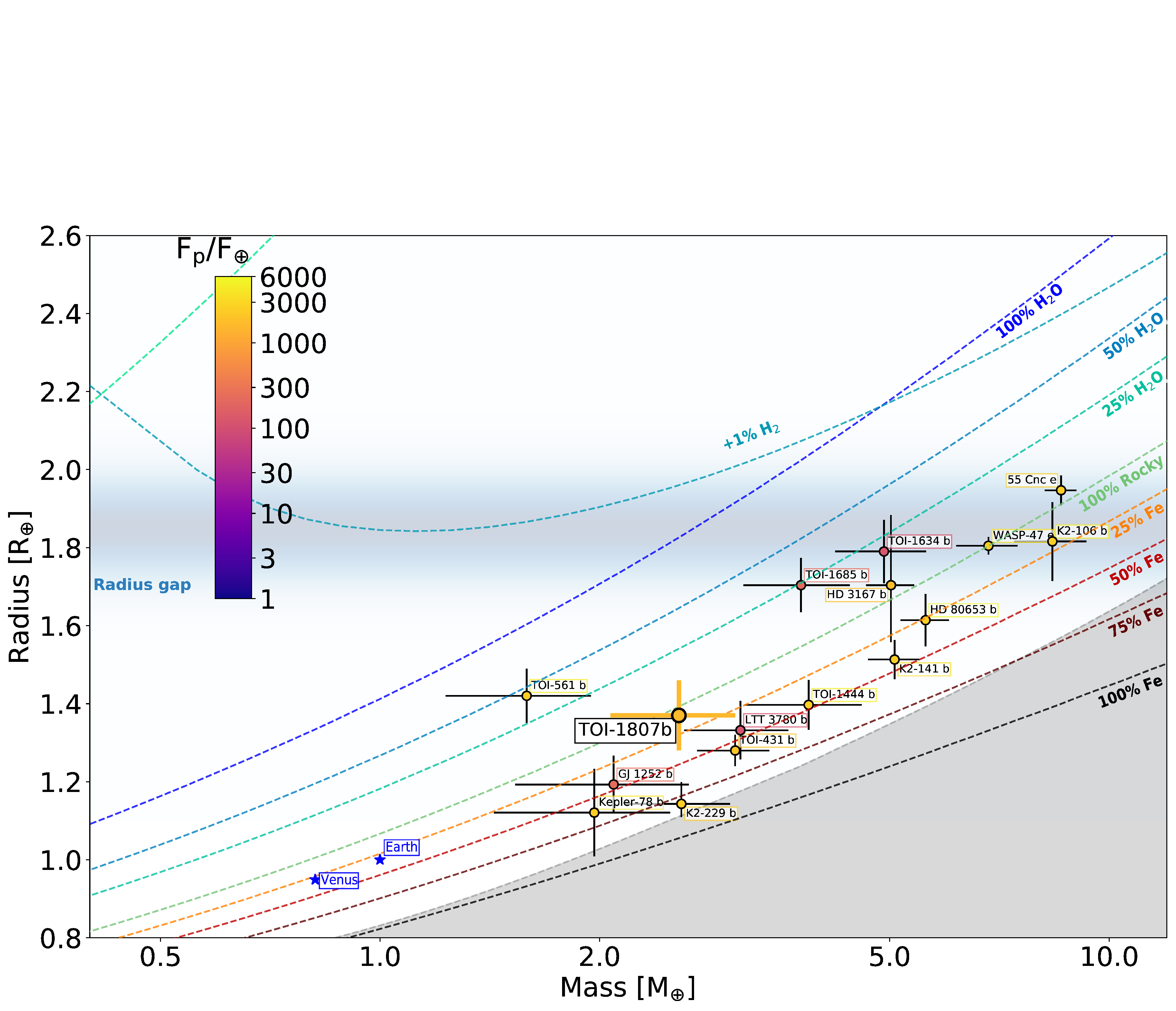}
\caption{Mass-radius diagram for known ultra-short
    period planets with mass and radius measurements more precise
  than 30 per cent and having Earth/Neptune sizes and masses. Points
  are color-coded according to the incidental flux (in Earth units)
  received by the planet. TOI-1807~b is highlighted with a larger
  point and thick contours. For this plot we used the
    TOI-1807~b's mass obtained in the case 1. The dashed colored
  lines are the theoretical mass-radius curves for different chemical
  compositions according to \citet{2019PNAS..116.9723Z}. The shaded
  area represents the maximum value of iron content predicted by
  collisional stripping (\citealt{2010ApJ...719L..45M}). Data from The
  Extrasolar Planets Encyclopaedia (http://exoplanet.eu/, updated to
  April 2022).
\label{fig:10}}
\end{figure*}

\subsection{The expected relativistic apsidal precession of TOI-1807~b}

Due to the closeness of TOI-1807~b to its hosting star, it is logical
asking whether some general relativistic effects are
expected\footnote{Special relativity can be ruled out because, even if
  the planet orbits very fast, $\beta= v/c \simeq 7.93\times
  10^{-4}$}.  If we compare the case of this planet to the well
studied case of Mercury (whose measured apsidal line precession led
to a confirmation of the validity of General relativity) we
see that TOI-1807~b actually orbits in a regime of stronger
gravitational field than Mercury does.  Indeed, the ratio of its
pericenter to the Schwarzschild radius ($r_{\rm S}$) of its hosting
star divided by the same quantity for Mercury is $\simeq 0.05$, so if
for Mercury General Relativity matters, this is a fortiori true also for
TOI-1807~b.

Adopting the data of Table~\ref{tab:2} and \ref{tab:4} for the
hosting star and planet, respectively, we calculated the expected 1st
order general relativistic precession of TOI-1807~b and compared it to
the semi-major axis precession of planet Mercury, which is $\delta
\varphi_{\rm Mer} \sim 0.1$~arcsec per period.

We calculated the $\delta \varphi$ of TOI-1807~b by using the formula:
\begin{equation}
  \delta \varphi = \delta \varphi_{\rm Mer} \frac{M_{\star}+M_{\rm P,b}}{M_\odot + M_{\rm Mer}}\frac{a_{\rm Mer}(1-e_{\rm Mer}^2)}{a_{\rm b}},
\end{equation}

where $M_{\star}$, $M_{\rm P,b}$ and $a_{\rm b}$ indicate the mass of
the hosting star BD+39~2643 and the mass and semi-major axis of the
hosted TOI-1807~b planet ($e_{\rm b}$ is assumed zero, see
Table~\ref{tab:4}), respectively, while $M_{\rm Mer}$, $a_{\rm
  Mer}$, and $e_{\rm Mer}$ refer to the mass, the semi-major axis and
orbital eccentricity of Mercury, respectively.  We obtained
$\delta\varphi \simeq 2.35 \delta\varphi_{\rm Mer} \simeq 0.23$ arcsec
per orbit, which is a significantly large value. On the other side,
gravitational redshift $z_{\rm b}$ of the TOI-1807~b planet would be
impossible to detect, being the ratio:

\begin{equation}
  \frac{1+z_{\rm b}}{1+z_{\rm Mer}}=\frac{\sqrt{1-(r/r_{\rm S})_{\rm Mer}}}{\sqrt{1-(r/r_{\rm S})_{\rm b}}} \sim 1.000000065,
\end{equation}
i.e., having evaluated Mercury's $r/r_{\rm S}$ at pericenter, TOI-1807~b  
shows almost the same very small gravitational redshift of Mercury around the Sun.

\begin{figure*}
\centering
\includegraphics[width=0.95\textwidth, bb=27 357 553 690]{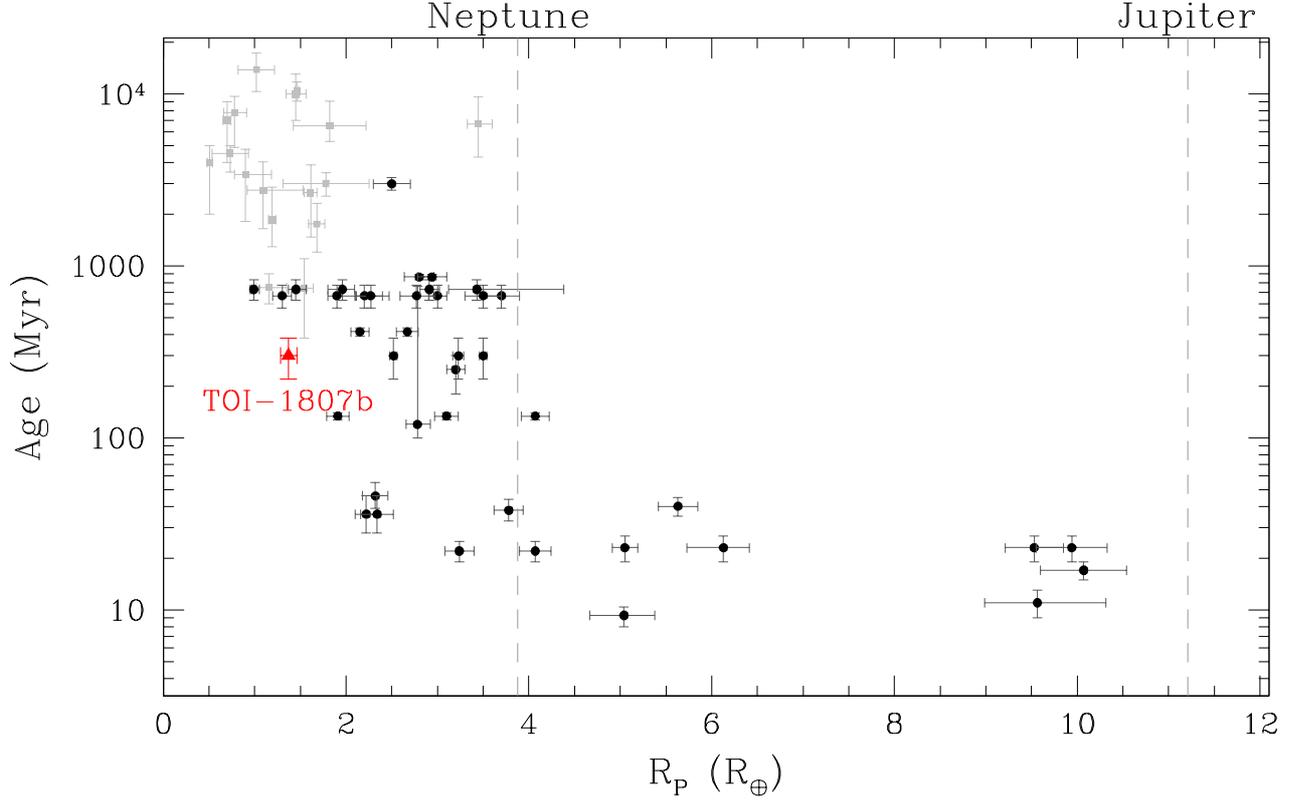}\\
\caption{The Age versus Planetary radius distribution for planets in
  stellar clusters, associations, and moving groups, whose ages are
  well constrained (black circles). Grey squares represent the USP
  planets whose age is measured with an uncertainty lower than 50 per
  cent. TOI-1807~b is represented as a red triangle. See text for
  details.  \label{fig:11}}
\end{figure*}

\begin{table*}
  \renewcommand{\arraystretch}{1.75}
  \caption{Confirmed and candidate close-in ($P_{\rm orb}<100$~d) exoplanets with well measured ages}
  \resizebox{0.99\textwidth}{!}{
    \begin{tabular}{l c c c l | l c c c l}
\hline
Object & Cluster/Association   & Age & $R_{\rm P}$  & Reference & Object & Cluster/Association   & Age & $R_{\rm P}$  & Reference\\
       &                       & (Myr) & ($R_{\oplus }$) &          &        &                       & (Myr) & ($R_{\oplus }$) &         \\
\hline
K2-33~b           &  Upp-Sco      & $9.3^{+1.1}_{-1.1}$          & $5.04    ^{+0.34}_{-0.37}$            & \citet{2016AJ....152...61M} & TOI-2076~c        &  TOI-1807 MG  & $300^{+80 }_{-80}$           & $3.50    ^{+0.04}_{-0.04}$            & \citet{2022arXiv220303194O}  \\
TOI-1227~b        &  Low Cen~Crux OB  & $11 ^{+2  }_{-2 }$       & $9.6^{+0.8}_{-0.6}$                  &  \citet{2022AJ....163..156M}  & TOI-2076~d        &  TOI-1807 MG  & $300^{+80 }_{-80}$           & $3.23    ^{+0.06}_{-0.06}$            & \citet{2022arXiv220303194O}  \\                                 
HIP\,67522~b      &  Sco-Cen      & $17 ^{+2  }_{-2 }$           & $10.07   ^{+0.47}_{-0.47}$            & \citet{2020AJ....160...33R}   &    HD\,63433~b      &  UMa          & $414^{+23 }_{-23}$           & $2.15    ^{+0.10}_{-0.10}$            & \citet{2020AJ....160..179M} \\                               
AU~Mic~b         &  AU~Mic       & $22 ^{+3  }_{-3 }$           & $4.07    ^{+0.17}_{-0.17}$            & \citet{2021AA...649A.177M}       &   HD\,63433~c      &  UMa          & $414^{+23 }_{-23}$           & $2.67    ^{+0.12}_{-0.12}$            & \citet{2020AJ....160..179M} \\                              
AU~Mic~c         &  AU~Mic       & $22 ^{+3  }_{-3 }$           & $3.24    ^{+0.16}_{-0.16}$            &  \citet{2021AA...649A.177M}     &     K2-95~b           & Praesepe      & $670 ^{+100}_{-100}$        & $3.7 ^{+0.2 }_{-0.2 }$               & \citet{2017AJ....153...64M} \\                              
V~1298~Tau~b     &  Tau          & $23 ^{+4  }_{-4 }$           & $9.53   ^{+0.32}_{-0.32}$            & \citet{2022ApJ...925L...2F}      &     K2-100~b          & Praesepe      & $670 ^{+100}_{-100}$        & $3.8 ^{+0.2 }_{-0.2 }$                & \citet{2019MNRAS.490..698B} \\                             
V~1298~Tau~c     &  Tau          & $23 ^{+4  }_{-4 }$           & $5.05    ^{+0.14}_{-0.14}$            & \citet{2022ApJ...925L...2F}    &      K2-101~b          & Praesepe      & $670 ^{+100}_{-100}$        & $3.0 ^{+0.1 }_{-0.1 }$                & \citet{2017AJ....153...64M} \\                             
V~1298~Tau~d     &  Tau          & $23 ^{+4  }_{-4 }$           & $6.13    ^{+0.28}_{-0.28}$            & \citet{2022ApJ...925L...2F}     &     K2-102~b          & Praesepe      & $670 ^{+100}_{-100}$        & $1.3 ^{+0.1 }_{-0.1 }$                & \citet{2017AJ....153...64M} \\                             
V~1298~Tau~e     &  Tau          & $23 ^{+4  }_{-4 }$           & $9.94    ^{+0.39}_{-0.39}$            & \citet{2022ApJ...925L...2F}      &    K2-103~b          & Praesepe      & $670 ^{+100}_{-100}$        & $2.2 ^{+0.2 }_{-0.1 }$                & \citet{2017AJ....153...64M} \\                             
KOI-7368~b       &  Cep-Her      & $36 ^{+10  }_{-8 }$           & $2.22 ^{+0.12}_{-0.12}$               & \citet{2022arXiv220501112B}      &   K2-104~b          & Praesepe      & $670 ^{+100}_{-100}$        & $1.9 ^{+0.2 }_{-0.1 }$                & \citet{2017AJ....153...64M} \\                             
KOI-7913~b       &  Cep-Her      & $36 ^{+10  }_{-8 }$           & $2.34 ^{+0.12}_{-0.18}$               & \citet{2022arXiv220501112B}       &  K2-264~b          & Praesepe      & $670 ^{+100}_{-100}$        & $2.27^{+0.20}_{-0.16}$                & \citet{2018AJ....156..195R} \\                             
Kepler-1627~b    & $\delta$~Lyr  & $38 ^{+6  }_{-5 }$           & $3.78 ^{+0.16}_{-0.16}$               & \citet{2022AJ....163..121B}           & K2-264~c          & Praesepe      & $670 ^{+100}_{-100}$        & $2.77^{+0.20}_{-0.18}$                & \citet{2018AJ....156..195R} \\                           
DS~TUcA~b        &  Tuc-Hor      & $40 ^{+5  }_{-5 }$           & $5.63    ^{+0.22}_{-0.21}$            & \citet{2019AA...630A..81B}      &       HD\,283869~b      & Hyades        & $730 ^{+100}_{-100}$        & $1.96^{+0.13}_{-0.16}$               & \citet{2018AJ....156...46V} \\                            
Kepler-1643~b    &   Cep-Her     & $46 ^{+9  }_{-7 }$           & $2.32    ^{+0.14}_{-0.14}$            & \citet{2022arXiv220501112B}       &      K2-25~b           & Hyades        & $730 ^{+100}_{-100}$        & $3.43^{+0.95}_{-0.31}$               & \citet{2016ApJ...818...46M}   \\                         
K2-284~b          &  Cas-Tau      & $120^{+640}_{-20}$           & $2.78    ^{+0.14}_{-0.12}$            & \citet{2018AJ....156..302D}       &     K2-136A~b         & Hyades        & $730 ^{+100}_{-100}$        & $0.99^{+0.06}_{-0.04}$                & \citet{2018AJ....155....4M}  \\                         
TOI-451~b         &  Psc-Eri      & $134^{+6.5}_{-6.5}$          & $1.91^{+0.12}_{-0.12}$                & \citet{2021AJ....161...65N}        &    K2-136A~c         & Hyades        & $730 ^{+100}_{-100}$        & $2.91^{+0.11}_{-0.10}$                & \citet{2018AJ....155....4M}  \\                         
TOI-451~c         &  Psc-Eri      & $134^{+6.5}_{-6.5}$          & $3.10^{+0.13}_{-0.13}$                & \citet{2021AJ....161...65N}         &   K2-136A~d         & Hyades        & $730 ^{+100}_{-100}$        & $1.45^{+0.11}_{-0.08}$                & \citet{2018AJ....155....4M}  \\                         
TOI-451~d         &  Psc-Eri      & $134^{+6.5}_{-6.5}$          & $4.07^{+0.15}_{-0.15}$                & \citet{2021AJ....161...65N}           &  K-66~b            & NGC\,6811     & $863 ^{+30   }_{-30 }$       & $2.80^{+0.16}_{-0.16}$                & \citet{2013Natur.499...55M}  \\                       
TOI-1098~b        &  Melange-1    & $250^{+50}_{-70}$            & $3.2^{+0.1}_{-0.1}$                   & \citet{2021AJ....161..171T}            & K-67~b            & NGC\,6811     & $863 ^{+30   }_{-30 }$       & $2.94^{+0.16}_{-0.16}$                & \citet{2013Natur.499...55M}  \\                       
TOI-2076~b        &  TOI-1807 MG  & $300^{+80 }_{-80}$           & $2.52    ^{+0.04}_{-0.04}$            & \citet{2022arXiv220303194O}            &    K2-231~b         &  Ruprecht\,147 & $3000^{+250  }_{-250}$    &  $2.5 ^{+0.2 }_{-0.2 }$                & \citet{2018AJ....155..173C} \\                          
\hline
\end{tabular}

}
      \label{tab:5}
\end{table*}

\section{Discussion and Conclusions}
\label{sec:sum}
In this work we confirmed the planetary nature and presented
the  characterization of the youngest USP planet
discovered so far, TOI-1807~b, detected in the \tess light curve of
the active K dwarf star BD+39 2643.

We derived the properties of the host star (effective temperature,
mass, radius, lithium content, etc.) and we analyzed its activity by
using ground-based photometric and spectroscopic observations; in this
way we also obtained a first estimate of the rotation period of the
star ($P_{\rm rot} \sim 8.8$~d). We identified all the stars comoving
with TOI-1807 (and TOI-2076), and through a gyrochronological analysis
performed by using \tess light curves and literature results, we
obtained an age of $300 \pm 80$ Myr for TOI-1807.

Combining the light curves obtained with \tess in two sectors with the
exquisite spectroscopic data-set collected with HARPS-N at TNG (161
measurements in two years, collected with a USP-specific observational
strategy), we modeled the activity of the star by using a GP
approach. We considered different approaches for the modeling of the
stellar activity: the first trained on photometry (case 1) and the
second trained on spectroscopy (case 2). In the two cases, we obtained
measurements of the stellar rotation in agreement, i.e., $P_{\rm rot}
= 8.83 \pm 0.08$~d and
$P_{\rm rot} = 8.84 \pm 0.08$~d for case 1 and 2,
respectively. The results of the stellar activity modeling are
reported in Tables~~\ref{tab:3} and ~\ref{tab:4} and illustrated in
Figs.~\ref{fig:7} and ~\ref{fig:9}.

Simultaneously to the activity, we modeled the USP signal well
detectable both in the light curve and in the RV series after the
removal of the signal due to stellar activity.  The modeling of the planet signal both in the
photometric and spectroscopic series is illustrated in
Figs.~\ref{fig:8} and \ref{fig:9}. As reported in Tables~\ref{tab:3}
and \ref{tab:4}, the USP planet TOI-1807~b orbits its host star in
$P_{\rm b}= 13.1849 \pm 0.0002$~hours; we measured a planetary radius
$R_{\rm P,b}=1.37 \pm 0.09$~R$_{\oplus}$ and a mass $M_{\rm P,b}=2.57
\pm 0.50~M_{\oplus}$ in the case 1 and $M_{\rm P,b}=2.67
\pm 0.43~M_{\oplus}$ in the case 2.

Because of the closeness of TOI-1807~b to the host star, we expect a
relativistic apsidal precession of its orbit. We calculated it by
using the well known relativistic apsidal precession of Mercury as a
proxy. We found an apsidal precession $\delta\varphi \sim 0.23$~arcsec
per orbit, that is twice the apsidal precession measured for Mercury.

TOI-1807~b belongs to the small sample of USP planets with masses and
radii measurements more precise than 30\%. We report this sample in
Fig.~\ref{fig:10}, limiting the analysis to the planets with masses
and radii between Earth and Neptune values. In the mass-radius
  diagram we report the theoretical mass-radius curves for different
  chemical composition; we also report the iron content limit
  predicted by \citet[gray area]{2010ApJ...719L..45M}.  The density
of TOI-1807~b is  $\rho_{\rm b}=1.00\pm
    0.29~\rho_{\oplus}(= 5.5 \pm 1.6$~g\,cm$^{-3}$) considering the
  case 1, or $\rho_{\rm b}=1.04\pm 0.28~\rho_{\oplus}(= 5.7
  \pm 1.5$~g\,cm$^{-3}$) in the case 2, and it is consistent with a
rocky terrestrial composition (silicates and iron), probably with an
iron core between 25\% and 50\% of the total mass, in line with the
large part of USP planets that follow the Earth-like composition
line. On the basis of the distance of the planet from the host star
($a \sim 0.012$ AU), the high incident flux ($F_{\rm P} \simeq
1600~F_{\oplus}$), and the analysis of the mass-radius diagram, we
excluded the presence of any thick envelope composed by H/He; it means
that, in its 300~Myr lifetime, TOI-1807~b has probably already lost
most of its atmosphere via photo-evaporation processes, as expected
for this kind of exoplanets (\citealt{2017MNRAS.472..245L}).

We analyzed the age versus planetary radius distribution for close-in
planets ($P_{\rm orb}$<100~d) as already done in
\citet{2021MNRAS.505.3767N}. The results are shown in
Fig.~\ref{fig:11}: we considered both confirmed and candidate
exoplanets orbiting stars in stellar clusters, associations, and
moving groups (black circles), whose ages are well constrained by
using different methods to derive the stellar age (isochrone fitting,
gyrochronology, etc.). These objects (and the associated references)
are listed in Table~\ref{tab:5}. We also considered USP
planets ($P_{\rm orb}<1$~d) in literature\footnote{Selected from the
  NASA Exoplanet Archive, https://exoplanetarchive.ipac.caltech.edu/}
having a mass $M<10~M_{\rm J}$ or no mass measurements, and age
estimates better than 50 percent: they are shown as gray squares in
Fig.~\ref{fig:11}.  We highlight that TOI-1807~b is the youngest
USP planet discovered so far, and that all the other USP planets are
older than 0.5-1.0~Gyr.  As already reported in
\citet{2021MNRAS.505.3767N}, for close-in exoplanets there is a trend
of the radius as a function of the age: objects having $R_{\rm P}
\lesssim 3.5$-$4~R_{\oplus}$ are more concentrated at ages >100
Myr\footnote{However, the lack of young planets in this radius
  interval could be linked to an observational bias due to the
  difficulty of identifying shallow transits around very active
  stars.}; planets having radii $4 ~R_{\oplus} \lesssim R_{\rm P}
\lesssim 12~R_{\oplus}$ are distributed at ages <100~Myr and the lack
of planets at older ages can not be due to observational bias: on the
contrary, they should be more easily detectable. This trend can be
interpreted in the context of atmosphere loss of low mass close-in
planets linked to photoevaporation mechanisms that happen on
time-scales <100-200~Myr
(\citealt{2019AREPS..47...67O,2020SSRv..216..129O}).  As we have
already reported in the analysis of the mass-radius diagram,
TOI-1807~b has no extended atmosphere. This is in agreement with its
location in the age-planetary radius diagram: indeed, it is located in
the region of small-radii ($R_{\rm P} \lesssim 4 R_{\oplus}$),
"middle-age" (100-1000~Myr), close-in, low-mass planets that have
probably already lost their atmospheres during their early life.  The
host star is characterized by an X-ray emission $L_X$ close to the
median value for stars of the same spectral type and age
(e.g. \citealt{2002ASPC..269..107M, 2005ApJS..160..390P}), therefore
we expect that the planet may have suffered a significant mass loss
due to its small distance from the host star and to its small
mass. Indeed, low gravity planets are subject to more intense
evaporation than the most massive planets
(\citealt{2008A&A...479..579P}). After a few hundred million years the
planet may have lost all of its atmosphere and simultaneously
contracted to its present size. Its radius puts the planet at the low
size peak of the bimodal distribution of planetary radii observed in
planets with orbital periods less than 100 days
(\citealt{2017AJ....154..109F}). \citet{2020ApJ...891..158M}
demonstrated that evaporation due to energy radiation can produce the
observed bimodal distribution. They predict also that the 90 $\%$ of
the planets at the $\sim$1.3~$R_{\oplus}$ peak completely lost their
envelope, consistently with the hypothesis that TOI-1807~b is today
without atmosphere.

Because many USP planets are hosted by multi-planet systems, finally
we also checked for the presence of a second non-transiting planet
orbiting TOI-1807 (Case 3). Even if we obtained some significant
detection (in some cases with a significance $>5\sigma$) in the RV
series, we found that period and semi-amplitude of a hypothetical
planet c changed drastically on the basis of the activity model used
for the analysis (while almost nothing changes for planet
b). Comparing the Bayesian evidence of the case 1/2 with case 3 and
analyzing the facts previously reported, we concluded that any
detection obtained in this analysis is due to a false positive
(probably linked to the stellar activity modeling) and that new data
are mandatory for the detection of an hypothetical second planet
orbiting TOI-1807.


\begin{acknowledgements}
  The authors thank the anonymous referee for carefully reading the
  paper and for the useful comments and suggestions that have
  contributed to improving the quality of the manuscript.  DN
  acknowledges the support from the French Centre National d'Etudes
  Spatiales (CNES).  This work has been supported by the PRIN-INAF
  2019 "Planetary systems at young ages (PLATEA)". We acknowledge
  financial support from the ASI-INAF agreement n.2018-16-HH.0.  This
  paper includes data collected by the {\it TESS} mission. Funding for
  the {\it TESS} mission is provided by the NASA Explorer
  Program. This work has made use of data from the European Space
  Agency (ESA) mission {\it Gaia}
  (\url{https://www.cosmos.esa.int/gaia}), processed by the {\it Gaia}
  Data Processing and Analysis Consortium (DPAC,
  \url{https://www.cosmos.esa.int/web/gaia/dpac/consortium}). Funding
  for the DPAC has been provided by national institutions, in
  particular the institutions participating in the {\it Gaia}
  Multilateral Agreement.  This work makes use of observations
  collected at the Asiago Schmidt 67/92cm telescope (Asiago, Italy) of
  the INAF - Osservatorio Astronomico di Padova.  This research has
  made use of the Keck Observatory Archive (KOA), which is operated by
  the W. M. Keck Observatory and the NASA Exoplanet Science Institute
  (NExScI), under contract with the National Aeronautics and Space
  Administration (Prog. ID H6aH, PI Boesgaard; Prog. ID N038Hr, PI
  Fischer; Prog. ID H198Hr, H212Hr, H170Hr, PI Shkolnik; Prog. ID
  A297Hr, PI Fischer; Prog. ID C103Hr, PI Bowler) Based on data
  obtained from the ESO Science Archive Facility (Prog. ID
  091.C-0216(A), PI Rodriguez; Prog. ID 090.D-0133(A), PI Datson;
  Prog. ID 072.C-0488(E), PI Mayor, and 183.C-0972(A), PI Udry).
  Based on data retrieved from the SOPHIE archive at Observatoire de
  Haute-Provence (OHP), available at atlas.obs-hp.fr/sophie (Prog. ID
  09B.PNP.CONS, 11A.PNP.CONS, 17A.PNCG.SOUB). Data at IRTF/iSHELL were
  obtained by A.Sohani, F. Zohrabi, D. Vermilion, S. Arnez,
  C. Geneser, A. Tanner, K. Collins, J. Berberian, I. Helm, and
  P. Newman.
\end{acknowledgements}


\bibliographystyle{aa}
\bibliography{biblio}

\begin{appendix}

\section{Light curve correction}
\label{sec:light_curve_corr}
\begin{figure*}
  \centering
  \includegraphics[width=0.9\textwidth, bb=19 520 571 717]{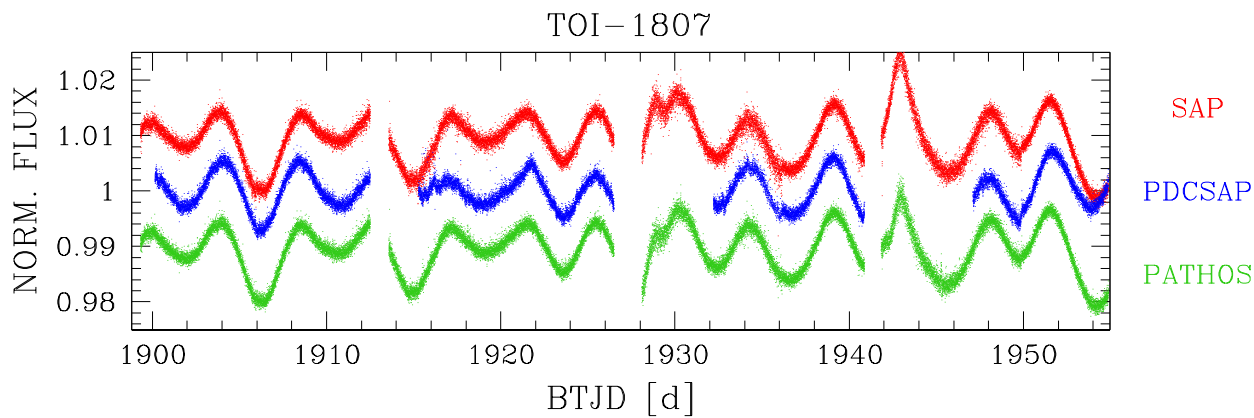} \\
  \caption{Comparison between the raw SAP light curve of TOI-1807 (in
    red) and the same light curve corrected by using the PDC (in blue)
    and the PATHOS (in green) pipelines. \label{fig:A1}}
\end{figure*}

In Fig.~\ref{fig:A1} we reported a comparison between the raw SAP
light curve (red points), the light curve corrected by the \tess team
(PDCSAP, blue points) and the light curve corrected using the PATHOS
pipeline and adopted in this work (green points).

\section{Comoving objects}
\label{sec:comoving}

\subsection{Search for comoving objects}

In order to improve the age constraints for our targets and better
understand their dynamical environment, we searched for wide physical
companions and comoving objects.

We first noticed, as also done by \citet{2021AJ....162...54H}, that
the planet hosts TOI-1807 and TOI-2076 have very similar space
velocity, having a difference of about 0.5 km\,s$^{-1}$ considering
all the three coordinates, and consistent age diagnostics.  Therefore,
they are likely comoving and coeval.  In order to look for additional
comoving objects, we exploited the catalog by \citet{smart2021}, which
includes the U, V, and W space velocities for Gaia stars within 100 pc
with available radial velocity.  We searched over the whole sky for
targets within 60 pc which have a space velocity $\delta$UVW
(considering the three components) which differ by less than 4
km\,s$^{-1}$ with respect to the mean of TOI-1807 and TOI-2076; 76
objects fulfilling these criteria were identified, some of which
previously known as being young and/or active.  Few additional objects
can be considered being physical companions of one of the selected
targets.  We noticed that the space velocities of TOI-1807 and
TOI-2076 somewhat resemble those of B3 subgroup of the Local
Association identified by \citet{asiain1999}. An age of 300$\pm$120
Myr was estimated in their study from isochrone fitting of early type
stars.

As well known in the literature \citep[see, e.g.][]{lopezantiago2009}
interlopers with similar kinematics but different ages are likely to
exist.  For intermediate age groups (few hundreds Myr old), the
existence itself of some groups is questioned \citep[see, e.g.][for
  Castor moving group]{Zuckerman2013}.

To shed further light on the possibility of a group of coeval objects
we considered the selected targets individually, deriving age
indicators from the literature and public data, when available.

\subsection{{\it TESS} data reduction}
Out of 79 stars among wide companions and comoving objects, 55 were
observed by \tess in one or more Sectors.  We obtained the light
curves of these stars from the Full Frame Images (FFIs) by using the
PATHOS pipeline described in detail by \cite{2019MNRAS.490.3806N} and
also reported in Section~\ref{sec:tessphot}.  We selected the best
aperture for each target comparing their mean rms distributions, as
described in detail by \cite{2020MNRAS.495.4924N} and
\citet{Messina2022}.

\subsection{Rotation period}

We analyzed the \tess light curves of all 55 stars to measure the
rotation period. Details on the procedure can be found in
\citet{Messina2022}. Briefly, we used three different methods:
Generalized Lomb-Scargle (GLS; \citealt{2009A&A...496..577Z}), CLEAN
(\citealt{Roberts87}) and AutoCorrelation Function (ACF;
\citealt{Mcquillan13}) in order to provide a "grade" of confidence on
the correctness of the measured rotation periods ("A" if the three
methods provided a similar value; "B" when only two methods found a
similar value; "C" when period estimates differed in all three
methods). We selected only rotation periods with grade "A" and "B" and
with False Alarm Probability (FAP) $<$ 0.1\%. We followed the method
used by \cite{Lamm04} to compute the error associated with the period
determination.  On a total sample of 55 stars, we measured 34 periods
with grade A and 7 periods with grade B (see Table\,\ref{tab:b1}).
Data from the literature have been considered as well.  Five targets
have the rotation measured by HATNET \citep{Hartman11}, two by Mearth
\citep{West15} one by ASAS \citep{kiraga2012}, two by ASAS-SN
\citep{Jayasinghe18}, five by KELT \citep{Oelkers18}, and one by
Mascara \citep{Burggraaff18}.  Overall, the rotation period is
measured for 47 targets.

\subsection{Other age indicators}

Additional age diagnostics were derived from the literature or from
available high-resolution spectra.  For TOI-2076 we exploited the
spectra gathered by the GAPS program, analyzed as those of TOI-1807,
being published in a forthcoming paper in preparation.  We also
analyzed reduced spectra available in public archives:
SOPHIE\footnote{\url{http://atlas.obs-hp.fr/sophie/}} (3 objects),
FEROS \footnote{\url{http://archive.eso.org/cms.html}} (1 object), and
HIRES/Keck \footnote{\url{https://koa.ipac.caltech.edu/cgi-bin/KOA/}}
(6 objects)

The resulting measurements of Li 6708\AA~doublet are listed in
Table~\ref{tab:b1}.

\subsection{A group of coeval and comoving objects?}

From the comparison of the age indicators, it results that TOI-1807,
TOI-2076, and additional 24 comoving stars have a similar age,
intermediate between Pleiades and Hyades.  Using Group X \citep[age
  300$\pm$60 Myr,][]{Messina2022} and the open cluster NGC\,3532
\citep[age 415$\pm$30 Myr][]{dias2021} as comparisons, the estimated
age for these stars is of 300$\pm$80 Myr. Other 32 stars among the
comoving objects appear to have different age (older in most cases),
and then are classified as kinematic interlopers.  Finally, 18 stars
have ambiguous status (mostly because of lack of information beside
kinematics, or because of the uncertainty in the age determination).

\begin{figure*}
  \centering
  \includegraphics[width=0.75\textwidth, bb=19 427 583 714]{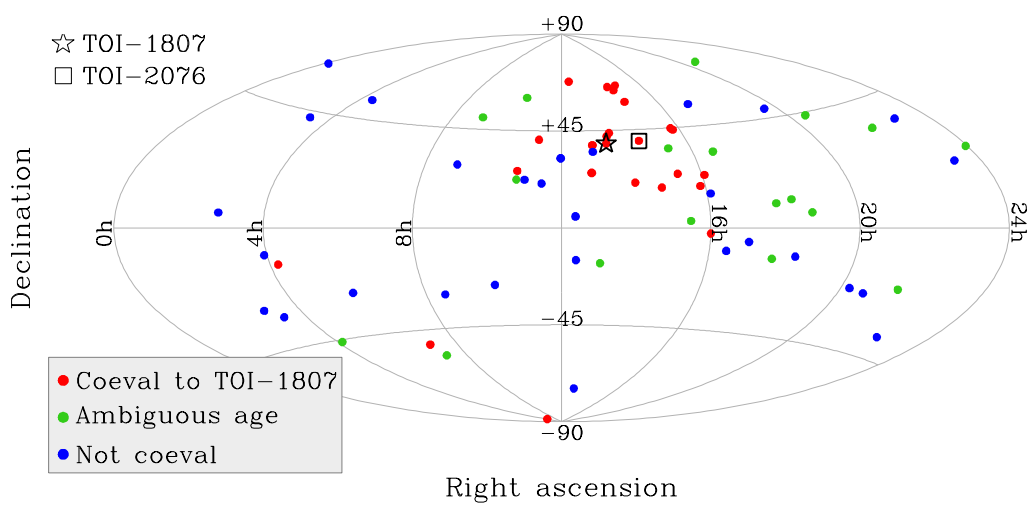} \\
  \caption{ \label{fig:B1} Right ascension and declination for the
    kinematically selected stars with U,V,W close to that of TOI-1807
    and TOI-2076. Stars coeval to TOI-1807 and TOI-2076, not coeval,
    or with undetermined age are plotted with different
    colors. TOI-1807 and TOI-2076 are indicated with a star
      symbol and a square, respectively. }
\end{figure*}

The 23 kinematically selected stars which result to have an age
similar to TOI-1807 and TOI-2076 are moderately clustered on the sky
(Fig. \ref{fig:B1}) and are at a similar distance (median 42.1 pc, rms
7.4 pc).  The other three stars (ASAS J041255-1418.6, HD 34652, and
UCAC2 9643914) are located in a different region of the sky. The stars
classified with ambiguous or discrepant age are instead scattered over
the whole sky.

\subsection{Individual objects}

Description of individual systems is provided below

\begin{description}

\item {\bf HD 123 = V640 Cas A = TIC 604446831} Triple system, formed
  by a G2 + G8 star separated by 1.44$^{\prime\prime}$, with the
  secondary being itself a spectroscopic binary.  After deblending of
  2MASS magnitude, the star lies above similar-color members of Group
  X in the period-color diagram, implying an age slightly older than
  300\,Myr and similar to Hyades.  Li EW \citep{takeda2005} and log
  R$^{\prime}_{HK}$ \citep{borosaikia2018} are also compatible with
  this age assignment.  \citet{fuhrmann2008} noted instead a lower
  activity level for the secondary, suggesting that the primary is
  somewhat rejuvenated by accretion of angular momentum. Independently
  on this speculation, the system results older than TOI-1807.

\item {\bf UCAC4 180-001659 = TIC 231019115} M1 dwarf, proposed as a
  member of Tuc-Hor association by \citet{riedel2017}.  However, in
  the period-color diagram, this target lies significantly above
  similar-color members of Group X, implying an age older than
  300\,Myr.  Also, Li non detection and gravity from 7699 \AA\, line
  \citep{riedel2017} are compatible with a main sequence object, not
  particularly young.  RV by \citet{riedel2017} and Gaia EDR3 are
  discrepant at 2.5$\sigma$ level, suggesting the possibility that the
  star is an SB.  BANYAN group assignment using Gaia data also
  supports a field object (87\% probability). We conclude that the
  star is a field M dwarf somewhat older than TOI-1807.

\item {\bf HD 17250C = TIC 318841149} Isolated component (separation
  494$^{\prime\prime}$) of a triple system.  The other two components
  (HD17250 A and B) form a tight pair (separation
  1.9$^{\prime\prime}$) and do not satisfy our kinematic selection
  criteria for objects comoving with TOI-1807 and TOI-2076.  The
  system is considered in the literature as a possible member of
  Tuc-Hor association \citep[e.g.,][]{zuckerman2011}.  Our own
  kinematic analysis with BANYAN $\Sigma$ and Gaia data yields
  membership probabilities for Tuc-Hor of 98.7\% and 89\% for HD
  17250\,A and C, respectively.  HD 17250\,C is expected to be a M2.5
  star from its colors.  Its position on color-magnitude diagram (CMD) is
  well above the main sequence and compatible with the age of Tuc-Hor,
  unless it is itself an unresolved binary.  Instead, in the
  period-color diagram, HD 17250\,C, with period graded B, is well
  above the sequence of Group X members, suggesting an age much older
  than 300\,Myr.  Considering the uncertainty in the period
  determination and the other evidence supporting the membership to
  Tuc-Hor association, we conclude that this system is likely much
  younger than TOI-1807.

\item {\bf HD 18414 = TIC 91556441} \tess data show a significant but
  irregular non-periodic variability. There are no indication of youth
  in the literature beside kinematics.

\item {\bf Wolf 140 = TIC 416098108} M3.5 star. The long rotation
  period reported by \citet{newton2016} and the low levels of
  H$\alpha$ activity \citep{lu2019,cifuentes2020} indicate that this
  object is an old interloper.

\item {\bf HD 20776 = TIC 44628969} The \tess light curve of this
  target appeared non variable during the time span of
  observations. The low $v \sin i$ of the star \citep[1.0
    km\,s$^{-1}$, ][]{nordstrom2004} and lack of X-ray detection also
  support an old age.

\item {\bf 2MASS J03550477-1032415 = TIC 55441420} Ultra-cool object
  (spectral type M8.5) with signatures of low gravity
  \citep{gagne2015}. A very young age is also supported by the strong
  Li line and activity \citep{shkolnik2017}.  The \tess light curve of
  this target appeared non variable during the time span of
  observations.  The star is flagged as possible member of $\beta$ Pic
  MG in the above studies while our own BANYAN analysis with Gaia EDR3
  parameters supports membership to Columba association (92\%).
  Independently on the membership assignment, the object is much
  younger than TOI-1807.

\item {\bf HD 26413 = TIC 152473055} Triple system, formed by a F3
  primary with one companion (B) at 0.8-1.5$^{\prime\prime}$ and
  another one (C) at 18$^{\prime\prime}$. From absolute mag, B and C
  are expected to be a K2/K3 and a M3 star, respectively.  The RV of
  the primary from Gaia DR2 and \citet{nordstrom1985} differ by 13.2
  km\,s$^{-1}$, suggesting the presence of a fourth component.  The
  very short photometric periodicity (P = 0.395\,d) likely refers to
  the brightest F3 component and arises from flux rotational
  modulation (it is compatible with the observed $v\sin{i}$ and
  stellar radius) rather than from pulsations, considering the
  significant evolution of the light curve amplitude during the \tess
  observations; the last likely due to active region growth and decay.
  The position on the CMD is slightly above main sequence implying an age
  older than TOI-1807, unless the photometry is significantly
  contaminated by the unknown spectroscopic component. The position on
  the period-color diagram is roughly consistent with that of Group X,
  although the period distribution is poorly defined at these blue
  colors.

\item {\bf ASAS J041255-1418.6 = TIC 332660818} It is a M1 star.  We
  measured from \tess photometry a rotation period P= 7.3$\pm$1.0\,d
  in agreement with the literature value P = 7.30\,d from ASAS
  \citep{kiraga2012}. The rotation period is fully compatible with
  that of similar-color members of Group X, implying an age of about
  300\,Myr.  No Lithium is seen in the archive FEROS spectrum we
  analyzed, as also found by \citet{bowler2019}, indicating age older
  than about 100 Myr.  High levels of chromospheric and coronal
  activity have been measured \citep{bowler2019,zerjal2017}.

\item {\bf HD 34652 = TIC 317383399} We measured from \tess photometry
  a rotation period P= 1.753$\pm$0.026\,d and a secondary period (of
  comparable power in both GLS and CLEAN periodograms) P= 1.527\,d
  likely arising from surface differential rotation.  In the
  period-color diagram, this lies among similar-color members of Group
  X, then compatible with an age of about 300\,Myr.  The isochrone
  fitting \citep[using PARAM, ][]{dasilva2006} yields an inconclusive
  value of 970$\pm$800Myr.  The X-ray emission (R$_X$=$-$4.89) is
  close to the median value of Hyades members of similar color, but
  compatible with fairly broad range of ages.  We conclude that this
  star may have a similar age to TOI-1807 but the limited sensitivity
  of age diagnostics for mid-F stars does not allow us firm
  conclusions.

\item {\bf HD 27857 = TIC 470480305} Moderately young G5 star.  Both
  the rotation period from \tess photometric time series and Li EW by
  \citet{kim2016} indicate an age much older than TOI-1807 and
  slightly older than the Hyades.

\item {\bf UCAC2 20545888 = TIC 33062967} Emission-line M dwarf. In
  the period-color diagram, TIC 33062967 lies significantly above
  similar-color members of Group X, implying a likely older age than
  300\,Myr.

\item {\bf UCAC2 9643914 = TIC 238088359} M dwarf with large RUWE in
  Gaia EDR3 (11.0) indicating a spatially unresolved binary.  We
  measured from \tess photometry a rotation period P= 2.82$\pm$0.10\,d
  superimposed to a secondary period of about 20\,d. The position in
  the period-color diagram is consistent with similar-color members of
  Group X.  The star has also prominent X-ray emission (R$_X$=$-$3.2).

\item {\bf GSC 08558-01964 = TIC 294154590} Close binary star
  \citep[M1.5+M4, projected separation 1.1$^{\prime\prime}$,
  ][]{bergfors2010} The star results quite active in $H{\alpha}$,is
  X-ray bright \citep{riaz2006} and with fast rotation from $v\sin{i}$
  measurement \citep{malo2014}.  Instead, there are no determinations
  of the rotation period.  Therefore, the star may have similar age to
  TOI-1807 and Group X but considering the spectral type and the
  additional uncertainties due to binarity we are not able to put
  tight constraints on system age.

\item {\bf HD 72687 = TIC 283397452} We measured from \tess photometry
  a rotation period P= 3.87$\pm$0.31\,d in agreement with the
  literature value P = 3.84\,d from KELT
  (\citealt{Oelkers18}). Considering also additional age diagnostics
  such as lithium \citep{torres2006}, this stars appears younger than
  similar-color members of Group X.

\item {\bf HD 76332 = TIC 126312524} Moderately active star and binary
  system, as revealed by the large RUWE in Gaia EDR3 and, more
  recently, by the RV + astrometry analysis \citep[a=78 mas, mass of
    companion 0.47 M$_{\odot}$][]{dalal2021}.  We measured from \tess
  photometry a rotation period P= 10.7$\pm$1.1\,d in agreement with
  the literature value P = 13.25\,d from KELT \citep{Oelkers18}. The
  position in the color-period diagram suggests an age significantly
  older than 300\,Myr. This is also supported by the Li EW measured by
  us on SOPHIE spectra (49.9$\pm$1.0 m\AA), which is below that of
  Hyades stars of similar color.

\item {\bf TYC 3430-1344-1 = TIC 456344030} Its position in the
  period-color diagram is slightly above similar-color members of
  NGC\,3532, implying an age slightly older than 300\,Myr.  There are no
  additional data on age indicators to further constrain the system
  age.

\item {\bf TYC 6621-759-1 = TYC 309024402} Star with large RUWE in
  Gaia EDR3 (7.30) and then probable tight binary system.  Its
  position in the period-color diagram is significantly above
  similar-color members of NGC\,3532, implying an age much older than
  300\,Myr.  There are no other indication of youth in the literature
  beside kinematics.

\item {\bf TYC 4144-944-1 = TIC 355868825} We measured from \tess
  photometry a rotation period P= 11.4$\pm$2.9\,d in agreement with
  the literature value P = 12.75\,d from KELT
  (\citealt{Oelkers18}). The rotation period is slightly slower than
  similar-color members of Group X, indicating a slightly older age
  than 300\,Myr.

\item {\bf 40 LMi = TIC 241197867} It is a A4Vn star identified as
  close binary by \citet{galicher2016}.  It also has significant
  $\Delta \mu$ signature.  From the H band magnitude, the companion is
  expected to be a M2.5/M3 star from Mamajek tables.  The system has
  also an X-ray counterpart from CHANDRA observations. Assuming this
  is entirely originating from the late type companion, we estimated
  R$_X$=$-$2.8, compatible with both Pleiades and Hyades early M
  dwarfs.  Isochrone age is 300$\pm$200 Myr.  We conclude that the
  star has an age of few hundreds Myr, then compatible with the age of
  TOI-1807, but with significant uncertainty.

\item {\bf UCAC4 562-049479 = TIC 95776155} Its position in the
  period-color diagram is consistent with similar-color members of
  NGC\,3532, implying an age of about 300\,Myr.

\item {\bf LP 373-35 = TIC 84925818} We measured from \tess photometry
  a rotation period P= 18$\pm$6\,d in rough agreement with the
  literature values P = 23.8\,d from MEarth (\citealt{West15};
  \citealt{newton2016}). Its position in the period-color diagram
  suggests an age much older than that of Group X.

\item {\bf 2MASS J11155546+4049569 = TIC 450332591} We measured from
  \tess photometry a rotation period P= 4.72$\pm$0.41\,d in agreement
  with the literature values P = 4.764\,d from HATNET
  (\citealt{Hartman11}).  Its position in the period-color diagram
  suggests an age similar to that of Group X.  The X ray luminosity
  (R$_X$ =$-$2.8) and the moderate H${\alpha}$ emission are consistent
  with this age estimate.

\item {\bf HD 99419 = TIC 3901189} Rotation period is not
  available. Indicators from archive SOPHIE spectra (EW Li well below
  the Hyades locus, slow \vsini, low RV scatter, modest chromospheric
  emission) indicate an age significantly older than TOI-1807.

\item {\bf HD 103928 = TIC 99302268} and {\bf HD 103928B = TIC
  99302269} Binary system with components separated by
  7$^{\prime\prime}$.  The primary is a late A/early F star, while the
  estimated spectral type of the secondary from absolute magnitude and
  colors is around M1-M2.5.  A photometric period of 0.70 d is
  reported by \citet{koen2002}, possibly due to the pulsations of the
  primary.  Assuming that all the X-ray flux is originating from the
  secondary, the X-ray luminosity is below the loci of Hyades and
  Pleiades M dwarfs, suggesting an older age, although possibly
  compatible with the lower edge of their distributions.  Isochrone
  fitting of the primary is also inconclusive (550$\pm$450 Myr).
  There is then some indication that the system is older than
  TOI-1807, although the same age can not be ruled out considering the
  large uncertainties.

\item {\bf 17 Vir = TIC 377227654} and {\bf 17 VirB} Visual binary,
  formed by a F8 and a K5 star. The low levels of chromospheric and
  coronal activity \citep[e.g., ][]{wright2004} and the 3.1$\pm$0.7
  Gyr isochrone age \citep{aguileragomez2018} show that this pair is
  an old kinematic interloper.

\item {\bf TYC 4394-114-1 = TIC 148910632} K star flagged as a
  candidate member of the AB Dor association by \citet{schlieder2012};
  however our own BANYAN analysis with Gaia parameters rules out
  membership in any known association.  The position on the
  color-period diagram is consistent with similar-color members of
  Group X, implying an age of about 300\,Myr.

\item {\bf UCAC4 376-064366 = TIC 40560697} The periodogram analysis
  of \tess data did not provide any reliable rotation period. There is
  no indication of youth in the literature beside kinematic. This M
  dwarf is then probably older than TOI-1807.

\item {\bf LX Com = HIP 62758 = HD 111813 = TIC 450335652} and {\bf
  BD+26 2401 = HIP 62794 = TIC 156514310} They form a very wide common
  proper motion pair (separation 373$^{\prime\prime}$ = 14150 au).
  BD+26 2401 is an SB2 with orbital solution \citep[period
    19.436$\pm$0.001d][]{halbwachs2012}.  This is likely responsible
  for the significant RUWE.  Both stars are X-ray sources, with RX of
  -4.59 and -4.55 for LX Com and BD+26 2401, respectively.  For LX Com
  we measured on \tess data a rotation period of P =
  3.9$\pm$0.25. Since the light curve is clearly double dip, it
  implies that the rotation period is P $\times$ 2 = 7.8\,d in good
  agreement with \citet{strassmeier2000} who derived a period of
  7.74d.  For BD+26 2401, we measured a period of P= 20.3 $\pm$ 3\,d
  using GLS and CLEAN methods, which may be related to the orbital
  period of the binary. We then do not consider it for dating
  purposes.  We instead rely on indicators of LX Com, which are all
  compatible with an age intermediate between Hyades and Pleiades.

\item {\bf L68-145 = TIC 361397288} This star is moderately active
  \citep{zerjal2017} but it lies well above the period distribution of
  Group X members, being likely significantly older (most likely with
  age similar to Hyades and Praesepe).

\item {\bf GJ 490A = TIC 950125199} Quadruple system formed by two
  pairs of M dwarfs separated by 16$^{\prime\prime}$.  GJ 490A and GJ
  490B have projected separations of 0.10$^{\prime\prime}$ and
  0.17$^{\prime\prime}$, respectively \citep{bowler2015}.  We measured
  from \tess photometry a primary rotation period P= 3.37$\pm$0.21\,d
  in agreement with the literature values P = 3.17\,d from
  \citet{Wright11} and P = 3.36\,d from \citet{Norton07}, as well as a
  secondary period P = 0.4681\,d, which likely belongs to one of the
  unresolved secondary components.  This star lies well inside the
  period distribution of Group X members. However, a quite broad range
  of ages are possible considering the rotation distribution of M
  dwarfs with measured ages.  The system was also proposed to be a member of
  younger groups, such as Tuc-Hor \citep{shkolnik2012} and AB Dor
  \citep{malo2013}.  BANYAN analysis using Gaia EDR3 parameters
  indicates a field object.  Kinematic parameters and group assignment
  should be taken with caution considering the multiplicity of the
  system.  The marginal detection of Lithium on HIRES/Keck spectra,
  although uncertain because of the difficulty in continuum placement
  for such cool object, supports an age significantly younger than
  that of TOI-1807.

\item {\bf HD 112733 = TIC 17740825} and {\bf HD 112733B = TIC
  17740827} They form a wide pair (36.0$^{\prime\prime}$ $\sim$ 1500
  au), with the secondary being itself a SB in short-period orbit
  \citep{galvez2006,strassmeier2012}.  The system was flagged as
  possible member of Hercules-Lyra association by
  \citet{eisenbeiss2013}.  They note a small (about 2 km\,s$^{-1}$)
  discrepancy in W velocity with respect to the bulk of the
  association.  The age indicators discussed by \citet{eisenbeiss2013}
  are in agreement with the adopted age for Her-Lyr.  The system was
  considered in several works on young stars and planet searches with
  ages ranging from 100 Myr \citep{meshkat2017}, 250 Myr
  \citep{vigan2017}, 260 Myr \citep{brandt2014}. From \tess time
  series, where the two components are unresolved, we measured a
  photometric period of 5.15$\pm$0.48\,d days, while
  \citet{strassmeier2012} reported a tentative H${\alpha}$ period of
  2.3 days.  The \tess light curve is less noisy for an aperture
  centered on the secondary, suggesting that the periodicity belongs
  to this component.  Assuming the photometric period belongs to the
  secondary, the position on period-color diagram would imply an age
  younger than that of the Pleiades. An age similar to Group X and
  TOI-1807 would be possible if the rotational curve is double-dip and
  the rotation period is actually two times the observed one.  Even
  more likely, the presence of the close companion should have altered
  the rotation of the star through tidal effects and indeed our
  photometric period is marginally consistent with the orbital period
  by \citet{griffin2010} and \citet{strassmeier2012}.  Focusing on
  Lithium EW, we estimate a most probable age of 250 Myr but
  compatible with the 300 Myr of Group X within errors.

\item {\bf HD 113414 = TIC 1727745} Li EW by \citet{waite2011} and
  X-ray emission (R$_X$ = $-$4.23) are close to the locus of Pleiades,
  suggesting a similar age.  In the period-color diagram, the star is
  close to the sequence of members of Group X, but the age sensitivity
  of rotation period is quite limited for F7/F8 stars and the
  measurement is also compatible with Pleiades.  We conclude that the
  star is likely younger than Group X.

\item {\bf HIP 65775 = HD 117378 = TIC 288405352} F9.5 star, with
  astrometric acceleration and significant proper motion differences
  \citep{makarov2005}.  RV from \cite{nordstrom2004} (2 epochs) and
  Gaia DR2 agree within errors.  The star is active, as resulting from
  X-ray and chromospheric emission \citep{gray2003}.  Our rotation
  period from \tess light curve (4.38 d) is fully compatible with the
  expectation from the activity level and projected rotational
  velocity.  The position on the color-period diagram is consistent
  with similar-color members of Group X, implying an age of about
  300\,Myr.  The Li EW measured by us on HIRES/Keck spectra is
  intermediate between Hyades and Pleiades median values for stars of
  similar color, fully supporting the 300 Myr age.

\item {\bf TYC 3032-368-1 = TIC 288487378} Triple system, with an
  isolated primary at 20$^{\prime\prime}$ from a close pair of very
  low mass stars ({\bf Gaia 1501788886175060608 } and {\bf Gaia
    1501788881878538496}). Only the primary is included in the
  \citet{smart2021} catalog.  We derived a rotation period of 12.2
  days, which most likely belongs to the primary considering the
  faintness of the low mass companions ($\Delta G > 7$).  The position
  on the color-period diagram is consistent with similar-color members
  of Group X, implying an age of about 300\,Myr.  The X-ray luminosity
  is intermediate between Hyades and Pleiades of similar color and
  compatible with the proposed age.

\item {\bf HIP 68732 = TIC 135154868} Triple system, as the comoving
  object {\bf StKM 1-1119} at 62$^{\prime\prime}$ is itself a close
  pair (sep 1$^{\prime\prime}$).  The position of the primary on the
  color-period diagram is consistent with similar-color members of
  Group X, implying an age of about 300\,Myr.  The X-ray source 1RXS
  J140409.4+204449 lies at 18$^{\prime\prime}$ from the secondary and
  45$^{\prime\prime}$ from the primary. The X-ray luminosity is a bit
  lower than expected from the rotation age.

\item {\bf RX J1419.0+6451 = TIC 166087190} M3 star proposed as a
  member of AB Dor MG by \citet{malo2013}.  Membership analysis with
  BANYAN including Gaia astrometry instead leads to field object
  classification.  The position on the color-period diagram is
  consistent with similar-color members of Group X. The star has also
  prominent X-ray emission (R$_X$=$-$3.03).

\item {\bf TOI-2076 = TIC 27491137} Host of a multi-planet system
  \citep{2021AJ....162...54H,2022arXiv220303194O}. It will be
  discussed in a forthcoming paper. For the purpose of this work, we
  note that in the period-color diagram, TIC 27491137 lies in the same
  position of similar-color members of Group X, implying a likely age
  of 300\,Myr. The other age indicators are also compatible with an
  age close to that of TOI-1807.

\item {\bf HD 127821 = TIC 166178883} Mid F-type star with IR excess
  due to debris disk as a separation of about 210 au \citep{rhee2007}.
  The position on CMD is compatible with TOI-1807 and Group X age.
  The short photometric period (0.57 days) could also be compatible
  with those of Group X members, although the origin (pulsation vs
  rotation) can not be conclusively determined from available data.
  If due to rotation, coupled with the \vsini = 55.6 km\,s$^{-1}$
  \citep{reiners2006}, a stellar inclination of 29 deg is derived.
  The X-ray emission from ROSAT and $\log R_{HK}$
  \citep{schroeder2009} are also compatible with the Group X age
  although these have limited age sensitivity for mid-F type stars.

\item {\bf HD 129425 = TIC 158496710} The position on the color-period
  diagram is consistent with similar-color members of Group X,
  implying an age of about 300\,Myr.  The Li EW we measured on SOPHIE
  spectra (79.8 m\AA) is also consistent with this age assignment.

\item {\bf HD 130460 = TIC 282855338} The highly significant
  photometric period is most likely due to rotational modulations,
  although pulsations can not be ruled out, considering the spectral
  type of the star.  If due to rotation, it would imply an inclination
  of about 45 deg when coupled with the observed $v\sin{i}$ \citep[28
    km\,s$^{-1}$, ][]{nordstrom2004}.  The position on the
  color-period diagram is consistent with similar-color members of
  Group X, implying an age of about 300\,Myr.  The star was observed
  with Spitzer, with no IR excess found \citep{trilling2008}. The
  isochrone age is inconclusive.  The X-ray emission (R$_X$=$-$4.69)
  is slightly above the median values of Hyades members of similar
  colors.

\item {\bf BD+18 2930 = TIC 345299634} There is no rotation period
  available.  The X-ray emission (R$_X$=$-$4.44) is intermediate
  between Hyades and Pleiades and \vsini \citep{nordstrom2004} is
  compatible with the Group X rotation sequence for an orientation
  close to edge-on.  Therefore, we consider the star as a possible but
  unconfirmed object coeval to TOI-1807.

\item {\bf TYC 2569-1485-1 = TIC 272610678} The three rotation period
  search methods provide different values (grade C); however, a visual
  inspection of the light curve shows the P = 9.89\,d the most
  convincing and in fair agreement with the literature value P =
  11.82\,d from HATNET (\citealt{Hartman11}), which position this
  target consistently with the similar-color members of Group X.

\item {\bf TYC 2025-810-1 = TIC 357500019} The rotation period is
  slightly slower than similar-color members of Group X whereas fits
  well into the distribution of the NGC 3532 members.

\item {\bf HD 137897 = TIC 371267644} Visual binary (separation
  5$^{\prime\prime}$, $\Delta$G 5.63 mag).  There are no specific
  indications of youth beside kinematics.

\item {\bf UCAC4 675-059372 = TIC 156000373} We measured from \tess
  photometry a rotation period P= 2.20$\pm$0.04\,d in very good
  agreement with the literature values P = 2.20\,d from HATNET
  (\citealt{Hartman11}) and P = 2.2047\,d from ASAS-SN
  (\citealt{Jayasinghe18}). This star appears slightly faster than
  similar-color members of Group X, whereas it may belong to those
  stars transiting from the fast to the slow rotation sequence.

\item {\bf TYC 3059-299-1 = TIC 282941472} We measured from \tess
  photometry a rotation period P= 6.58$\pm$0.40\,d in good agreement
  with the literature values P = 6.61\,d from ASAS-SN
  (\citealt{Jayasinghe18}) and P = 6.46\,d from HATNET
  (\citealt{Hartman11}).  This star appears marginally faster than
  similar-color members of Group X, whereas it may belong to those
  stars transiting from the fast to the slow rotation sequence.  The
  star is also very bright in X ray (R$_X$=$-$3.26).

\item {\bf UCAC4 544-056450 = TIC 307915958} The position on the
  color-period diagram is consistent with similar-color members of
  Group X, implying an age of about 300\,Myr.

\item {\bf 2MASS J16001203-0230594 = TIC 168457615} The activity and
  gravity indicators \citep{bowler2019,zerjal2017}, the lithium
  non-detection and the position in the CMD close
  to MS are consistent with an age of few hundreds Myr but we do not
  have a determination of rotation period to fully confirm and refine
  it.

\item {\bf LSPM J1604+2331 = TIC 445889890} It is a M5 star.  We
  measured from \tess photometry a rotation period P=
  0.756$\pm$0.005\,d in good agreement with the literature value P =
  0.7564\,d from MEarth (\citealt{West15}).  With respect to
  similar-color members of Group X, this star appears slower and
  therefore likely older.  However, it is possible that the observed
  colors are altered by binarity, as the star has a very large RUWE
  (9.8) in Gaia EDR3. ROBO-AO observations by \citet{lamman2008} did
  not resolve the binary.  The position on CMD is slightly above the
  main sequence, possibly due to the contribution by the unresolved
  companion.  The star has also high levels of magnetic and coronal
  activity.  Overall, we consider it as a possible but unconfirmed
  coeval object to TOI-1807 and Group X.

\item {\bf HD 144489 = TIC 172712253} Rotation period is not
  available.  X-ray emission and \vsini (from SOPHIE archive spectra)
  suggest an age older than the Hyades, while Li EW is similar to
  Hyades member of similar color. Therefore, the star is likely older
  than TOI-1807.

\item {\bf HD 148319 = TIC 163915173} The Li EW measured by us is
  similar to that of Hyades members and below the expectations for a
  300 Myr object. Activity measurements span fairly broad range, from
  $\log R_{HK}$ = $-$4.49 to $-$4.62
  \citep{jenkins2008,butler2017,gomesdasilva2021}, which is consistent
  with an age close to Hyades. Finally the application of chemical
  abundances clock provides an age of 800 Myr \citep{casali2020}.
  Therefore, the star is likely older than TOI-1807.

\item {\bf 2MASS J16371518+3331426 = TIC 57031688} We measured from
  \tess photometry a rotation period P = 11.5 $\pm$1.2\,d in agreement
  with the literature value P = 10.99\,d from HATNET survey
  \citep{Hartman11}. Its position on the period-color diagram is
  significantly above similar-color members of Group X, implying an
  older age than 300\,Myr.  The star has a large RUWE (3.34)
  suggesting unresolved multiplicity.

\item {\bf BD-05 4394 = TIC 41038121} The low levels of chromospheric
  activity \citep{astudillo2017} and RV jitter (rms 2.5 m\,s$^{-1}$
  from archive HARPS spectra) make it likely that it is significantly
  older than TOI-1807.

\item {\bf $\mu$ Dra = TIC 198355687} Quadruple system, formed by a
  pair of F6 stars (one of which is itself a SB) separated by
  2$^{\prime\prime}$ and an additional component (M3) at
  12$^{\prime\prime}$.  The periodogram analysis of \tess data shows
  significant but irregular variability, likely because of
  superposition of variability of the components, which are blended in
  \tess observations.  The system is considered part of the Castor MG
  by \citet{caballero2010}.  Only the primary matches our kinematic
  selection criteria, and there are significant proper motion
  difference with respect to the other components.  Considering the
  parameters of the more isolated component C, less affected by
  orbital motion, we infer that this object is not comoving with
  TOI-1807 and the other targets identified here, although it may have
  similar age (likely slightly older).

\item {\bf TYC 5668-239-1 = TIC 418673682} There are no specific
  indications of youth in the literature apart for kinematics.

\item {\bf HD 162199 = TIC 446245637} There are no indication of youth
  in the literature apart for the photometric period of 0.93d reported
  by \citet{Oelkers18}.

\item {\bf 2MASS J18153959+1152077 = TIC 391453943} Star with high
  RUWE, indicating binarity. There are no available \tess data and
  additional indications of youth in the literature. The age remains
  then unconstrained.

\item {\bf HD 168746 = TIC 18097734} RV planet host
  \citep{2002A&A...388..632P}. Confirmed old star from low levels of
  chromospheric and coronal activity and isochrone age.

\item {\bf TYC 460-624-1 = TIC 449263348} The rotation period reported
  by \citet{kiraga2012}, 16.18d, is longer than Group X members of
  similar color but within the distribution of NGC 3532 members.
  Chromospheric emission \citep{jeffers2018} and X-ray emission are
  compatible with an age intermediate between Group X and the Hyades.

\item {\bf KIC 11087368 = TIC 26960092} The periodogram analysis of
  \tess data did not provided any reliable rotation period, and a
  grade C was assigned.  The low activity level in $H{\alpha}$
  \citep{lu2019} supports the interpretation of an old interloper.

\item {\bf 2MASS J20144598-2306214 = TIC 71480177} Active M3-M4 dwarf,
  considered for membership in the Tuc-Hor association by
  \citet{kraus2014} but rejected on the basis of the discrepant radial
  velocity.  Our own BANYAN analysis adopting Gaia ERD3 parameters
  also indicates a field object. The activity level and UV excess are
  compatible with a very young age, with the Li non detection being
  non conclusive considering the spectral type of the star.  Position
  in CMD is above main-sequence, indicating a young age, similar to
  Tuc-Hor or slightly older.  While a tidally-locked binary with
  similar components might also explain the observed characteristics,
  the object is likely much younger than TOI-1807.

\item {\bf 1RXS J203300.7+435147 = TIC 188452312 } The high RUWE
  indicates that the star is likely an unresolved binary.  The
  periodogram analysis of \tess data did not reveal any reliable
  rotation period.  The X-ray luminosity is at lower bound of Pleiades
  members of similar colors. This is consistent with a 300 Myr age but
  with large uncertainties.

\item {\bf EM* StHA 182 = 2MASS J20434114-2433534 = TIC 269940990 }
  This star is identified as a member of $\beta$ Pic MG in several
  literature sources \citep[e.g.,][]{bell2015}. Application of BANYAN
  $\Sigma$ on-line tool including Gaia EDR3 data supports membership.
  The star is known to be a close binary (projected separation
  1.45$^{\prime\prime}$, $\Delta G$ 0.07 mag); individual spectral
  types M3.7 and M4.1).  We derived from \tess data a rotation period
  of 0.98 days (Grade A), similar to that derived by
  \citet{gunther2020} on Sectors 1-2 only. A longer period (1.610d)
  was instead provided by \citet{messina2017}.  Its position on the
  period-color diagram is consistent with similar-color members of a
  variety of objects such as IC2391, Pleiades, and Group X. The
  rotation period is faster than the typical ones for $\beta$ Pic MG
  members, but possibly compatible with few other very fast rotators
  members (mostly binaries).  A marginally significant Li EW is
  measured by us on HIRES/Keck spectra ($5.3\pm2.5$ m\AA). If
  confirmed, this would strongly support the very young age of $\beta$
  Pic MG members and rule out older ages \citep{messina2016}.  The
  position on CMD (after correction for binarity when applicable) is
  well above main sequence and consistent with other $\beta$ Pic MG
  members.  We then conclude that this object is significantly younger
  than TOI-1807 and TOI-2046.

\item {\bf HD 198767 = TIC 387511797 } It is a G0 star with very high
  RUWE in Gaia EDR3 suggesting unresolved binarity.  We measured from
  \tess photometry a rotation period P= 6.19$\pm$0.07\,d in agreement
  with the literature values P = 6.55 from KELT (\citealt{Oelkers18})
  and P = 6.21\,d from Mascara (\citealt{Burggraaff18}).  This star
  appears slightly slower and therefore a bit older than similar-color
  members of Group X.  The Li EW measurement \citep{wichmann2003} is
  intermediate between Hyades and Pleiades loci, closer to the latter
  one.  The X-ray luminosity is intermediate between Pleiades and
  Hyades.

\item {\bf 2MASS J21364848-2200541 = TIC 441026957 } The rotation
  period is slower than similar-color members of Group X whereas still
  compatible with members of NGC\,3532.

\item {\bf BD+34 4580 = TIC 236671835 } We measured from \tess
  photometry a rotation period P= 10.1$\pm$1.0\,d. We note that the
  rotation period P = 1.238\,d reported by \citet{Oelkers18} is likely
  a beat of the rotation period measured by us.  In the period-color
  diagram this star appears marginally older than 300\,Myr, possibly
  compatible within errorbars.

\item {\bf UCAC4 260-199238 = TIC 152889010 } The periodogram analysis
  of \tess data did not provide any reliable rotation period. There is
  no other indication of youth in the literature beside the
  kinematics.

\item {\bf BD+36 4976 = TIC 418960381} This star is slower than
  similar-color members of Group X and therefore, likely older than
  300\,Myr.

\item {\bf UCAC4 558-143573 = TIC 436525094 } Low activity in
  $H{\alpha}$ is reported by \citet{lu2019}, suggesting a moderately
  old age.

\item {\bf BD+24 4863 = TIC 269786865 } There are no \tess data
  available and no other indication of youth in the literature beside
  kinematics. The age of this star is then unknown.

\end{description}
 
\longtab[1]{
\begin{landscape}
\begin{longtable}{lrrrrcccccl}
\caption{
 {Our measurements of rotation periods, Li EW, and age classification of the kinematically selected targets comoving with TOI-1807}} \label{tab:b1}\\
\hline
\hline
\multicolumn{1}{c}{ID}   &  \multicolumn{1}{c}{TIC}   &   \multicolumn{1}{c}{$\alpha$(J2000.0)}   &    \multicolumn{1}{c}{$\delta$(J2000.0)}   &  \multicolumn{1}{c}{$G$}      &   \multicolumn{1}{c}{($G-K_{\rm 2MASS}$)$_0$}       &  $P$       &      $\sigma$  &   grade & EW Li & Coeval  \\
 \multicolumn{1}{c}{  }   &  \multicolumn{1}{c}{   }   &   \multicolumn{1}{c}{(deg.)}              &    \multicolumn{1}{c}{(deg.)}              &  \multicolumn{1}{c}{(mag)}    &   \multicolumn{1}{c}{(mag) }                        &  (d)       &      (d)       &         & m\AA  &         \\
\hline
\endfirsthead
\caption{Continued.} \\
\hline
\hline
\multicolumn{1}{c}{ID}   &  \multicolumn{1}{c}{TIC}   &   \multicolumn{1}{c}{$\alpha$(J2000.0)}   &    \multicolumn{1}{c}{$\delta$(J2000.0)}   &  \multicolumn{1}{c}{$G$}      &   \multicolumn{1}{c}{($G-K_{\rm 2MASS}$)$_0$}       &  $P$       &      $\sigma$  &   grade & EW Li & Coeval  \\
 \multicolumn{1}{c}{  }   &  \multicolumn{1}{c}{   }   &   \multicolumn{1}{c}{(deg.)}              &    \multicolumn{1}{c}{(deg.)}              &  \multicolumn{1}{c}{(mag)}    &   \multicolumn{1}{c}{(mag) }                        &  (d)       &      (d)       &         & m\AA  &         \\
\hline
\endhead
\hline
\endfoot
\hline
\endlastfoot
HD 123                  &  604446831 &         1.565889 &    58.4367   &        6.196 &   1.860   &   9.67 &  0.21  & A &                 &  N?  \\  
UCAC4 180-001659        &  231019115 &        27.274584	&   -54.1992   &       12.209 &   3.353   &        &        &   &                 &  N?  \\
HD 17250C               &  318841149 &        41.469222 &     5.4901   &       12.588 &   3.400   &     20 &  4     & B &                 &  N?  \\
HD 18414                &   91556441 &        44.175443 &   -30.8602   &        8.756 &   1.583   &    --- & ---    & C &                 &  N?  \\
Wolf 140                &  416098108 &        49.529217	&    42.6692   &       12.178 &   3.739   &        &        &   &                 &  N  \\
HD 20776                &   44628969 &        49.912190 &   -34.0237   &        9.077 &   1.692   &    --- & ---    & C &                 &  N?  \\
2MASS J03550477-1032415 &   55441420 &        58.769894 &   -10.5449   &       17.529 &   5.550   &    --- & ---    & C &                 &  N  \\
HD 26413                &  152473055 &        62.141159 &   -45.8648   &        6.540 &   0.945   &  0.395 &  0.003 & A &                 &  ?  \\
ASAS J041255-1418.6     &  332660818 &        63.232301 &   -14.3164   &       11.914 &   3.276   &   7.34 &  1.05  & A &  $0$            &  Y?  \\ 
HD 34652                &  317383399 &        65.319224 &   -87.8129   &        7.384 &   0.952   &  1.753 &  0.026 & A &                 &  Y?  \\
HD 27857                &  470480305 &        66.609138 &    52.5048   &        7.893 &   0.659   &   12.9 &  3.3   & A &                 &  N  \\
UCAC2 20545888          &   33062967 &        88.514907 &   -27.3237   &       12.496 &   3.493   &     21 &  9     & B &                 &  N  \\
UCAC2 9643914           &  238088359 &        105.01363 &   -51.3543   &       12.568 &   3.467   &   2.82 &  0.10  & A &                 &  Y?  \\
GSC 08558-01964         &  294154590 &        107.74969	&   -56.5499   &        8.108 &   1.397   &        &        &   &                 &  ?  \\
HD 72687                &  283397452 &        128.31408 &   -29.9566   &        8.107 &   1.395   &   3.87 &  0.31  & A &                 &  N?  \\
HD 76332                &  126312524 &        134.09563 &    28.6679   &        8.416 &   1.569   &   10.8 &  2.1   & A &  49.9$\pm$1.0   &  N?  \\
TYC 3430-1344-1         &  456344030 &        136.24858 &    50.4177   &        9.854 &   1.941   &   10.6 &  2.9   & A &                 &  ?  \\
TYC 6621-759-1          &  309024402 &        151.10760 &   -26.2086   &       10.889 &   2.557   &     23 &  8     & B &                 &  N?  \\
TYC 4144-944-1          &  355868825 &        157.00903 &    60.0963   &       10.304 &   2.177   &   11.4 &  2.9   & A &                 &  ?  \\
40 LMi                  &  241197867 &        160.75784 &    26.3256   &        5.483 &   0.432   &        &        &   &                 &  Y?  \\
UCAC4 562-049479        &   95776155 &        160.78268 &    22.3020   &       12.197 &   3.178   &   13.0 &  3.0   & A &                 &  ?  \\
LP 373-35               &   84925818 &        164.26578 &    22.2889   &       14.797 &   4.449   &     18 &  6     & B &                 &  N  \\
2MASS J11155546+4049569 &  450332591 &        168.98113 &    40.8325   &       13.686 &   3.655   &   4.72 &  0.41  & A &                 &  Y  \\
HD 99419                &    3901189 &        171.61328	&    20.5181   &        7.788 &   1.302   &        &        &   &  54.8$\pm$1.6   &  N  \\  
HD 103928               &   99302268 &        179.52989 &    32.2739   &        6.344 &   0.744   &        &        &   &                 &  N?  \\
HD 103928B              &   99302269 &        179.52800	&    32.2728   &       12.154 &   3.466   &        &        &   &                 &  N?  \\
17 Vir                  &  377227654 &        185.63349 &    05.3054   &        6.326 &   1.169   &        &        &   &                 &  N  \\
17 Vir B                &  377227655 &        185.63126 &    05.3108   &        9.016 &   2.274   &        &        &   &                 &  N  \\
TYC 4394-114-1          &  148910632 &        185.83058 &    67.9022   &       10.827 &   2.437   &   10.5 &  0.2   & A &                 &  Y?  \\
TIC 40560697            &   40560697 &        185.91647 &   -14.9889   &       12.261 &   3.481   &    --- & ---    & C &                 &  N?  \\
LX Com                  &  450335652 &        192.91004 &    25.5088   &        8.840 &   1.846   &   7.80 &  0.56  & A &  78.8$\pm$4.7   &  Y?  \\ 
BD+26 2401              &  156514310 &        193.00904 &    25.4566   &        9.436 &   2.489   &     20 &  3     & B &   5.6$\pm$0.7   &  Y?  \\  
L68-145                 &  361397288 &        193.62125 &   -74.5023   &       11.021 &   3.293   &     20 &  3     & A &                 &  N?  \\
GJ 490A                 &  950125199 &        194.41755 &    35.2250   &        9.765 &   3.210   &   3.37 &  0.21  & A &  25.8$\pm$105   &  N?  \\ 
HD 112733               &   17740825 &        194.63321 &    38.2788   &        8.472 &   1.620   &   5.15 &  0.48  & A &                 &  Y?  \\
HD 112733B              &   17740827 &        194.64583 &    38.2801   &        9.002 &   2.125   &   5.15 &  0.48  & A & 111.9$\pm$1.1   &  Y?  \\ 
HD 113414               &    1727745 &        195.91254 &   -16.3366   &        7.598 &   1.147   &   3.18 &  0.20  & A &                 &  ?  \\
TOI-1807                &  180695581 &        201.28332 &    38.9225   &        9.675 &   2.105   &   8.68 &  0.17  & A & 104.0$\pm$1.0   &  Y  \\ 
HD 117378               &  288405352 &        202.26368 &    42.2383   &        7.491 &   1.244   &   4.40 &  0.17  & A & 102.2$\pm$1.5   &  Y  \\ 
TYC 3032-368-1          &  288487378 &        203.85894 &    43.7658   &       10.489 &   2.591   &   12.2 &  0.3   & A &                 &  Y?  \\
HIP 68732               &  135154868 &        211.04165 &    20.7591   &        9.787 &   2.271   &   10.0 &  1.0   & B &                 &  Y?  \\
RX J1419.0+6451         &  166087190 &        214.76373 &    64.8629   &       13.173 &   3.614   &  0.425 &  0.001 & A & $0$             &  Y?  \\  
TOI-2076                &   27491137 &        217.39268 &    39.7904   &        8.910 &   1.820   &   7.29 &  0.12  & A &  89.4$\pm$3.0   &  Y  \\  
HD 127821               &  166178883 &        217.69196 &    63.1858   &        5.991 &   0.976   &  0.570 &  0.001 & A &                 &  Y?  \\
HD 129425               &  158496710 &        220.04402 &    57.7132   &        7.739 &   1.159   &   3.78 &  0.05  & A &  79.8$\pm$3.0   &  Y?  \\  
HD 130460               &  282855338 &        221.31925 &    65.3300   &        7.092 &   1.037   &  1.595 &  0.005 & A &                 &  Y?  \\
BD+18 2930              &  345299634 &        221.73805 &    18.3003   &        9.251 &   1.715   &        &        &   &                 &  Y?  \\
TYC 2569-1485-1         &  272610678 &        229.68640 &    35.9487   &       11.070 &   2.749   &    --- & ---    & C &                 &  ?  \\
TYC 2025-810-1          &  357500019 &        229.81312 &    24.3271   &       10.576 &   2.574   &   11.6 &  2.6   & A &                 &  Y?  \\
HD 137897               &  371267644 &        232.13727	&    03.1104   &        9.014 &   1.505   &        &        &   &                 &  ?  \\
UCAC4 675-059372        &  156000373 &        235.92797 &    44.8477   &       12.386 &   3.372   &   2.20 &  0.04  & A &                 &  Y?  \\
TYC 3059-299-1          &  282941472 &        236.59651 &    44.0541   &       11.226 &   2.806   &   6.58 &  0.40  & A &                 &  Y?  \\
UCAC4 544-056450        &  307915958 &        237.91306 &    18.6732   &       12.138 &   3.237   &   4.05 &  0.31  & A &                 &  Y?  \\
2MASS J16001203-0230594 &  168457615 &        240.05012 &   -02.5165   &       11.675 &   3.254   &        &        &   &                 &  Y?  \\
LSPM J1604+2331         &  445889890 &        241.05512 &    23.5274   &       13.317 &   4.218   &  0.756 &  0.005 & A &                 &  Y?  \\
HD 144489               &  172712253 &        241.39598 &    15.2575   &        8.524 &   1.422   &        &        &   &  83.3$\pm$5.2   &  N?  \\  
HD 148319               &  163915173 &        246.87854 &   -10.1429   &        8.411 &   1.348   &        &        &   &  74.0$\pm$7.8   &  N?  \\  
2MASS J16371518+3331426 &   57031688 &        249.31323 &    33.5285   &       13.405 &   3.499   &   11.5 &  1.2   & A &                 &  ?  \\
BD-05 4394              &   41038121 &        255.70661	&   -06.0684   &       10.052 &   3.107   &        &        &   &                 &  N?  \\
$\mu$ Dra               &  198355687 &        256.33385	&    54.4700   &        5.513 &   1.756   &        &        &   &                 &  N?  \\
TYC 5668-239-1          &  418673682 &        266.11879 &   -13.0744   &       10.095 &   3.107   &        &        &   &                 &  ?  \\
HD 162199               &  446245637 &        267.34646	&    10.3388   &        8.174 &   1.697   &        &        &   &                 &  ?  \\
2MASS J18153959+1152077 &  391453943 &        273.91499 &    11.8688   &       12.597 &   3.517   &        &        &   &                 &  ?  \\
HD 168746               &   18097734 &        275.45743 &   -11.9227   &        7.772 &   1.519   &        &        &   &                 &  N  \\
TYC 460-624-1           &  449263348 &        281.29279 &    06.3378   &       10.004 &   3.198   &        &        &   &                 &  ?  \\
KIC 11087368            &   26960092 &        293.46911 &    48.6578   &       13.447 &   3.683   &    --- & ---    & C &                 &  N?  \\
2MASS J20144598-2306214 &   71480177 &        303.69160	&   -23.1060   &       12.792 &   3.847   &        &        &   &                 &  N?  \\
1RXS J203300.7+435147   &  188452312 &        308.25105 &    43.8633   &         ---  &    ---    &        &        &   &                 &  ?  \\
EM* StHA 182            &  269940990 &        310.92140	&   -24.5648   &         ---  &    ---    &        &        &   &    5.3$\pm$2.5  &  N?  \\  
HD 198767               &  387511797 &        312.07519 &    69.1418   &        7.817 &   1.265   &   6.19 &  0.07  & A &                 &  ?  \\
2MASS J21364848-2200541 &  441026957 &        324.20204 &   -22.0151   &       12.363 &   3.382   &   14.5 &  3.9   & A &                 &  ?  \\
BD+34 4580              &  236671835 &        329.90301 &    35.6840   &        9.256 &   1.955   &   10.0 &  1.0   & A &                 &  ?  \\
UCAC4 260-199238        &  152889010 &        337.59242 &   -38.1578   &       12.392 &   3.119   &    --- & ---    & C &                 &  N?  \\
BD+36 4976              &  418960381 &        345.03034 &    36.9331   &        9.423 &   1.954   &     17 &  6     & B &                 &  N?  \\
UCAC4 558-143573        &  436525094 &        349.12987 &    21.5425   &       12.087 &   3.426   &        &        &   &                 &  N?  \\
BD+24 4863              &  269786865 &        359.36201	&    25.1090   &        9.250 &   1.658   &        &        &   &                 &  ?  \\
\hline\noalign{\smallskip}
\end{longtable}
\tablefoot{
Sources of data for Li EW measurements: 
ASAS J041255-1418.6: FEROS, Prog. ID 091.C-0216(A), PI Rodriguez; 
HD 76332: FEROS, Prog. ID 090.D-0133(A), PI Datson;
HD 99419: SOPHIE, Prog ID 09B.PNP.CONS;
LX Com: HIRES/Keck, Prog. ID H6aH, PI Boesgaard;
BD+26 2401: HIRES/Keck, Prog. ID N038Hr, PI Fischer; 
GJ 490A: HIRES/Keck, Prog. ID H198Hr, H170Hr, PI Shkolnik;
HD 112733B: HIRES/Keck Prog. ID H6aH, PI Boesgaard;
TOI-1807: HARPS-N (this paper, Sect. \ref{sec:atmo_param})
HD 117378 : HIRES/Keck, Prog. ID A297Hr, PI Fischer;
RX J1419.0+6451: HIRES/Keck, Prog. ID C103Hr, PI Bowler;
TOI-2076: HARPS-N (GAPS team, priv. comm.);
HD 129425: SOPHIE, Prog. ID 17A.PNCG.SOUB, PI Soubiran;
HD 144489: SOPHIE, Prog. ID 11A.PNP.CONS;
HD 148319: HARPS, Prog. ID 072.C-0488(E), PI Mayor and 183.C-0972(A), PI Udry;
EM* StHA 182: HIRES/Keck, Prog. ID H198Hr, H212Hr, H170Hr, PI Shkolnik
}
\end{landscape}
}

%
%

\section{Combined analysis of HARPS-N and iSHELL data sets}
\begin{figure*}[t!]
  \centering
  \includegraphics[width=0.9\textwidth, bb=50 313 555 631]{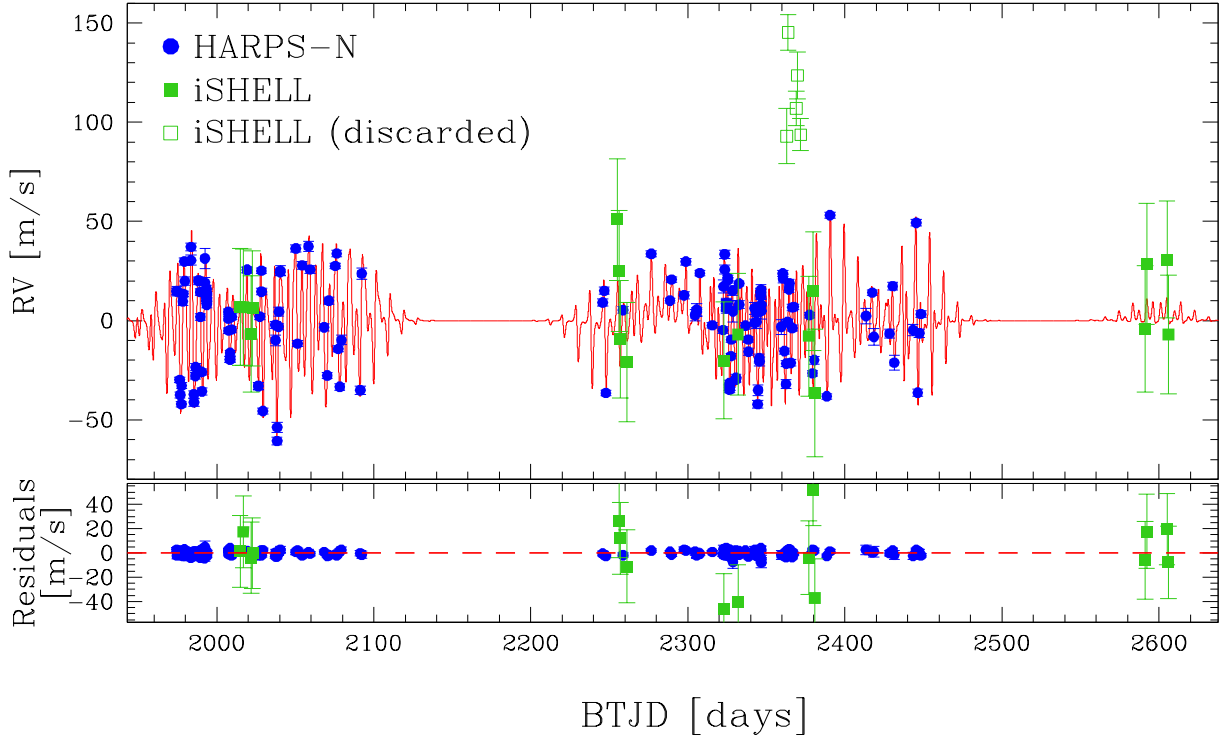} \\
  \caption{Top panel shows the activity modeling of HARPS-N
      (blue points) and iSHELL (green squares) data by using the same
      configuration used for case 1 described in
      Sect.~\ref{sec:case1}. Empty squares represent the iSHELL
      measurements we discarded (see text for details). Bottom panel
      illustrates the residual after subtracting the model (shown in
      red in top panel). \label{fig:C1}}
\end{figure*}

\subsection{Observations and data reduction of iSHELL data} 
We observed TOI-1807 with the iSHELL spectrograph
  \citep{2016SPIE.9908E..84R} on the NASA Infrared Telescope
  Facility. 232 exposures were recorded in KGAS mode (2.18 -- 2.47
  $\mu$\,m) with an integration time of 300 seconds on 23 nights
  between June 14th, 2020 -- January 26th, 2022. The median S/N per
  spectral pixel is $\approx$30 and ranges from 20--40 due to variable
  weather conditions at Maunakea. Spectra are reduced and RVs are
  computed using updated methods to those described in
  \citet{caleetal2019}. RVs from individual exposures and echelle
  orders are co-added within a night to generate one RV per night
  TOI-1807 was observed. In doing so, we discard two exposures from UT
  date February 9th, 2021, all exposures from June 10th, 2021, and two
  exposures from June 15th, 2021 due to low S/N ($<$ 15) from poor
  weather conditions at Maunakea. The median error bar of the co-added
  iSHELL RVs is 8.8 m\,s$^{-1}$.

\subsection{Analysis of HARPS-N and iSHELL RVs} 

We run \texttt{PyORBIT} both on the HARPS-N and iSHELL data,
  by using the same configuration and priors described in
  Sect.~\ref{sec:case1} (case 1), but excluding the modeling of the
  light curve. In this analysis we excluded the iSHELL RV measurements
  obtained between May 28th, 2021 and June 6th, 2021, because
  characterized by anomalous high values ($\sim 100$ \ms above the
  mean iSHELL RVs); we also compared these measurements with HARPS-N
  RVs collected in the same days and we have had confirmation that
  they are affected by some kind of systematic. The results of the
  modeling are reported in Table~\ref{tab:C1}. The adopted iSHELL RVs
  follow the trend due to the stellar activity, as shown in
  Fig.~\ref{fig:C1}; because of the large errors, the iSHELL RVs did
  not contribute to the measurement of the planet parameters, and for
  this reason we have decided not to use them for the analysis of our
  work.

\begin{table*}
  \renewcommand{\arraystretch}{1.25}
  \caption{Priors and results of the model of planet b from the analysis of combined HARPS-N and iSHELL spectroscopic series}
  \begin{tabular}{l l l l}
\hline
\hline
\multicolumn{4}{c}{{\bf Stellar activity}} \\
\hline
Parameter    &  Unit  &    Prior      &       Value  \\
\hline
Rotational period ($P_{\rm rot}$)           & days       & $\mathcal{U}( 8.0,  10.0)$    &  $ 8.89 \pm 0.09$                    \\
Decay Timescale of activity ($P_{\rm dec}$)                & days       & $\mathcal{U}(10.0, 100.0)$    &  $14.3^{+1.2}_{-1.2}$                    \\
Coherence scale ($w$)                 &            & ...                           &  $0.26^{+0.03}_{-0.02}$                 \\
Amplitude of the RV signal (HARPS-N)                  &  m/s       & $\mathcal{U}(0.01, 1500.00)$  &  $25.2^{+2.6}_{-2.2}$            \\
Amplitude of the RV signal (iSHELL)                  &  m/s       & $\mathcal{U}(0.01, 1500.00)$  &  $50.1^{+19.2}_{-23.6}$            \\
Uncorrelated HARP-N RV jitter ($\sigma_{\rm jitter}^{\rm RV, HARPS-N}$) &  m~s$^{-1}$ & ... & $0.69_{-0.45}^{+0.53}$ \\
HARPS-N RV offset ($\gamma^{\rm RV, HARPS-N}$) &  m~s$^{-1}$ & ... & $-6838.3 \pm 4.2$ \\
Uncorrelated iSHELL RV jitter ($\sigma_{\rm jitter}^{\rm RV, iSHELL}$) &  m~s$^{-1}$ & ... & $42.9_{-17.8}^{+20.00}$ \\
iSHELL RV offset ($\gamma^{\rm RV, iSHELL}$) &  m~s$^{-1}$ & ... & $-1.0 \pm 17.6$ \\
\hline
\hline
\multicolumn{4}{c}{{\bf Planet b}} \\
\hline
Parameter    &  Unit  &    Prior      &       Value  \\
\hline
Orbital Period ($P_{\rm b}$)                           & days   &  $\mathcal{U}(0.549, 0.550)$  & $0.549418_{-0.000046}^{+0.000043}$ \\
Central time of the first transit ($T_{\rm 0,b}$)      & BJD    &  $\mathcal{U}(2458899.3, 2458899.4)$  & $2458899.330875_{-0.02}^{+0.03    }$ \\
Orbital eccentricity ($e_{\rm b}$)  &   & fixed & 0 \\
Semi-major-axis-to-stellar-radius ratio ($(a_{\rm b}/R_{\star})$) &  & ... & $3.7\pm 0.2$ \\
Orbital Semi-major axis ($a_{\rm b}$) & au & ... & $0.0120 \pm 0.0003$ \\
RV semi-amplitude ($K_{\rm b}$)                              & m~s$^{-1}$ & $\mathcal{U}(0.01, 10)$ & $2.31_{-0.49}^{+0.46}$ \\
Planetary mass ($M_{\rm P,b}$, $i=82\pm 2$~deg) & ${\rm M_{\oplus}}$ & ... & $2.45 \pm 0.52$  \\
\hline
\end{tabular}

  \label{tab:C1}
\end{table*}

\section{GP framework results with different tools}

We compared the results obtained with the GP framework
  modeling implemented in \texttt{PyORBIT} (Sect.~\ref{sec:case2}),
  with those obtained (by using the same initial conditions) with the
  original software by \citet{2015MNRAS.452.2269R} and with the
  modeling obtained with \texttt{pyaneti}
  (\citealt{2022MNRAS.509..866B}). The comparison is reported in
  Table~\ref{tab:D1}. All the results are in agreement within the
  errors.

\begin{table*}
  \renewcommand{\arraystretch}{1.25}
  \caption{Comparison between the results obtained with GP framework modeling implemented in different software.}
  \begin{tabular}{l l l l l}
\hline
\hline
Parameter    &  Unit  &   \citet{2015MNRAS.452.2269R}     &   \texttt{pyaneti} & \texttt{PyORBIT} \\
\hline
Uncorrelated RV jitter ($\sigma_{\rm jitter,0}^{\rm RV}$)    &  m~s$^{-1}$  &   $0.23^{+0.01}_{-0.26}$   &   $0.70^{+0.60}_{-0.50}$                    & $0.79_{-0.49}^{+0.51}$ \\
RV offset ($\gamma_{\rm 0}^{\rm RV}$)                        &  m~s$^{-1}$  &   $-6839.4 \pm 0.8$            &      $-6839.5 \pm 1.2$   & $-6839.6 \pm 1.1$ \\
Uncorrelated BIS jitter ($\sigma_{\rm jitter,0}^{\rm BIS}$)  &  m~s$^{-1}$  &   $12.31_{-0.84}^{+0.71}$           &   $12.30_{-0.8}^{+0.8}$                     & $12.30_{-0.76}^{+0.83}$ \\
BIS offset ($\gamma_{\rm 0}^{\rm BIS}$)                      &  m~s$^{-1}$  &   $33.8 \pm 1.1$           &        $34.0 \pm 1.5$                & $34.0 \pm 1.4$ \\
Uncorrelated \logrhk jitter ($\sigma_{\rm jitter,0}^{\rm \log{R'_{HK}}}$) & &   $0.0133_{-0.0010}^{+0.0009}$           &    $0.0130_{-0.0010}^{+0.0010}$                    & $0.0132_{-0.0009}^{+0.0010}$ \\
\logrhk offset ($\gamma_{\rm 0}^{\rm \log{R'_{HK}}}$)        &              &   $-4.359 \pm 0.001$           &          $-4.360 \pm 0.003$              & $-4.360 \pm 0.003$ \\
$V_c$  & m~s$^{-1}$ &  $0.9_{-1.7}^{+1.7}$   &  $-1.4_{-2.2}^{+2.3}$ & $1.2_{-1.9}^{+1.9}$ \\
$V_r$  & m~s$^{-1}$ &  $25.0_{-3.1}^{+2.4}$   & $28.5_{-3.7}^{+4.5}$  & $27.8_{-3.4}^{+4.3}$ \\
$B_c$  & m~s$^{-1}$ &  $-1.7_{-1.5}^{+1.5}$   & $1.9_{-1.8}^{+1.9}$  & $-1.9_{-1.8}^{+1.7}$ \\
$B_r$  & m~s$^{-1}$ &  $-21.2_{-2.3}^{+2.8}$   & $-24.4_{-4.1}^{+3.4}$  & $-23.8_{-3.9}^{+3.2}$ \\
$L_c$  &            &  $-0.0010_{-0.0015}^{+0.0018}$   &  $0.0110_{-0.0020}^{+0.0020}$ & $-0.0110_{-0.0020}^{+0.0019}$ \\
Rotational period ($P_{\rm rot}$)           & days       &  $ 8.85 \pm 0.07$  &  $ 8.84 \pm 0.08$  &  $ 8.84 \pm 0.08$                    \\
Decay Timescale of activity ($P_{\rm dec}$)  & days       &  $12.41^{+1.10}_{-0.99}$ &   $12.90^{+1.20}_{-1.20}$   &  $12.85^{+1.15}_{-1.12}$                    \\
Coherence scale ($w$)             &            & $0.40_{-0.03}^{+0.03}$  &  $0.44_{-0.04}^{+0.04}$   &  $0.43_{-0.04}^{+0.04}$\\
Orbital Period ($P_{\rm b}$)       & days   &  $0.549381_{-0.000015}^{+0.000015}$  &  $0.549388_{-0.000024}^{+0.000024}$   & $0.549380_{-0.000016}^{+0.000015}$ \\
Central time of the first transit ($T_{\rm 0,b}$)      & BJD    & $2458899.3449_{-0.0008}^{+0.0008}$  & $2458899.3449_{-0.0008}^{+0.0008}$ & $2458899.3449_{-0.0008}^{+0.0008}$ \\
RV semi-amplitude ($K_{\rm b}$)         & m~s$^{-1}$ & $2.53_{-0.36}^{+0.36}$ & $2.49_{-0.39}^{+0.38}$ & $2.48_{-0.39}^{+0.38}$ \\
\hline
\end{tabular}

  \label{tab:D1}
\end{table*}

\section{Spectroscopic time series}
\label{sec:rvdata}
The spectroscopic time series (RV, BIS, \logrhk, see tables \ref{tab:yo41rv1},\ref{tab:yo41rv2}, and \ref{tab:ishell} ) will be available in
electronic format as supplementary material of the paper.

\begin{table*}
  \caption{First season of HARPS-N RV, BIS, and \logrhk measurements of TOI-1807}
  \resizebox{0.99\textwidth}{!}{
    \begin{tabular}{l c c c c r || l c c c c r}
\hline
\hline
JD$-$2450000   &        RV  &   $\sigma_{\rm RV}$ & log$R'_{\rm HK}$ &  $\sigma_{{\rm log} R'_{\rm HK}}$ & \multicolumn{1}{c}{BIS} &  JD$-$2450000   &        RV  &   $\sigma_{\rm RV}$ & log$R'_{\rm HK}$ &  $\sigma_{{\rm log} R'_{\rm HK}}$ & \multicolumn{1}{c}{BIS} \\
               &    (km~s$^{-1}$) &   (km~s$^{-1}$) &        &           &  (km~s$^{-1}$)  &                              &    (km~s$^{-1}$)  & (km~s$^{-1}$) &        &                  & (km~s$^{-1}$)          \\
\hline
8974.46441595 &    -6.8233 &     0.0015 &         -4.3814 &     0.0029     &   0.0181 & 9008.51021374 &    -6.8543 &     0.0010 &         -4.3707 &     0.0018 & 0.0658 \\
8974.61600577 &    -6.8231 &     0.0023 &         -4.3827 &     0.0054	   &  -0.0133 & 9008.58962957 &    -6.8576 &     0.0010 &         -4.3651 &     0.0019 & 0.0617 \\
8975.51827612$^*$ &    -6.7910 &     0.0011 &         -4.3606 &     0.0020	   &  -0.0020 & 9009.40084674 &    -6.8425 &     0.0013 &         -4.3648 &     0.0025 & 0.0595 \\
8975.63956639$^*$ &    -6.8383 &     0.0035 &         -4.3627 &     0.0088	   &  -0.0520 & 9009.50183513 &    -6.8361 &     0.0013 &         -4.3519 &     0.0026 & 0.0541 \\
8976.44347286 &    -6.8679 &     0.0012 &         -4.3754 &     0.0022	   &   0.0555 & 9019.38956783 &    -6.8125 &     0.0011 &         -4.3674 &     0.0021 & 0.0153 \\
8976.65122857 &    -6.8755 &     0.0014 &         -4.3606 &     0.0029	   &   0.0595 & 9019.45377624 &    -6.8125 &     0.0013 &         -4.3702 &     0.0025 & 0.0095 \\
8977.41639792 &    -6.8801 &     0.0011 &         -4.3669 &     0.0020	   &   0.0504 & 9026.38768335 &    -6.8710 &     0.0015 &         -4.3662 &     0.0032 & 0.0609 \\
8977.62983603 &    -6.8708 &     0.0013 &         -4.3537 &     0.0026	   &   0.0366 & 9026.50468897 &    -6.8710 &     0.0015 &         -4.3731 &     0.0031 & 0.0664 \\
8978.44811542 &    -6.8284 &     0.0017 &         -4.3570 &     0.0035	   &   0.0066 & 9027.38664056 &    -6.8360 &     0.0012 &         -4.3645 &     0.0022 &  0.0352\\
8978.59219355 &    -6.8245 &     0.0011 &         -4.3514 &     0.0019	   &   0.0024 & 9028.40158054 &    -6.8129 &     0.0012 &         -4.3650 &     0.0022 &  0.0120\\
8979.39425788 &    -6.8082 &     0.0017 &         -4.3456 &     0.0036	   &   0.0235 & 9028.49611091 &    -6.8234 &     0.0014 &         -4.3630 &     0.0028 &  0.0121\\
8979.62786819 &    -6.8181 &     0.0020 &         -4.3453 &     0.0046	   &   0.0236 & 9029.39434628 &    -6.8836 &     0.0013 &         -4.3695 &     0.0025 &  0.0781\\
8983.51984327 &    -6.8076 &     0.0024 &         -4.3816 &     0.0062	   &   0.0310 & 9037.38815359 &    -6.8402 &     0.0016 &         -4.3701 &     0.0033 &  0.0339\\
8983.56943540 &    -6.8009 &     0.0019 &         -4.3787 &     0.0045	   &   0.0229 & 9037.50453450 &    -6.8479 &     0.0026 &         -4.3887 &     0.0071 &  0.0379\\
8985.41144889 &    -6.8752 &     0.0017 &         -4.3694 &     0.0037	   &   0.0544 & 9038.38668478 &    -6.8987 &     0.0019 &         -4.3281 &     0.0039 &  0.0860\\
8985.55519063 &    -6.8792 &     0.0017 &         -4.3677 &     0.0037	   &   0.0600 & 9038.52708029 &    -6.8918 &     0.0025 &         -4.3599 &     0.0072 &  0.0798\\
8986.40951622 &    -6.8659 &     0.0015 &         -4.3630 &     0.0031	   &   0.0412 & 9039.38863040 &    -6.8411 &     0.0013 &         -4.3842 &     0.0025 &  0.0225\\
8986.57451820 &    -6.8615 &     0.0017 &         -4.3742 &     0.0038	   &   0.0317 & 9039.51469822 &    -6.8335 &     0.0021 &         -4.3750 &     0.0052 &  0.0258\\
8988.42992979 &    -6.8176 &     0.0013 &         -4.3363 &     0.0025	   &   0.0111 & 9040.37477881 &    -6.8136 &     0.0014 &         -4.3773 &     0.0028 &  0.0050\\
8988.58227026 &    -6.8182 &     0.0014 &         -4.3312 &     0.0028	   &   0.0224 & 9040.47274672 &    -6.8129 &     0.0024 &         -4.3680 &     0.0062 &  0.0195\\
8989.43032134 &    -6.8237 &     0.0014 &         -4.3383 &     0.0027	   &   0.0430 & 9050.37744444 &    -6.8017 &     0.0017 &         -4.3496 &     0.0033 &  0.0036\\
8989.58069423 &    -6.8362 &     0.0015 &         -4.3403 &     0.0031	   &   0.0429 & 9051.39733395 &    -6.8496 &     0.0013 &         -4.3434 &     0.0023 &  0.0736\\
8990.45362751 &    -6.8639 &     0.0012 &         -4.3639 &     0.0024	   &   0.0716 & 9054.37401342 &    -6.8102 &     0.0017 &         -4.3593 &     0.0037 &  0.0001\\
8990.60241476 &    -6.8737 &     0.0015 &         -4.3649 &     0.0031	   &   0.0817 & 9058.40701370 &    -6.8006 &     0.0023 &         -4.3720 &     0.0056 &  0.0117\\
8992.38380996 &    -6.8188 &     0.0025 &         -4.3809 &     0.0066	   &   0.0320 & 9059.37938790 &    -6.8123 &     0.0017 &         -4.3399 &     0.0029 &  0.0161\\
8992.46549471 &    -6.8067 &     0.0050 &         -4.3921 &     0.0175	   &   0.0205 & 9068.36923554 &    -6.8414 &     0.0015 &         -4.3244 &     0.0024 &  0.0299\\
8993.40261228 &    -6.8217 &     0.0014 &         -4.3883 &     0.0030	   &   0.0173 & 9070.37216797 &    -6.8657 &     0.0013 &         -4.3772 &     0.0025 &  0.0511\\
8993.44472770 &    -6.8237 &     0.0013 &         -4.3943 &     0.0027	   &   0.0181 & 9071.38883149 &    -6.8280 &     0.0015 &         -4.3821 &     0.0034 &  0.0282\\
8993.48428541 &    -6.8244 &     0.0025 &         -4.3996 &     0.0061	   &   0.0174 & 9075.37245446 &    -6.8105 &     0.0012 &         -4.3551 &     0.0021 &  0.0198\\
8993.52620408 &    -6.8265 &     0.0015 &         -4.3910 &     0.0031	   &   0.0153 & 9076.36724852 &    -6.8042 &     0.0015 &         -4.3246 &     0.0030 & -0.0004\\
8993.56836579 &    -6.8280 &     0.0013 &         -4.3813 &     0.0027	   &   0.0214 & 9077.36882563 &    -6.8523 &     0.0013 &         -4.3138 &     0.0022 &  0.0600\\
8993.60875678 &    -6.8300 &     0.0013 &         -4.3839 &     0.0028	   &   0.0262 & 9078.36645711 &    -6.8714 &     0.0014 &         -4.3125 &     0.0028 &  0.0657\\
9007.40029000 &    -6.8337 &     0.0012 &         -4.3539 &     0.0021	   &   0.0399 & 9079.36222621 &    -6.8478 &     0.0023 &         -4.3685 &     0.0059 &  0.0469\\
9007.50066506 &    -6.8371 &     0.0009 &         -4.3339 &     0.0015	   &   0.0405 & 9091.35641808 &    -6.8731 &     0.0020 &         -4.3801 &     0.0055 &  0.0456\\
9007.58233770 &    -6.8431 &     0.0010 &         -4.3243 &     0.0018	   &   0.0411 & 9092.35503638 &    -6.8143 &     0.0027 &         -4.4001 &     0.0082 &  0.0040\\
9008.39959635 &    -6.8577 &     0.0013 &         -4.3700 &     0.0025	   &   0.0612 &                &            &            &                 &           &         \\
\hline									       
\end{tabular}

  }
  $^*$ Not used during planet and stellar activity modelling.
  \label{tab:yo41rv1}
\end{table*}

\begin{table*}
  \caption{Second season of HARPS-N RV, BIS and \logrhk measurements of TOI-1807}
  \resizebox{0.99\textwidth}{!}{
    \begin{tabular}{l c c c c r || l c c c c r}
\hline
\hline
JD$-$2450000   &        RV  &   $\sigma_{\rm RV}$ & log$R'_{\rm HK}$ &  $\sigma_{{\rm log} R'_{\rm HK}}$ & \multicolumn{1}{c}{BIS} &  JD$-$2450000   &        RV  &   $\sigma_{\rm RV}$ & log$R'_{\rm HK}$ &  $\sigma_{{\rm log} R'_{\rm HK}}$ & \multicolumn{1}{c}{BIS} \\
               &    (km~s$^{-1}$) &   (km~s$^{-1}$) &        &           &  (km~s$^{-1}$)  &                              &    (km~s$^{-1}$)  & (km~s$^{-1}$) &        &                  & (km~s$^{-1}$)          \\
\hline
9245.71820728 &    -6.8289 &     0.0016 &         -4.3773 &     0.0034 &   0.0159 & 9338.52325852 &    -6.8537 &     0.0015 &         -4.3464 &     0.0030 &    0.0514\\
9246.75235034 &    -6.8230 &     0.0016 &         -4.3667 &     0.0033 &   0.0208 & 9342.47056341 &    -6.8392 &     0.0017 &         -4.3420 &     0.0029 &    0.0168\\
9247.80531298 &    -6.8745 &     0.0012 &         -4.3688 &     0.0023 &   0.0672 & 9342.60351957 &    -6.8318 &     0.0028 &         -4.3332 &     0.0059 &    0.0193\\
9258.72515565 &    -6.8327 &     0.0015 &         -4.3703 &     0.0031 &   0.0075 & 9343.48228209 &    -6.8341 &     0.0019 &         -4.3467 &     0.0036 &    0.0209\\
9276.70613765 &    -6.8044 &     0.0012 &         -4.3845 &     0.0024 &   0.0042 & 9343.61577051 &    -6.8386 &     0.0033 &         -4.3407 &     0.0073 &    0.0298\\
9288.63117172 &    -6.8279 &     0.0010 &         -4.3590 &     0.0018 &   0.0244 & 9344.52135958 &    -6.8802 &     0.0018 &         -4.3358 &     0.0030 &    0.0554\\
9289.62913153 &    -6.8173 &     0.0011 &         -4.3547 &     0.0019 &   0.0280 & 9344.65394516 &    -6.8731 &     0.0026 &         -4.3459 &     0.0057 &    0.0372\\
9297.53293888 &    -6.8252 &     0.0017 &         -4.3708 &     0.0037 &   0.0332 & 9345.42856827 &    -6.8570 &     0.0021 &         -4.3486 &     0.0040 &    0.0347\\
9298.67051675 &    -6.8083 &     0.0015 &         -4.3560 &     0.0029 &   0.0181 & 9345.58234475 &    -6.8586 &     0.0020 &         -4.3369 &     0.0037 &    0.0233\\
9304.55013936 &    -6.8352 &     0.0027 &         -4.3583 &     0.0067 &   0.0272 & 9346.40817978 &    -6.8236 &     0.0018 &         -4.3406 &     0.0033 &    0.0080\\
9305.55635483 &    -6.8336 &     0.0020 &         -4.3813 &     0.0043 &   0.0145 & 9346.45210101 &    -6.8256 &     0.0015 &         -4.3494 &     0.0027 &    0.0131\\
9305.56656304 &    -6.8325 &     0.0030 &         -4.3726 &     0.0073 &   0.0066 & 9346.52170370 &    -6.8243 &     0.0017 &         -4.3483 &     0.0031 &    0.0000\\
9307.57306441 &    -6.8141 &     0.0010 &         -4.3444 &     0.0017 &   0.0092 & 9346.61142104 &    -6.8224 &     0.0025 &         -4.3638 &     0.0068 &    0.0091\\
9315.56461272 &    -6.8404 &     0.0011 &         -4.3611 &     0.0020 &   0.0468 & 9346.64270406 &    -6.8301 &     0.0030 &         -4.3455 &     0.0087 &    0.0086\\
9322.42066058 &    -6.8428 &     0.0014 &         -4.3685 &     0.0030 &   0.0518 & 9346.69584931 &    -6.8331 &     0.0040 &         -4.3310 &     0.0117 &    0.0061\\
9322.71609101 &    -6.8208 &     0.0033 &         -4.3364 &     0.0098 &   0.0300 & 9359.40602561 &    -6.8412 &     0.0037 &         -4.3681 &     0.0108 &    0.0493\\
9323.41920484 &    -6.8046 &     0.0019 &         -4.3668 &     0.0042 &  -0.0030 & 9360.45664208 &    -6.8142 &     0.0013 &         -4.3472 &     0.0026 &    0.0157\\
9323.54739509 &    -6.8123 &     0.0020 &         -4.3716 &     0.0046 &  -0.0037 & 9360.57386843 &    -6.8168 &     0.0020 &         -4.3255 &     0.0038 &    0.0051\\
9324.38355936 &    -6.8294 &     0.0018 &         -4.3548 &     0.0039 &   0.0439 & 9361.39599194 &    -6.8534 &     0.0013 &         -4.3316 &     0.0022 &    0.0523\\
9324.43312945 &    -6.8304 &     0.0022 &         -4.3662 &     0.0050 &   0.0374 & 9362.45681500 &    -6.8700 &     0.0021 &         -4.3457 &     0.0045 &    0.0381\\
9324.47732934 &    -6.8319 &     0.0022 &         -4.3420 &     0.0050 &   0.0309 & 9362.57358987 &    -6.8597 &     0.0013 &         -4.3401 &     0.0025 &    0.0319\\
9324.52385554 &    -6.8300 &     0.0022 &         -4.3605 &     0.0053 &   0.0336 & 9363.42179882 &    -6.8388 &     0.0011 &         -4.3549 &     0.0019 &    0.0379\\
9324.56852995 &    -6.8310 &     0.0023 &         -4.3544 &     0.0053 &   0.0344 & 9363.56161612 &    -6.8389 &     0.0014 &         -4.3580 &     0.0028 &    0.0362\\
9324.61286873 &    -6.8312 &     0.0016 &         -4.3519 &     0.0034 &   0.0351 & 9364.42230071 &    -6.8222 &     0.0014 &         -4.3580 &     0.0026 &    0.0231\\
9324.65635104 &    -6.8310 &     0.0015 &         -4.3464 &     0.0029 &   0.0295 & 9364.55892371 &    -6.8192 &     0.0017 &         -4.3530 &     0.0038 &    0.0166\\
9324.70296982 &    -6.8292 &     0.0023 &         -4.3290 &     0.0054 &   0.0318 & 9365.54275951 &    -6.8593 &     0.0015 &         -4.3560 &     0.0030 &    0.0767\\
9325.40108234 &    -6.8165 &     0.0013 &         -4.3506 &     0.0024 &   0.0101 & 9366.45758326 &    -6.8418 &     0.0012 &         -4.3744 &     0.0024 &    0.0490\\
9325.53454990 &    -6.8177 &     0.0012 &         -4.3523 &     0.0022 &   0.0147 & 9366.56998321 &    -6.8314 &     0.0016 &         -4.3723 &     0.0035 &    0.0444\\
9326.36722874 &    -6.8696 &     0.0012 &         -4.3543 &     0.0022 &   0.0763 & 9367.45813011 &    -6.8311 &     0.0011 &         -4.3736 &     0.0021 &    0.0158\\
9326.40955361 &    -6.8722 &     0.0011 &         -4.3535 &     0.0021 &   0.0743 & 9377.43036413 &    -6.8352 &     0.0011 &         -4.3547 &     0.0020 &    0.0388\\
9326.45132294 &    -6.8718 &     0.0010 &         -4.3532 &     0.0018 &   0.0801 & 9379.40317741 &    -6.8646 &     0.0012 &         -4.3651 &     0.0023 &    0.0484\\
9326.49485147 &    -6.8730 &     0.0012 &         -4.3552 &     0.0023 &   0.0804 & 9380.44381032 &    -6.8579 &     0.0018 &         -4.3564 &     0.0040 &    0.0347\\
9326.53833370 &    -6.8716 &     0.0012 &         -4.3565 &     0.0021 &   0.0785 & 9388.45962108 &    -6.8762 &     0.0011 &         -4.3763 &     0.0020 &    0.0416\\
9326.60411837 &    -6.8717 &     0.0012 &         -4.3575 &     0.0022 &   0.0805 & 9390.47008063 &    -6.7849 &     0.0013 &         -4.3364 &     0.0023 &   -0.0165\\
9326.64815613 &    -6.8717 &     0.0011 &         -4.3612 &     0.0019 &   0.0742 & 9413.46189904 &    -6.8357 &     0.0038 &         -4.3483 &     0.0094 &    0.0475\\
9327.42357917 &    -6.8559 &     0.0010 &         -4.3484 &     0.0018 &   0.0411 & 9417.42186840 &    -6.8239 &     0.0010 &         -4.3543 &     0.0019 &    0.0229\\
9327.55162999 &    -6.8476 &     0.0010 &         -4.3444 &     0.0018 &   0.0306 & 9418.44355285 &    -6.8462 &     0.0042 &         -4.3708 &     0.0134 &    0.0252\\
9328.47268336 &    -6.8229 &     0.0022 &         -4.3551 &     0.0053 &  -0.0092 & 9428.43009929 &    -6.8446 &     0.0021 &         -4.3450 &     0.0050 &    0.0409\\
9328.65304685 &    -6.8333 &     0.0053 &         -4.3556 &     0.0181 &  -0.0054 & 9430.41405431 &    -6.8207 &     0.0018 &         -4.3389 &     0.0040 &    0.0136\\
9330.49354193 &    -6.8679 &     0.0012 &         -4.3614 &     0.0022 &   0.0733 & 9431.40611978 &    -6.8592 &     0.0036 &         -4.3230 &     0.0099 &    0.0317\\
9330.63536494 &    -6.8669 &     0.0016 &         -4.3577 &     0.0033 &   0.0654 & 9443.37940582 &    -6.8428 &     0.0027 &         -4.3659 &     0.0074 &    0.0440\\
9332.54051802 &    -6.8193 &     0.0010 &         -4.3569 &     0.0017 &   0.0113 & 9445.36272734 &    -6.7887 &     0.0015 &         -4.3678 &     0.0032 &   -0.0049\\
9332.66888067 &    -6.8300 &     0.0018 &         -4.3477 &     0.0033 &   0.0084 & 9446.37889092 &    -6.8744 &     0.0016 &         -4.3618 &     0.0037 &    0.0816\\
9336.61763016 &    -6.8406 &     0.0015 &         -4.3646 &     0.0031 &   0.0267 & 9447.39404862 &    -6.8444 &     0.0019 &         -4.3696 &     0.0048 &    0.0091\\
9338.43760288 &    -6.8476 &     0.0011 &         -4.3291 &     0.0020 &   0.0495 & 9448.38183570 &    -6.8347 &     0.0013 &         -4.3624 &     0.0027 &    0.0240\\
\hline
\end{tabular}

  }
  \label{tab:yo41rv2}
\end{table*}

\begin{table*}
  \caption{iSHELL RV measurements of TOI-1807}
    \begin{tabular}{l c c }
\hline
\hline
JD$-$2450000   &        RV  &   $\sigma_{\rm RV}$ \\
               &    (m~s$^{-1}$) &   (m~s$^{-1}$)  \\
\hline
      9014.835956239 &        -24.7 &          7.2 \\
      9016.809708646 &        -24.7 &          6.8 \\
      9021.801696615 &        -38.2 &          6.5 \\
      9022.801223677 &        -25.3 &          5.3 \\
      9255.144664628 &         50.9 &         11.3 \\
      9256.160684256 &         -6.5 &         10.5 \\
      9257.103522918 &        -40.9 &          7.3 \\
      9261.168549077 &        -52.4 &          9.6 \\
      9322.949459804 &        -51.6 &          7.1 \\
      9331.867578560 &        -38.4 &         11.0 \\
      9362.909338106$^*$ &         93.0 &         13.9 \\
      9363.866398826$^*$ &        145.3 &          9.0 \\
      9368.917426278$^*$ &        107.0 &          8.6 \\
      9369.863219064$^*$ &        123.6 &         11.8 \\
      9371.816766591$^*$ &         93.7 &          8.1 \\
      9376.921081772 &        -39.3 &          9.8 \\
      9379.854432749 &        -16.6 &          8.5 \\
      9380.852652085 &        -68.0 &         14.7 \\
      9591.139044181 &        -35.5 &         14.0 \\
      9592.159119182 &         -2.9 &         10.8 \\
      9605.085212320 &         -0.7 &          7.1 \\
      9606.087809338 &        -38.4 &          8.6 \\
\hline									       
\end{tabular}

  $^*$ Not used during planet and stellar activity modeling.
  \label{tab:ishell}
\end{table*}

\end{appendix}

%
%


\end{document}